\documentclass[reprint,showpacs,superscriptaddress,twocolumn,english,notitlepage,longbibliography,nofootinbib]{revtex4-1}
\usepackage[utf8]{inputenc}
\usepackage{color}
\usepackage{array}
\usepackage{verbatim}
\usepackage{multirow}
\usepackage{amsmath}
\usepackage{amssymb}
\usepackage{dsfont}
\usepackage{graphicx}
\usepackage{esint}
\usepackage[unicode=true,
 bookmarks=true,bookmarksnumbered=true,bookmarksopen=false,
 breaklinks=true,pdfborder={0 0 0},backref=false,colorlinks=true,linkcolor=blue]
 {hyperref}
\usepackage{enumitem}
\usepackage{titlesec}
\usepackage{cleveref}
\usepackage{longtable}
\usepackage{bbm}
\usepackage[normalem]{ulem}
\usepackage{wasysym}
\usepackage{bm}% bold math
\usepackage{siunitx}

\newcommand{\PreserveBackslash}[1]{\let\temp=\\#1\let\\=\temp}
\newcolumntype{C}[1]{>{\PreserveBackslash\centering}p{#1}}
\newcolumntype{R}[1]{>{\PreserveBackslash\raggedleft}p{#1}}
\newcolumntype{L}[1]{>{\PreserveBackslash\raggedright}p{#1}}

\usepackage{cleveref}
\crefname{equation}{Eq.}{Eqs.}
\crefname{figure}{Fig.}{Figs.}
\crefname{table}{Table}{Tables}
\crefname{section}{Section}{Sections}
\renewcommand{\paragraph}[1]{\vspace{0.2cm}{\bf \textit{#1}}}

\usepackage{cleveref}
\crefname{equation}{Eq.}{Eqs.}
\crefname{figure}{Fig.}{Figs.}
\crefname{table}{Table}{Tables}
\crefname{section}{Section}{Sections}

\def\beq#1\eeq{\begin{equation}#1\end{equation}}
\def\beqs#1\eeqs{\begin{align}#1\end{align}}

\def\kk{\mathbf{k}}
\def\qq{\mathbf{q}}

\def\GG{\mathbf{G}}
\def\QQ{\mathbf{Q}}

\makeatother

\begin{document}

\title{Twisted bilayer graphene I. Matrix elements, approximations, perturbation theory and a $k\cdot p$ 2-Band model}

\author{B. Andrei Bernevig \thanks{\href{bernevig@princeton.edu}{bernevig@princeton.edu}}}
\affiliation{Department of Physics, Princeton University, Princeton, New Jersey 08544, USA}
\author{Zhi-Da Song}
\affiliation{Department of Physics, Princeton University, Princeton, New Jersey 08544, USA}
\author{Nicolas Regnault}
\affiliation{Department of Physics, Princeton University, Princeton, New Jersey 08544, USA}
\affiliation{Laboratoire de Physique de l'Ecole normale superieure, ENS, Universit\'e PSL, CNRS,
Sorbonne Universit\'e, Universit\'e Paris-Diderot, Sorbonne Paris Cit\'e, Paris, France}
\author{Biao Lian}
\affiliation{Department of Physics, Princeton University, Princeton, New Jersey 08544, USA}

\begin{abstract}
We investigate the Twisted Bilayer Graphene (TBG) model of Bistritzer and MacDonald (BM) \cite{bistritzer_moire_2011} to obtain an analytic understanding of its energetics and wavefunctions needed for many-body calculations. We provide an approximation scheme for the wavefunctions of the BM model, which first elucidates why the BM $K_M$-point centered original calculation containing only $4$ plane-waves provides a good analytical value for the first magic angle ($\theta_M\approx 1^\circ$). The approximation scheme also elucidates why most of the  many-body matrix elements in the Coulomb Hamiltonian projected to the active bands can be neglected. By applying our approximation scheme at the first magic angle to a $\Gamma_M$-point  centered model of 6 plane-waves, we analytically understand the reason for the small $\Gamma_M$-point gap between the active and passive bands in the isotropic limit $w_0=w_1$. Furthermore, we analytically calculate the group velocities of the passive bands in the isotropic limit, and show that they are \emph{almost} doubly degenerate, even away from the $\Gamma_M$-point, where no symmetry forces them to be. Furthermore, moving away from the $\Gamma_M$ and $K_M$ points, we provide an explicit analytical perturbative understanding as to why the TBG bands are flat at the first magic angle, despite the first magic angle is defined by only requiring a vanishing $K_M$-point Dirac velocity. We derive analytically a connected ``magic manifold''  $w_1=2 \sqrt{1 + w_0^2} -\sqrt{2 + 3 w_0^2}$, on which the bands remain extremely flat as $w_0$ is tuned between the isotropic ($w_0=w_1$) and chiral ($w_0=0$) limits.  We analytically show why going away from the isotropic limit by making  $w_0$ less (but not larger) than $w_1$ increases the $\Gamma_M$- point gap between the active and the passive bands. Finally, by perturbation theory, we provide an analytic $\Gamma_M$ point $k\cdot p$ $2$-band model that reproduces the TBG band structure and eigenstates within a certain $w_0, w_1$ parameter range. Further refinement of this model are discussed, which suggest a possible faithful representation of the TBG bands by a $2$-band $\Gamma_M$ point $k\cdot p$ model in the full $w_0, w_1$ parameter range.

\end{abstract}

\date{\today}
\maketitle

%\tableofcontents

\section{\label{sec:introduction}Introduction}

The interacting phases in twisted bilayer  graphene (TBG) are one of the most important new discoveries of the last few years in condensed matter physics \cite{bistritzer_moire_2011,cao_correlated_2018,cao_unconventional_2018, lu2019superconductors, yankowitz2019tuning, sharpe_emergent_2019, saito_independent_2020, stepanov_interplay_2020, liu2020tuning, arora_2020, serlin_QAH_2019, cao_strange_2020, polshyn_linear_2019,  saito2020,das2020symmetry, wu_chern_2020,park2020flavour, xie2019spectroscopic, choi_imaging_2019, kerelsky_2019_stm, jiang_charge_2019,  wong_cascade_2020, zondiner_cascade_2020,  nuckolls_chern_2020, choi2020tracing, saito2020isospin,rozen2020entropic, lu2020fingerprints, burg_correlated_2019,shen_correlated_2020, cao_tunable_2020, liu_spin-polarized_2019, chen_evidence_2019, chen_signatures_2019, chen_tunable_2020, burg2020evidence, tarnopolsky_origin_2019, zou2018, fu2018magicangle, liu2019pseudo, Efimkin2018TBG, kang_symmetry_2018, song_all_2019,po_faithful_2019,ahn_failure_2019,Slager2019WL, lian2020, hejazi_multiple_2019, hejazi_landau_2019, padhi2020transport, xu2018topological,  koshino_maximally_2018, ochi_possible_2018, xux2018, guinea2018, venderbos2018, you2019,  wu_collective_2020, Lian2019TBG,Wu2018TBG-BCS, isobe2018unconventional,liu2018chiral, bultinck2020, zhang2019nearly, liu2019quantum,  wux2018b, thomson2018triangular,  dodaro2018phases, gonzalez2019kohn, yuan2018model,kang_strong_2019,bultinck_ground_2020,seo_ferro_2019, hejazi2020hybrid, khalaf_charged_2020,po_origin_2018,xie_superfluid_2020,julku_superfluid_2020, hu2019_superfluid, kang_nonabelian_2020, soejima2020efficient, pixley2019, knig2020spin, christos2020superconductivity,lewandowski2020pairing, xie_HF_2020,liu2020theories, cea_band_2020,zhang_HF_2020,liu2020nematic, daliao_VBO_2019,daliao2020correlation, classen2019competing, kennes2018strong, eugenio2020dmrg, huang2020deconstructing, huang2019antiferromagnetically,guo2018pairing, ledwith2020, repellin_EDDMRG_2020,abouelkomsan2020,repellin_FCI_2020, vafek2020hidden, fernandes_nematic_2020, Wilson2020TBG,wang2020chiral, ourpaper2,ourpaper3,ourpaper4,ourpaper5,ourpaper6}. The theoretical prediction that interacting phases would appear in this system was made based on the appearance of flat bands in the non-interacting Bistritzer-Macdonald (BM) Hamiltonian \cite{bistritzer_moire_2011}. This Hamiltonian is at the starting point of the understanding of every aspect of strongly correlated TBG (and other moir\'e systems) physics \cite{cao_correlated_2018,cao_unconventional_2018, lu2019superconductors, yankowitz2019tuning, sharpe_emergent_2019, saito_independent_2020, stepanov_interplay_2020, liu2020tuning, arora_2020, serlin_QAH_2019, cao_strange_2020, polshyn_linear_2019,  saito2020,das2020symmetry, wu_chern_2020,park2020flavour, xie2019spectroscopic, choi_imaging_2019, kerelsky_2019_stm, jiang_charge_2019,  wong_cascade_2020, zondiner_cascade_2020,  nuckolls_chern_2020, choi2020tracing, saito2020isospin,rozen2020entropic}. Remarkably, it even predicts quite accurately the so-called ``magic angles'' at which the bands become flat, and is versatile enough to accommodate the presence of different hoppings in between the $AA$ and the $AB$ stacking regions of the moir\'e lattice. The  BM Hamiltonian is in fact a large class of $k\cdot p$ models, which we will call BM-like models, where translational symmetry emerges at small twist angle even though the actual sample does not have an exact lattice commensuration. 

This paper is the first of a series of six papers on TBG \cite{ourpaper2,ourpaper3,ourpaper4,ourpaper5,ourpaper6}, for which we present a short summary here. In this paper, we investigate the spectra and matrix elements of the single-particle BM model by studying the $k\cdot p$ expansion of BM model at $\Gamma_M$ point of the moir\'e Brillouin zone. In TBG II~\cite{ourpaper2}, we prove that the BM model with the particle-hole (PH) symmetry defined in Ref.~\cite{song_all_2019} is always \emph{stable topological}, rather than fragile topological as revealed without PH symmetry \cite{po_origin_2018,song_all_2019,po_faithful_2019,ahn_failure_2019}. We further study TBG with Coulomb interactions in Refs.~\cite{ourpaper3,ourpaper4,ourpaper5,ourpaper6}. In TBG III~\cite{ourpaper3}, we show that the TBG interaction Hamiltonian projected into any number of bands is always a Kang-Vafek type \cite{kang_strong_2019} positive semi-definite Hamiltonian (PSDH), and generically exhibit an enlarged U(4) symmetry in the flat band limit due to the PH symmetry. This U(4) symmetry for the lowest 8 bands (2 per spin-valley) was previously shown in Ref.~\cite{bultinck_ground_2020}. We further reveal two chiral-flat limits, in both of which the symmetry is further enhanced into U(4)$\times$U(4) for any number of flat bands. The U(4)$\times$U(4) symmetry for the lowest 8 flat bands in the first chiral limit was first discovered in Ref.~\cite{bultinck_ground_2020}. With kinetic energy, the symmetry in the chiral limits will be lowered into U(4). TBG in the second chiral limit is also proved in TBG II~\cite{ourpaper2} to be a perfect metal without single-particle gaps \cite{Mora2019-graphene}. In TBG IV~\cite{ourpaper4}, under a condition called flat metric condition (FMC) which is defined in this paper (Eq.~\ref{eqn-condition-at-nuMT}), we derive a series of exact insulator ground/low-energy states of the TBG PSDH within the lowest 8 bands at integer fillings in the first chiral-flat limit and even fillings in nonchiral-flat limit, which can be understood as U(4)$\times$U(4) or U(4) ferromagnets. We also examine their perturbations away from these limits. In the first chiral-flat limit, we find exactly degenerate ground states of Chern numbers $\nu_C=4-|\nu|,2-|\nu|,\cdots, |\nu|-4$ at integer filling $\nu$ relative to the charge neutrality. Away from the chiral limit, we find the Chern number $0$ ($\pm1$) state is favored at even (odd) fillings. With kinetic energy further turned on, up to 2nd order perturbations, these states are intervalley coherent if their Chern number $|\nu_C|<4-|\nu|$, and are valley polarized if $|\nu_C|=4-|\nu|$. At even fillings, this agrees with the K-IVC state proposed in Ref.~\cite{bultinck_ground_2020}. At fillings $\nu=\pm1,\pm2$, we also predict a first order phase transition from the lowest to the highest Chern number states in magnetic field, which is supported by evidences in recent experiments \cite{nuckolls_chern_2020,choi2020tracing,saito2020,das2020symmetry, wu_chern_2020,saito2020isospin,rozen2020entropic}. In TBG V~\cite{ourpaper5}, we further derive a series of exact charge $0,\pm1,\pm2$ excited states in the (first) chiral-flat and nonchiral-flat limits. In particular, the exact charge neutral excitations include the Goldstone modes (which are quadratic). This allows us to predict the charge gaps and Goldstone stiffness. In the last paper of our series TBG VI~\cite{ourpaper6}, we present a full Hilbert space exact diagonalization (ED) study at fillings $\nu=-3,-2,-1$ of the projected TBG Hamiltonian in the lowest 8 bands. In the (first) chiral-flat and nonchiral-flat limits, our ED calculation with FMC verified that the exact ground states we derived in TBG IV~\cite{ourpaper4} are the only ground states at nonzero integer fillings. We further show that in the (first) chiral-flat limit, the exact charge $\pm1$ excitations we found in TBG V~\cite{ourpaper5} are the lowest excitations for almost all nonzero integer fillings. In the nonchiral case with kinetic energy, we find the $\nu=-3$ ground state to be Chern number $\pm1$ insulators at small $w_0/w_1$ (ratio of AA and AB interlayer hoppings, see Eq.~\ref{interlayermatrixelements1}), while undergo a phase transition to other phases at large $w_0/w_1$, in agreement with the recent density matrix renormalization group studies \cite{kang_nonabelian_2020,soejima2020efficient}. For $\nu=-2$, while we are restricted within the fully valley polarized sectors, we find the ground state prefers ferromagnetic (spin singlet) in the nonchiral-flat (chiral-nonflat) limit, in agreement with the perturbation analysis in Refs.~\cite{ourpaper4,bultinck_ground_2020}.

To date, most of our understanding of the BM-like models comes from numerical calculations of the flat bands, which can be performed in a momentum lattice of many moir\'e Brillouin zones, with a cutoff on their number. The finer details of the band structure so far seem to be peculiarities that vary with different twisting angles. However, with the advent of interacting calculations, where the Coulomb interaction is projected into the active, flat bands of TBG, a deeper, analytic understanding of the flat bands in TBG is needed. In particular, there is a clear need for an understanding of what quantitative and qualitative properties are not band-structure details. So far, the analytic methods have produced the following results: by solving a model with only $4$ plane waves (momentum space lattice sites, on which the BM is defined), Bistritzer and MacDonald \cite{bistritzer_moire_2011} found a value for the twist angle for which the Dirac velocity at the $K_M$ moir\'e point vanishes. This is called the magic angle. In fact, the full band away from the $K_M$ point is flat, a fact which is not analytically understood. A further analytic result is the discovery that, in a limit of vanishing $AA$-hopping, there are angles for which the band is \emph{exactly} flat. This limit, called the \emph{chiral limit} \cite{tarnopolsky_origin_2019}, has an extra chiral symmetry. However, it is not analytically known why the bands remain flat in the whole range of $AA$ coupling between the isotropic limit ($AA$=$AB$ coupling) and the chiral limit. We note that the realistic magic angle TBG is in between these two limits due to lattice relaxations \cite{Uchida_corrugation,Wijk_corrugation,dai_corrugation, jain_corrugation}. A last analytical result is the proof that, when particle-hole symmetry is maintained in the BM model \cite{song_all_2019}, the graphene active bands are topological \cite{po_origin_2018,song_all_2019,po_faithful_2019,ahn_failure_2019,lian2020,po_fragile_2018,cano_fragile_2018,Slager2019WL,kang_symmetry_2018}.

\begin{figure}
    \centering
    \includegraphics[width=\linewidth]{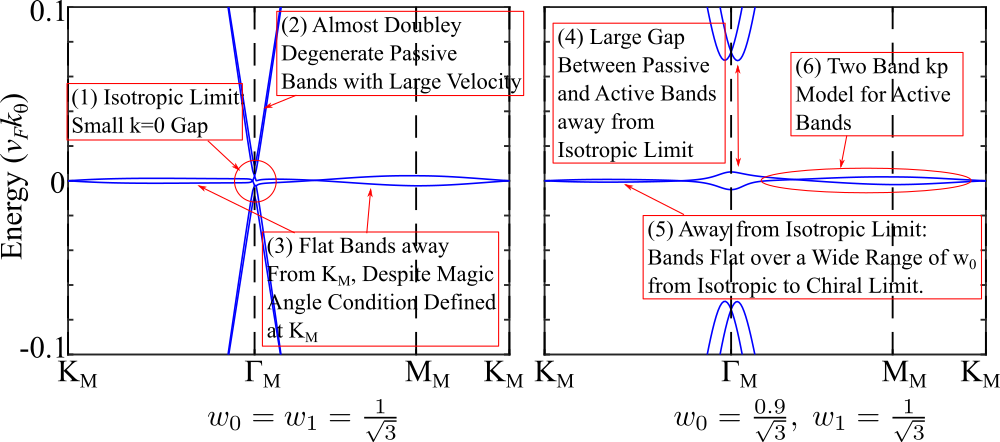}
    \caption{Several quantitative characteristics of the Bistritzer MacDonald model that require explanation. In particular, an analytic understanding of the  active band flatness is available only in the chiral limit $w_0=0$. However, the band is very flat far away from the chiral  limit. Several other features of the bands are pointed out.  }
    \label{fig:Questions1}
\end{figure}

This leaves a large series of un-answered questions. Rather than listing them in writing, we find it more intuitive to visualize the questions in a plot of the band-structure of TBG in the isotropic limit at the magic angle and away from it, towards the chiral limit. In Fig.~\ref{fig:Questions1}, we plot the TBG low-energy band structure in the moir\'e Brillouin zone, and the questions that will be answered in the current paper. To distinguish with the high symmetry points ($\Gamma,M,K,K'$) of the monolayer graphene Brillouin zone (BZ), we use a subindex $M$ to denote the high symmetry points ($\Gamma_M,M_M,K_M,K_M'$) of the moir\'e BZ (MBZ). Some salient feature of this band structure are: (1) In the isotropic limit, around the first magic angle,  it is hard to obtain two separate flat bands; it is hard to stabilize the gap to passive bands over a wide range of angles smaller than the first magic angle. In fact, Ref.~\cite{song_all_2019} computes the active bands separated regions as a function of twist angle, and finds a large region of gapless phases aroung the first magic angle. (2) The passive bands in the isotropic limit are \emph{almost} doubly degenerate, even away from the $\Gamma_M$-point, where no symmetry forces them to be. Moreover, their group velocities seem very high, i.e. they are very dispersive. (3) While the analytic calculation of the magic angle \cite{bistritzer_moire_2011} shows that the Dirac velocity vanishes in the isotropic limit at $AA$-coupling $w_0=1/\sqrt{3}$ (in the appropriate units, see below), it does not explain why the band is so flat even away from the Dirac point, for example on the $K_M -\Gamma_M-M_M-K_M$ line. (4) Away from the isotropic limit, while keeping $w_1=1/\sqrt{3}$, the gap between the active and passive bands increases immediately, while the bandwidth of the active bands does not increase. (5) The flat bands remain flat, over the wide range of $w_0 \in [0, 1/\sqrt{3}]$, from chiral to the isotropic limit. Also, our observation (6) in Fig.~\ref{fig:Questions1} shows that since the gap between the active and passive bands is large in the chiral limit compared to the bandwidth of active bands, a possible $k\cdot p$ Hamiltonian for the active bands might be possible. 

\begin{figure}
    \centering
    \includegraphics[width=0.7\linewidth]{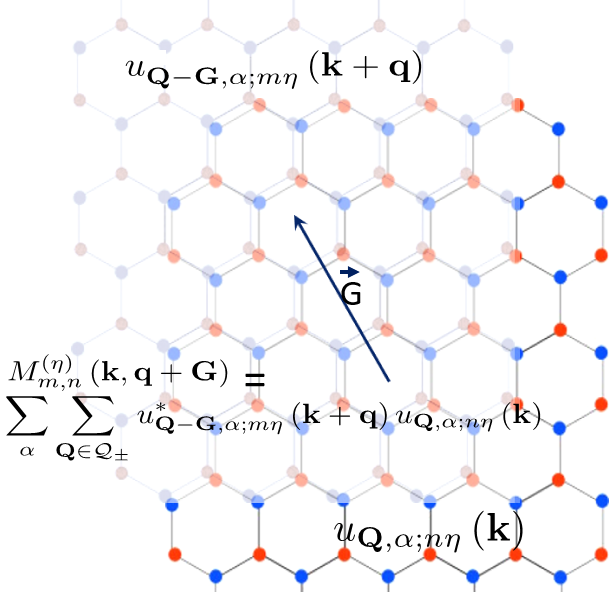}
    \caption{Matrix elements needed for the interacting problem. Specifically, the form factors  $M_{m,n}^{\left(\eta\right)}\left(\mathbf{k},\mathbf{q}+\mathbf{G}\right)=\sum_{\alpha}\sum_{{\mathbf{Q}\in\mathcal{Q}_{\pm}}} u_{\mathbf{Q}-\mathbf{G},\alpha;m\eta}^{*}\left(\mathbf{k}+\mathbf{q}\right)u_{\mathbf{Q},\alpha;n\eta}\left(\mathbf{k}\right)$ ,  of the Coulomb interaction are needed. They correspond to the overlap of the Bloch state  at momentum $\kk$, on the momentum lattice $\mathbf{Q}$,  $u_{\mathbf{Q},\alpha;n\eta}\left(\mathbf{k}\right)$ with the Bloch state at momentum $\qq+\kk$ on the momentum lattice $\mathbf{Q}+ \GG$, $u_{\mathbf{Q}-\mathbf{G},\alpha;m\eta}^{*}\left(\mathbf{k}+\mathbf{q}\right)$. Here $m,n$ are band indices, $\alpha=A,B$ is the graphene sublattice index, $\eta$ is the valley index, $\mathbf{G}$ is a reciprocal momentum, and $\mathbf{Q}$ is the honeycomb momentum lattice generated by the moir\'e reciprocal vectors shown in this figure.  }
    \label{fig:Questions2}
\end{figure}

A further motivation for the analytic investigation of the TBG Bistritzer-MacDonald model is to understand the behavior of the matrix elements $M_{m,n}^{\left(\eta\right)}\left(\mathbf{k},\mathbf{q}+\mathbf{G}\right)=\sum_{\alpha}\sum_{{\mathbf{Q}\in\mathcal{Q}_{\pm}}} u_{\mathbf{Q}-\mathbf{G},\alpha;m\eta}^{*}\left(\mathbf{k}+\mathbf{q}\right)u_{\mathbf{Q},\alpha;n\eta}\left(\mathbf{k}\right)$ as a function of $\GG$, which we call the \emph{form factor} (or \emph{overlap matrix}). These are the overlaps of different Bloch states in the TBG momentum space lattice (see Fig.~\ref{fig:Questions2}) and their behavior is important for the form factors of the interacting problem \cite{ourpaper3,ourpaper4}. These will be of crucial importance for the many-body matrix elements \cite{ourpaper2, ourpaper6} as well as for justifying the approximations made in obtaining exact analytic expressions for the many-body ground-states \cite{ourpaper4} and their excitations \cite{ourpaper5}.

We provide an analytic answer to all the above questions and observations. We will focus on the vicinity of the first magic angle. We first provide  an analytic perturbative framework in which to understand the BM model, and show that for the two flat bands around the first magic angle, only a very small number of momentum shells is needed. We justify our framework analytically, and check it numerically. This perturbative framework also shows that  $M_{m,n}^{\left(\eta\right)}\left(\mathbf{k},\mathbf{q}+\mathbf{G}\right)$ is negligible for $\GG$ more than $2$-times the moir\'e BZ (MBZ) momentum - at the first magic angle, irrespective of $\kk,\qq$. We then provide two approximate models involving a very small number of momentum lattice sites, the Tripod model ($K_M$ centered, also discussed in Ref.~\cite{bistritzer_moire_2011}), and a new, $\Gamma_M$ centered model. The Tripod model captures the physics around the $K_M$ point (but not around the $\Gamma_M$ point), and we show that the Dirac velocity vanishes when $w_1=1/\sqrt{3}$ irrespective of $w_0$. The $\Gamma_M$ centered model captures the physics around the $\Gamma_M$ point extremely well, as well as the physics around the $K_M$ point. Moreover, an approximation of the $\Gamma_M$ centered model with only 6 plane waves, which we call the Hexagon model, has an analytic $6$-fold exact degeneracy at the $\Gamma_M$ point in the isotropic limit $w_1=w_0=1/\sqrt{3}$, which is the reason for feature (1) in Fig.~\ref{fig:Questions1}. By performing a further perturbation theory in these $6$ degenerate bands away from the $\Gamma_M$ point, we obtain a model with an exact flatband at zero energy on the $\Gamma_M-K_M$ line, and almost flat bands on the $\Gamma_M-M_M$  line, answering (3) in Fig.~\ref{fig:Questions1}. In the same perturbative model, the velocity of the dispersive bands - which can be shown to be degenerate -  can be computed and found to be the same with the bare Dirac velocity (with some directional dependence), answering  (2) in Fig.~\ref{fig:Questions1}. Away from the isotropic limit, our perturbative model, which we still show to be valid for $w_0 \le w_1$ (but not for $w_0>> w_1$) allows for finding the analytic energy expressions at the $\Gamma_M$ point, and seeing a strong dependence on $w_0$ answering (3) in Fig.~\ref{fig:Questions1}. At the same time,  one can obtain \emph{all} the eigenstates of the Hexagon model at the $\Gamma_M$ point after tedious algebra, which can serve as the starting point of a perturbative $k\cdot p$ expansion of the $2$-active band Hamiltonians. With this, we provide an approximate $2$-band continuum model of the active bands, and find the manifold  $w_1(w_0)=2 \sqrt{1 + w_0^2} -\sqrt{2 + 3 w_0^2}$ with $w_0 \in [0, 1/\sqrt{3}]$, where the bandwidth of the active bands is the smallest, in this approximation.  The radius of convergence for the $k\cdot p$ expansion is great around the $\Gamma_M$-point but is not particularly good around the $K_M$ point for all $w_0, w_1$ parameters, but can be improved by adding more shells perturbatively, which we leave for further work. A series of useful matrix element conventions are also provided.

\section{New Perturbation Theory Framework for Low Energy States in $k\cdot p$ Continuum Models}\label{perturbationframework1}

\begin{figure}
\begin{centering}
\includegraphics[width=1\linewidth]{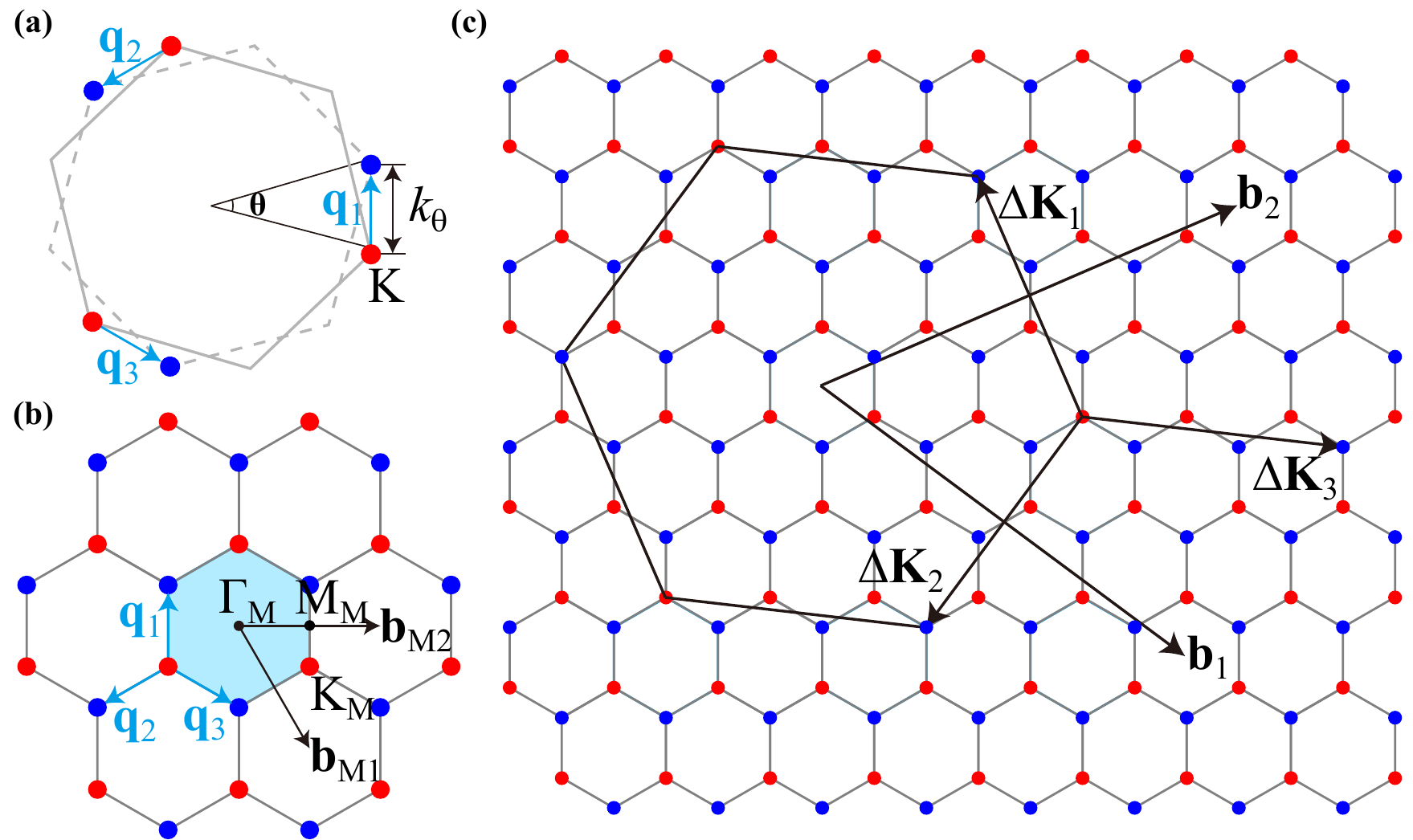}
\par\end{centering}

\protect\caption{\label{fig:MBZ}(a) The Brillouin Zones of two Graphene layers. 
The grey solid line and red dots represent the BZ and Dirac cones of top layer, and the grey dashed line and blue dots represent the
BZ and Dirac cones of the bottom layer. 
(b) The lattice formed by adding $\mathbf{q}_{1,2,3}$ iteratively. Red and blue circles represent
$\mathcal{Q}_{+}$ and $\mathcal{Q}_{-}$, respectively. (c) Relation
of graphene BZ and moir\'e BZ in commensurate case. 
Here we take the graphene BZ reciprocal vectors $\mathbf{b}_{1}=3{\mathbf{b}}_{M1}+2{\mathbf{b}}_{M2}$,
$\mathbf{b}_{2}=-2{\mathbf{b}}_{M1}+5{\mathbf{b}}_{M2}$.}
\end{figure}

In this section, we provide a general perturbation theory for the $k \cdot p$ BM-type Hamiltonians that exist in moir\'e lattices. We exemplify it in the TBG BM model,  but the general characteristics of this model allow this perturbation theory to be generalizable to other moir\'e system. The TBG BM Hamiltonian is defined on a momentum lattice of plane waves. Its symmetries and expressions have been extensively exposed in the literature (including in our paper \cite{ourpaper2}), and we only briefly mention them here for consitency. We first define $k_\theta=2|K|\sin(\theta/2)$ as the momentum difference between $K$ point of the lower layer and $K$ point of the upper layer of TBG, and denote the Dirac Fermi velocity of monolayer graphene as $v_F$. To make the TBG BM model dimensionless, we measure all the energies in units of $v_Fk_\theta$, and measure all the momentum in units of $k_\theta$. Namely, any quantity $E$ ($\kk$) with the dimension of energy (momentum) is redefined as dimensionless parameters 
\begin{equation}
E\rightarrow E/(v_Fk_\theta)\ ,\quad \kk\rightarrow \kk/k_\theta\ .
\end{equation}
We will then work with the dimensionless single particle Hamiltonian for the valley $\eta=+$, which in the second quantized form reads \cite{bistritzer_moire_2011,song_all_2019,ourpaper2}
\begin{equation}
\hat{H}_{0}^{\left(+\right)}=\sum_{\mathbf{k}\in\mathrm{MBZ}}\sum_{s\alpha\beta}\sum_{\mathbf{Q}\mathbf{Q}^{\prime}\in\mathcal{Q}_\pm}H_{\QQ\alpha,\QQ'\beta}\left(\mathbf{k}\right)c_{\mathbf{k},\mathbf{Q},+,\alpha s}^{\dagger}c_{\mathbf{k},\mathbf{Q}^{\prime},+,\beta s},
\end{equation}
where MBZ stands for moir\'e BZ, the momentum $\mathbf{k}$ is measured from the center ($\Gamma_M$ as shown in Fig. \ref{fig:MBZ}) point of the MBZ, $s=\uparrow,\downarrow$ is spin, and $\alpha,\beta$ denotes the 2 indices of $A,B$ sublattices. Here the dimensionless first quantized Hamiltonian $H_{\QQ\alpha,\QQ'\beta}\left(\mathbf{k}\right)$ is given by 
\beq\label{moireham1}
\begin{split}
H_{\QQ\alpha,\QQ'\beta}\left(\mathbf{k}\right)&=\delta_{\mathbf{Q},\mathbf{Q}^{\prime}}\left(\left(\mathbf{k}-\mathbf{Q}\right)\cdot\boldsymbol{\sigma}\right)_{\alpha\beta} \\& + \sum_{j=1}^3\left(\delta_{\mathbf{Q}-\mathbf{Q}^{\prime},\mathbf{q}_{j}}+\delta_{\mathbf{Q}^{\prime}-\mathbf{Q},\mathbf{q}_{j}}\right)(T_j)_{\alpha\beta},
\end{split}
\eeq
where
\begin{equation} 
\small T_j=w_{0}\sigma_{0}+w_{1}(\cos\frac{2\pi}{3}(j-1) \sigma_{x} + \sin\frac{2\pi}{3}(j-1)  \sigma_y), \label{interlayermatrixelements1}
\end{equation}
with $w_{0}$ being the interlayer $AA$-hopping and $w_{1}$ the interlayer
$AB$-hopping, $\bm{\sigma}=(\sigma_x,\sigma_y)$, and $\sigma_{0,x,y,z}$ stand for the identity and Pauli matrices in the 2-dimensional sublattice space. $\mathbf{k}$ takes value in MBZ, and $\mathbf{k=0}$
corresponds to the $\Gamma_M$ point in the moir\'e BZ. We define $\qq_1$ as the difference between the $K$ momentum of the lower layer of graphene and the rotated $K$ of the upper layer, and $\qq_2$ and $\qq_3$ as the $C_{3z}$ and $C_{3z}^{-1}$ rotations of $\qq_1$ (see Fig.~\ref{fig:MBZ}). The moir\'e reciprocal
lattice $\mathcal{Q}_{0}$ is then generated by the moir\'e reciprocal vectors ${\mathbf{b}}_{M1}=\mathbf{q}_{3}-\mathbf{q}_{1}$
and ${\mathbf{b}}_{M2}=\mathbf{q}_{3}-\mathbf{q}_{2}$, which contains the origin. We also define $\mathcal{Q}_{+}=\mathbf{q}_{1}+\mathcal{Q}_{0}$ and $\mathcal{Q}_{-}=-\mathbf{q}_{1}+\mathcal{Q}_{0}$ as the moir\'e reciprocal lattices shifted by $\mathbf{q}_{1}$ and $-\mathbf{q}_{1}$ respectively. $\QQ\in\mathcal{Q}_\pm$ is then in the combined momentum lattice $\mathcal{Q}_+ \oplus \mathcal{Q}_-$, which is a honeycomb lattice. For valley $\eta=+$, the fermion degrees of freedom $c_{\mathbf{k},\mathbf{Q},+,\alpha s}^{\dagger}$ with $\mathbf{Q}\in\mathcal{Q}_+$ and $\mathbf{Q}\in\mathcal{Q}_-$ are from layers $1$ and $2$, respectively. Since energy and momentum are measured in units of $v_F\kk_\theta$ and $k_\theta$, we have that $|\qq_i|=1$, and both $w_0$ and $w_1$ are dimensionless energies. It should be noticed
that, for infinite cutoff in the lattice $\mathbf{Q}$, we have $c_{\mathbf{k}+{\mathbf{b}}_{Mi},\mathbf{Q},\eta\alpha s}^{\dagger}=c_{\mathbf{k},\mathbf{Q}-{\mathbf{b}}_{Mi},\eta\alpha s}^{\dagger}\neq c_{\mathbf{k},\mathbf{Q},\eta\alpha s}^{\dagger}$, as proved in Refs.~\cite{song_all_2019,ourpaper2}. In practice, we always choose a finite cutoff $\Lambda_\mathbf{Q}$ for  $\mathbf{Q}$ ($\Lambda_\mathbf{Q}$ denotes the set of $\mathbf{Q}$ sites kept).

We note that in the Hamiltonian (\ref{moireham1}), we have adopted the zero angle approximation \cite{bistritzer_moire_2011,ourpaper2}, namely, we have approximated the Dirac kinetic energy $\mathbf{k}\cdot\boldsymbol{\sigma}_{\pm\theta/2}$ ($\pm$ for layers $1$ and $2$, respectively) as $\mathbf{k}\cdot\boldsymbol{\sigma}$, where $\boldsymbol{\sigma}_{\pm\theta/2}$ are the Pauli matrices $\boldsymbol{\sigma}$ rotated as a vector by angle $\pm\theta/2$ about the $z$ axis. With the zero angle approximation, the Hamiltonian (\ref{moireham1}) acquires a unitary particle-hole symmetry \cite{song_all_2019}, which is studied in details in another paper of us \cite{ourpaper2}. In the absence of the zero angle approximation, the particle-hole symmetry is only broken up to $1\%$ \cite{ourpaper2} near the first magic angle, and is exact in the (first) chiral limit $w_0=0$ \cite{wang2020chiral}. We also note that different variants of the TBG BM model exist in the literature, which further include  nonlocal tunnelings, interlayer strains or $\kk$ dependent tunnelings \cite{MacD2014Abinitio,carr2019_exact,koshino2020_effective,bi2019_designing}. However, we shall only focus on the BM model in Eq. (\ref{moireham1}) in this paper.

\emph{It is the cutoff $\Lambda_\mathbf{Q}$ that we are after}: we need to quantize what is the proper cutoff $\mathbf{Q}\in\Lambda_\mathbf{Q}$ in order to obtain a fast convergence of the Hamiltonian. We devise a perturbation theory which gives us the error of taking a given cutoff in the diagonalization of Hamiltonian in Eq.~\ref{moireham1}. For the first magic angle, we will see that this cutoff is particularly small, allowing for analytic results.

\subsection{Setting Up the Shell Numbering of the Momentum Lattice and Hamiltonian}

\begin{figure}
\begin{centering}
\includegraphics[width=1.0\linewidth]{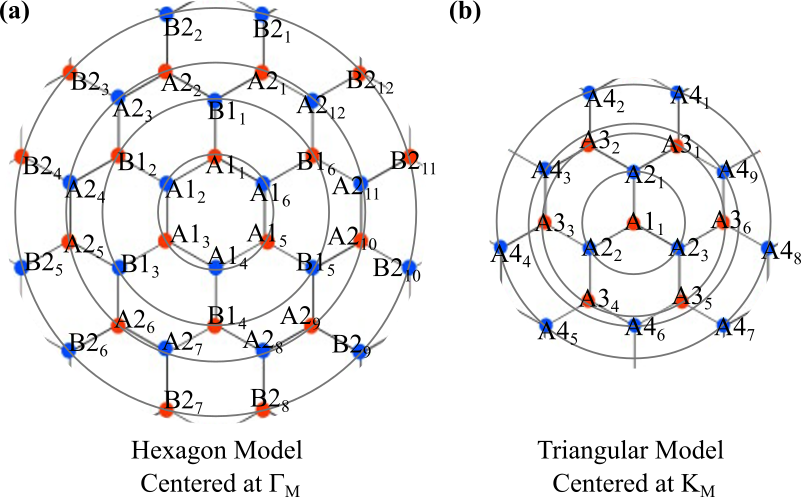}
\par\end{centering}
\protect\caption{Lattices centered around momentum $\kk_0$  on which one can calculate the TBG Hamiltonian.. a) The hexagon centered model ($\Gamma_M$-centered model, in which  we build ``shells'' by graph distance from the hexagon centered at the $\Gamma_M$ point. The circles denote the different shells, although going to larger graph distance will make the circles into hexagons. There are two different types of sub-shells in each shell, the $A$ and the $B$ sub-shells in this model. The $A$ shells connect to the $B$ shells, but the $A$ sites within a shells also contain hoppings within themselves. The $B$ sites hop only to $A$ sites.  b) The triangle centered  at the $K_M$- point model,  in which  we build ``shells'' by graph distance from the $K_M$-point centered at the origin. The circles denote the different shells, although going to larger graph distance will make the circles into triangles. There are only one type of shells, the $A$ shells in this model. The $A$ sites within a shell do not hop to other sites within each shell. \label{fig:MagicManifold}}
\end{figure}

We now consider the question of what momentum shell cutoff $\Lambda_\mathbf{Q}$ should we keep in performing a perturbation theory of the BM model. In effect, considering an infinite cutoff for the $\mathbf{Q}$ lattice,  we can build the BM model centered around any point $\mathbf{k}_0$ in the MBZ, by sending 
\begin{equation}\label{eq:shifted-kQ}
\kk \rightarrow \kk-\kk_0\ ,\qquad \QQ\rightarrow \QQ-\kk_0
\end{equation} in Eq.~\ref{moireham1}; however, it makes sense to pick $\kk_0$ as a high-symmetry point in the MBZ, and try to impose a finite cutoff $\Lambda_\mathbf{Q}$ in the shifted lattice $\mathbf{Q}$. Two important shifted lattices $\kk_0$ can be envisioned, see Fig.~\ref{fig:MagicManifold}. These lattices will be developed and analyzed in Sec.~\ref{StoryOfTwoLattices}; here we only focus on the perturbative framework of Eq.~\ref{moireham1}, which is the same for either of these two lattices (and in fact, on a lattice with any  $\kk_0$ center).

We introduce a numbering of the ``shells'' in momentum space $\QQ$ on this lattice.  In the $K_M$-centered lattice (Fig.~\ref{fig:MagicManifold}b) which is a set of Hexagonal lattices but centered at one of the ``sites'' (the $K_M$ point, corresponding to the choice $\kk_0=-\qq_1$), the sites of shells $n$ are denoted  $An_i$, with $n-1$ being the minimal graph distance (minimal number of bonds travelled on the honeycomb lattice from one site to another) from the center $A1_1$, while $i$ goes to the number of $\QQ$ sites with the same graph distance $n-1$. The truncation in $\QQ$ corresponds to a truncation in the graph distance $n-1$. In particular, with lattice $\QQ$ centered at the $K_M$ point, the momentum hopping $T_i$ in the BM Hamiltonian Eq.~\ref{moireham1} then \emph{only} happens between sites in two different shells $n \leftrightarrow n+1$ but not between sites in the same shell. The simplest version of this model, with a truncation at $n=2$, with sites  $A1_1$ and $A2_{1}, A2_{2}, A2_{3}$ was used by Bistritzer and MacDonald to show the presence of a ``magic angle''- defined as the angle for which the Dirac velocity vanishes. We call this the Tripod model. This  truncated  model (the Tripod model) does not respect the \emph{exact} $C_{2x}$ symmetry, although it becomes asymptotically good as more shells are added. The magic angle also does not explain analytically the flatness of bands, since it only considers the velocity vanishing at one point, $K_M$. However, the value obtained by BM \cite{bistritzer_moire_2011} for the  first magic angle is impressive: despite considering only two shells (4 sites), and despite obtaining this angle from the vanishing velocity of bands at only one point ($K_M$ in the BZ), the bands do not change much after adding more shells. Moreover they are flat throughout the whole BZ, not only around the $K_M$ point. The Dirac velocity also does not change considerably upon introducing more shells.

We now introduced a yet unsolved lattice, the $\Gamma_M$-centered model in Fig.~\ref{fig:MagicManifold}a, which corresponds to the choice $\kk_0=\mathbf{0}$ in Eq. (\ref{eq:shifted-kQ}). This model, which we call $\Gamma_M$-centered was not solved by BM, perhaps because of the larger Hilbert space dimension than the $K_M$-centered one. It however respects all the symmetries of the TBG (except Bloch periodicity, which is only fully recovered in the large cutoff $\Lambda_\mathbf{Q}$ limit) at any finite number of shells and not only in the large shell number limit. While not relevant for the perturbation theory described here, we find it useful to partition one shell $n$ in the $\Gamma_M$ centered lattice into two sub-shells $An$ and $Bn$, each of which has $6n$ sites. The first shell is $A1$ given by the 6 corners of the first MBZ; then we define $An$ as the shell with a minimal graph distance $2(n-1)$ to shell $A1$, and $Bn$ as the shell with a minimal graph distance $2n-1$ to shell $A1$. $An_i$ and $Bn_i$ where $i =1,\ldots, 6n$ is the index of sites in the sub-shell $An$ or $Bn$.  The partitioning in sub-shells is useful when we realize that the hopping $T_i$ in the BM Hamiltonian Eq.~\ref{moireham1} can only happen between $An$ and $Bn$ shells, between $Bn$ and $An+1$ shells, \emph{and within} an $An$ shell, but \emph{not} within the same $Bn$ shell. In App.~\ref{Appendix1} we provide an explicit efficient way of implementing the scattering matrix elements of the BM Hamiltonian Eq.~\ref{moireham1}, and provide a block matrix form of the BM Hamiltonian in the shell basis defined here.  Written compactly, the expanded matrix elements in App~.\ref{Appendix1}  read:

%TODO: Finish fixing the conventions for T and H_{\mathbf{k}}. -- not sure if it is worth it to switch from w_0 and w_1 to \alpha_0 and \alpha_1
%TODO: Define w
\begin{equation}
    (H_{An, An})_{\mathbf{Q}_1, \mathbf{Q}_2} = \begin{cases}
T_j  \text{ if }\mathbf{Q}_1 - \mathbf{Q}_2 = \pm \mathbf{q}_j \\
    0    \text{ otherwise}
    \end{cases}
\end{equation}
for the hopping terms, and similarly for $H_{An,Bn}$ where $\mathbf{Q}_1, \mathbf{Q}_2$ are the initial and final momenta in their respective shells. Finally for $\mathbf{k}-$dependent dispersion we take a linearized model:
\begin{equation}
    (H_{\mathbf{k}, An/Bn})_{\mathbf{Q}_1\mathbf{Q}_2} = (\mathbf{k} - \mathbf{Q}_1) \cdot \mathbf{\sigma} \delta_{\mathbf{Q}_1\mathbf{Q}_2}.\label{diagonal1}
\end{equation}
which is accurate in the small-angle low-energy approximations we make. Recall that the momentum is measured in units of  $k_\theta = 2 |K|\sin(\theta/2)$ with $\theta$ the twist angle, while the energy (and Hamiltonian matrix elements) are in units of $v_F k_\theta$. We may now write the  dimensionless BM Hamiltonian $H(\kk)$ in Eq.~\ref{moireham1} in block form as
\begin{align}
H &= 
\begin{pmatrix}
H_{\mathbf{k} A1}+ H_{A1,A1} &H_{A1,B1}          &0                             &\cdots \\
H_{A1,B1}^\dagger            &H_{\mathbf{k}B1}   &H_{B1, A2}                    &\cdots  \\
0                            &H_{B1, A2}^\dagger &H_{\mathbf{k}A2}+ H_{A2,A2}   &\ddots  \\
\vdots                       &0                  &\ddots                        &\ddots\\
\end{pmatrix} \nonumber \\
&\equiv \begin{pmatrix}
   M_1       & N_1 & 0 & 0 &  \dots & 0&0 \\
N_1^\dagger      & M_2 & N_2& 0 & \dots &0 & 0 \\
0     & N_2^\dagger & M_3& N_3 & \dots &0 & 0 \\
    \hdotsfor{7} \\
    0       & 0 & 0 & 0& \dots & M_{L-1} & N_{L-1}\\
        0       & 0 & 0 & 0& \dots & N^\dagger_{L-1} & M_{L}
\end{pmatrix} \label{perturbationHamiltonian1}
\end{align}
where $L$ is the shell cutoff that we choose. In the above equation, the $M, N$ block form of the matrix is a schematic, in the sense that both the $\Gamma_M$-centered model Fig.~\ref{fig:MagicManifold}a and the $K_M$-centered model Fig.~\ref{fig:MagicManifold}b can be written in this form, albeit with different $M_n,N_n$, $n=1,\ldots L$. Also, each $M_n$ depends on $\kk$, which for space purposes was not explicitly written in Eq.~\ref{perturbationHamiltonian1}.

\subsection{General Hamiltonian Perturbation for  Bands Close to Zero Energy with Ramp-Up Term}

In general, Eq.~\ref{perturbationHamiltonian1}, with generic matrices $M_i, N_i$  represents \emph{any} Hamiltonian with short range hopping (here on a momentum lattice), and not much progress can be made. However, for our BM-Hamiltonians, we \emph{know} several facts which render them special: 
\begin{itemize}

\item The Hamiltonian in Eq.~\ref{moireham1} has very flat bands, at close to zero energy $|E|\le 0.02 v_F k_\theta$. Numerically, the energy of the flat bands $\ll w_1$ and $w_0$, 
since numerically we know that the first magic angle happens at $w_1$ (or $w_0$) around $1/\sqrt{3}$.

\item The block-diagonal terms $M_n$ contain a ramping up diagonal term Eq.~\ref{diagonal1}, of eigenvalue $|\kk - \mathbf{Q}|$. The $\kk$ momentum runs in the first MBZ, which means that $|\kk| \le 1$. Since $Q$ for the $n$'th shell is proportional to $n$, higher order shells contribute larger terms to the diagonal of the BM Hamiltonian.

\end{itemize} We now show that, despite the higher shell diagonal terms being the largest in the BM Hamiltonian, they contribute exponentially little to the physics of the low energy (flat) bands. This should be a generic property of the moir\'e systems.

The $M_n, N_n$ Block Hamiltonian Eq.~\ref{perturbationHamiltonian1}  acts on the spinor wavefunction $(\psi_1, \psi_2, \psi_3, \ldots, \psi_{L-1}, \psi_L)$ where the $\Psi_n$'s are the components of the wavefunction on the shells $n=1,2,3, \ldots, L-1, L$, and $L$ is the cutoff shell. Notice that they likely have different dimensions: in the $\Gamma_M$-centered model, $\psi_1$ is a 12-dimensional spinor (6 vertices of the first Hexagon momentum $Q$ - for sub-shell $A1_i, i=1\ldots 6$- times $2$ for the $\alpha \beta$ indices ) $\psi_2$ is also a 12-dimensional spinor (6 legs coming out of the vertices of the first Hexagon momentum $\QQ$  - for sub-shell  $B1_i, i=1\ldots 6$- times $2$ for the $\alpha \beta$ indices),  $\psi_3$ is a 24-dimensional spinor (12 vertices of the momentum $\QQ$   - for sub-shell $A2_i, i=1\ldots 12$-  times $2$ for the $\alpha \beta$ indices) and $\psi_4$ is also a 24-dimensional spinor (12 legs coming out of the vertices of the previous momentum shell $\QQ$ -for sub-shell $B2_i, i=1\ldots 12$ times $2$ for the $\alpha \beta$ indices),  etc... To diagonalize $H$ we write down the action of $H$ in  Eq.~\ref{perturbationHamiltonian1} on the wavefunction $\psi = \left(\psi_1, \psi_2, \dots, \psi_L\right)$:
\begin{eqnarray}
& M_1\psi_1 + N_1 \psi_2= E\psi_1\nonumber \\  
& \ldots \nonumber \\  
& N_{n-1}^\dagger \psi_{n-1} + M_n \psi_n + N_n \psi_{n+1} = E \psi_n \nonumber \\  & \ldots \nonumber \\  
&N_{L-1}^\dagger \psi_{L-1} + M_L \psi_L= E \psi_L
\end{eqnarray}
and solve iteratively for $\psi_1$ starting from the \emph{last} shell. We find that
\begin{widetext}
\begin{align}
\psi_L &= (E- M_L)^{-1} N_{L-1}^\dagger \psi_{L-1} \nonumber \\ 
\psi_{L-1} &= (E- M_{L-1}- N_{L-1}(E-M_L)^{-1} N_{L-1}^\dagger)^{-1}  N_{L-2}^\dagger \psi_{L-2} \nonumber \\  \psi_{L-2} &= ( E- M_{L-2} - N_{L-2}  (E- M_{L-1}- N_{L-1}(E-M_L)^{-1} N_{L-1}^\dagger)^{-1} N_{L-2}^\dagger)^{-1} N_{L-3}^\dagger \psi_{L-3} \nonumber \\ 
& \ldots 
\end{align}
\end{widetext}

%Notice that $M_n$ is generically an invertable matrix with eigenvalues of the order $\pm n$ for the $n$'th shell: $M_n$ is just the momentum-dependent ramp-up term if the shell is of $B$ type and has the additional hopping term $H_{AnAn}$ if the shell is $A$ type. Because the magnitude of the momentum term increases linearly with $|\mathbf{k} - \mathbf{Q}| > |\mathbf{q}_1-\mathbf{q}_2|$ for momenta outside the first two shells, while the hopping term has constant magnitude, $H_{\mathbf{k} An}$ dominates. 

 We notice three main properties:
 \begin{itemize}
 
 \item $M_n \approx n $ for large shells $n\gg1$, is generically an invertable matrix with eigenvalues of the order $\pm n$ for the $n$-th shell. This is because $M_n$ is just the ramp-up term, block diagonal with the diagonal being $(\kk-\QQ)\cdot \bm{\sigma}$ for $Q$ in the $n$'th sub-shell of $B$ type; if the sub-shell is of $A$ type, then the matrix is still generically invertible, as it contains the diagonal term  $(\kk-\QQ)\cdot \bm{\sigma}$ plus the small (since $w_0, w_1 \approx 1/ \sqrt{3}$) hopping Hamiltonian $H_{An, An}$ (see App.~\ref{Appendix1}). Nonetheless, because the magnitude of the momentum term increases linearly with $|\mathbf{k} - \mathbf{Q}| \gg 1$ for momenta $\QQ$ outside the first two shells $n >2$, while the hopping term has constant magnitude, $H_{\mathbf{k} An}$ dominates the BM Hamiltonian. 

 \item Since we are interested in the flat bands  $E\approx 0$ ($E\approx 0.02$ in $v_F k_\theta$), we can expand in $E/M_n$ terms, especially after the first $n>2$ shells, and keep only the zeroth and first order terms. We use
 \beq
 (E- M)^{-1} \approx - M^{-1}- M^{-1} E M^{-1}
 \eeq if the eigenvalues of $E$ are smaller than those of $M$ $E\ll M$.
 
 \item For the first magic angle, the off-diagonal terms are also smaller than the diagonal terms, for the first magic angle, and for $|\QQ| \ge2$, we have that $N_{n-1}M_n^{-1}N_{n-1}^\dagger \ll 1$ for $n\ge 2$ and for $w_0,w_1 \approx 1/\sqrt{3}$ (more details on this will be given later).
 
   \end{itemize} 
With these approximations, we obtain that the general solution is 
\begin{equation}
    \psi_{n} = \left(E P_n - M_n+ R_n\right)^{-1} N_{n-1}^\dagger \psi_{n-1} \label{recurrencemacdonald}
\end{equation}
where $P_n$ is defined recursively as 
\begin{equation}\label{eq:P_recurrance}
    P_{L-n} = N_{L-n} M_{L-n+1}^{-1} P_{L-n+1} M_{L-n+1}^{-1} N_{L-n}^\dagger +1 
\end{equation}
subject to $P_L = 1$ and $R_n$ is
\begin{align}
    R_{L-n} = &N_{L-n} M_{L-n+1}^{-1} R_{L-n+1} M_{L-n+1}^{-1} N_{L-n}^\dagger \nonumber \\
    + &N_{L-n} M_{L-n+1}^{-1} N_{L-n}^\dagger \label{eq:M_recurrance}
\end{align}
with $R_{L}=0, R_{L-1}= N_{L-1} M_L^{-1} N_{L-1}^\dagger, P_L=1$. This continues until the first shell, where we have 
 \beq
 \psi_2=  [E P_2 - M_2+ R_2]^{-1} N_{1}^\dagger \psi_{1}
\eeq

\subsection{Form Factors and Overlaps from the General Perturbation Framework}
Notice that the wavefunction for the $E\approx 0$ bands decays \emph{exponentially} ($\psi_n \approx \frac{1}{n} \psi_{n-1}$) over the momentum space $\QQ$ as we go to larger and larger shells. This is due to the inverses in the linear ramp-up term $M_n \propto n$ of Eq.~\ref{recurrencemacdonald} (a consequence of the $\QQ$ term in Eq.~\ref{diagonal1}, This has immediate implications for the form factors. For example, in Refs.~\cite{ourpaper3,ourpaper4,ourpaper5} we have to compute

\begin{equation}
M_{m,n}^{\left(\eta\right)}\left(\mathbf{k},\mathbf{q}+\mathbf{G}\right)=\sum_{\alpha}\sum_{{\mathbf{Q}\in\mathcal{Q}_{\pm}}} u_{\mathbf{Q}-\mathbf{G},\alpha;m\eta}^{*}\left(\mathbf{k}+\mathbf{q}\right)u_{\mathbf{Q},\alpha;n\eta}\left(\mathbf{k}\right) \label{eq:form_factor}
\end{equation}
for $m,n$ the indices of the active bands, and for different $\GG\in\mathcal{Q}_0$. Notice that almost all $|\GG|\le |\QQ|$  change the shells (with the exception of $|\GG|=1$): if $\QQ$ is in the subshell $An/Bn$, while $\GG$ is of order $|\GG|\ge 2|\widetilde{\mathbf{b}}_1|$ with $\widetilde{\mathbf{b}}_1$ the moire reciprocal vector,  then $\QQ-\GG$ is \emph{not} in the subshell $An/Bn$. Hence, considering $|\QQ-\GG| >|\QQ|$ without loss of generality, we have, for $2|\widetilde{\mathbf{b}}_1| \le |\GG| \le |\QQ|$:
 \beq
 u_{\mathbf{Q}-\mathbf{G},\alpha;m\eta}^{*}\left(\mathbf{k}+\mathbf{q}\right) \le \frac{|\QQ|!}{|(\mathbf{Q}-\mathbf{G})|!}  u_{\mathbf{Q},\alpha;n\eta}^{*}\left(\mathbf{k}+\mathbf{q}\right) \label{decayofbands1}
 \eeq for any $m,n$. Since the wavefunctions of the active flat bands at (or close to) zero energy exponentially decay with the shell distance from the center we can approximate
 \begin{eqnarray}
&  M_{m,n}^{\left(\eta\right)}\left(\mathbf{k},\mathbf{q}+\mathbf{G}\right)\approx \sum_{\alpha}\sum_{{\mathbf{Q}\;  \text{or}\; \QQ-\GG \in An,\; Bn, \; n\le n_0 }} \nonumber \\ &u_{\mathbf{Q}-\mathbf{G},\alpha;m\eta}^{*}\left(\mathbf{k}+\mathbf{q}\right)u_{\mathbf{Q},\alpha;n\eta}\left(\mathbf{k}\right) \label{approximation2}
 \end{eqnarray} with $n_0$ a cutoff. For any $\kk,\qq$, the (maximum of any components of the) wavefunctions on the subshells  $A2, B2$ are of order $1/3!, 2!/4!$ times the components of the wavefunctions on the subshells  $A1, B1$. Hence we can restrict to small shell cutoff in the calculation of form factor matrices $n_0 =1$ (meaning only the subshells $A1, B1$ are taken into account), while paying at most a 15\% error.  Conservatively, we can keep $n_0=2$ and pay a much smaller error $<3$\%.

 Next, we ask for which $\GG$ momenta are the function $ M_{m,n}^{\left(\eta\right)}\left(\mathbf{k},\mathbf{q}+\mathbf{G}\right)$ considerably small. Employing Eq.~\ref{decayofbands1}, we see that $ M_{m,n}^{\left(\eta\right)}\left(\mathbf{k},\mathbf{q}+\mathbf{G}\right)$  falls off exponentially with increasing $\GG$, and certainly for $|\GG| >2 |\widetilde{\mathbf{b}}_1|$ they are negligible.  The largest contributions are for $\GG =0$ and  for $|\GG|=|\widetilde{\mathbf{b}}_1|$, i.e. for $\GG$ being one of the fundamental reciprocal lattice vectors. We hence make the approximation:
 
 \begin{eqnarray}
&  M_{m,n}^{\left(\eta\right)}\left(\mathbf{k},\mathbf{q}+\mathbf{G}\right)\approx \sum_{\alpha}\sum_{{\mathbf{Q}\;  \text{or}\; \QQ-\GG \in  A1,\; B1}} \nonumber \\ &u_{\mathbf{Q}-\mathbf{G},\alpha;m\eta}^{*}\left(\mathbf{k}+\mathbf{q}\right)u_{\mathbf{Q},\alpha;n\eta}\left(\mathbf{k}\right)(\delta_{\GG, 0} + \delta_{|\GG|, |\widetilde{\mathbf{b}}_1|})  \label{approximation2}
 \end{eqnarray}  
\begin{figure}[t]
\centering
\includegraphics[width=\linewidth]{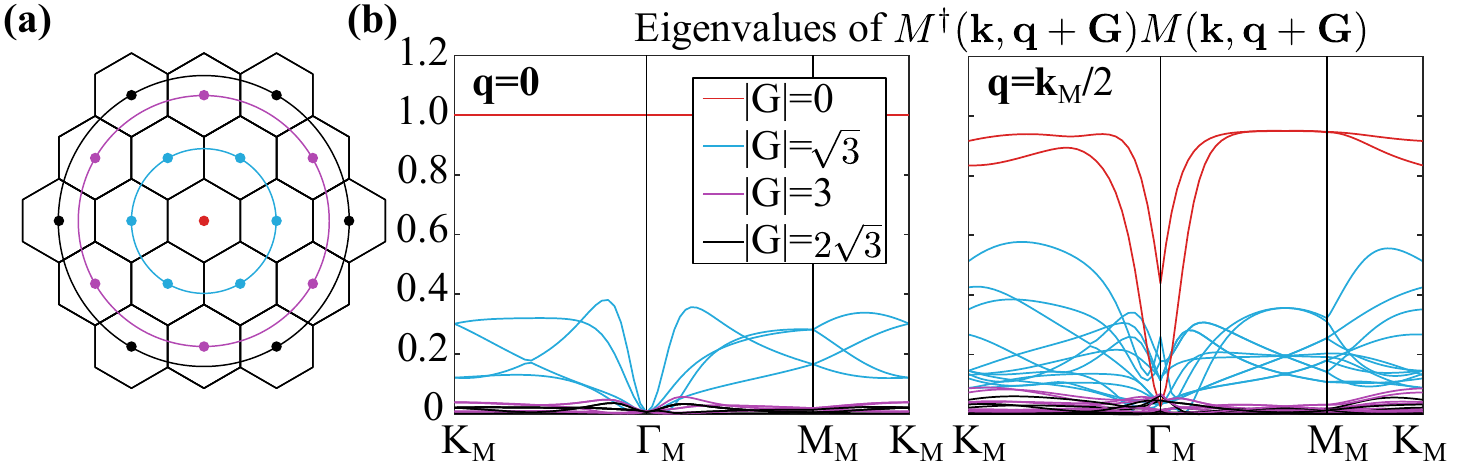}
\caption{The magnitude of the form factor (overlap matrix) $M^{(\eta=+)}(\mathbf{k},\mathbf{q+G})$, calculated for $w_0=0.4745$ and $w_1=0.5931$. 
(a) The colored dots are the $\mathbf{G}$ vectors we consider in $M^{(\eta=+)}(\mathbf{k},\mathbf{q+G})$. Different colors represent different length of $\mathbf{G}$.
(b) The eigenvalues of $M^{(\eta=+)\dagger}(\mathbf{k},\mathbf{q+G}) M^{(\eta=+)}(\mathbf{k},\mathbf{q+G})$ as functions of $\mathbf{k}$. In the left and right panels we choose $\mathbf{q=0}$ and $\mathbf{q}=\frac12\mathbf{k}_M$, respectively, where $\mathbf{k}_M$ is the $M_M$ momentum in the moir\'e BZ. 
}
\label{fig:MkqG}
\end{figure}

This is one of the most important results of our perturbative scheme. In Refs.~\cite{ourpaper3, ourpaper4,ourpaper5,ourpaper6} we employ heavily an approximation called the ``flat metric condition'' (see \cite{ourpaper5} for the link between this condition and the quantum metric tensor) to show that some exact eigenstates of the interacting Hamiltonian are in fact, ground-states. The flat metric condition requires that 

\begin{equation}\label{eqn-condition-at-nuMT}
\text{Flat Metric Condition:}\;\; M_{m,n}^{\left(\eta\right)}\left(\mathbf{k},\mathbf{G}\right)=\xi(\mathbf{G})\delta_{m,n}
\end{equation}  In light of our findings on the matrix elements Eq.~\ref{approximation2}, we see that the flat metric condition is satisfied for $|\GG| \ge 2 |\widetilde{\mathbf{b}}_1|$, as the matrix element vanishes $ M_{m,n}^{\left(\eta\right)}\left(\mathbf{k},\mathbf{G}\right) \approx 0 \rightarrow \xi(\mathbf{G})\approx 0$ for  $|\GG| \ge 2 |\widetilde{\mathbf{b}}_1|$. For $\GG=0$, the condition Eq.~\ref{eqn-condition-at-nuMT} is always satisfied, even without \emph{any} approximation Eq.~\ref{approximation2}, as it represents the block wavefunction orthonormality. Hence, the flat metric condition Eq.~\ref{eqn-condition-at-nuMT} is almost always satisfied, with one exception: the only requirement in the flat metric condition is $   M_{m,n}^{\left(\eta\right)}\left(\mathbf{k},\mathbf{G}\right)=\xi(\mathbf{G})\delta_{m,n}$ for $|\GG|= |\widetilde{\mathbf{b}}_1|$. There are $6$ $\GG$ vectors that satisfy this condition, namely $\GG =\pm \widetilde{\mathbf{b}}_1, \pm \widetilde{\mathbf{b}}_2, \pm (\widetilde{\mathbf{b}}_2- \widetilde{\mathbf{b}}_1)$. The overlaps are all related by symmetry.

In Fig. \ref{fig:MkqG}a we plot the eigenvalues at $\qq=0$ of the $M^\dagger M$ matrix. We see clearly that these eigenvalues are virtually negligible for $|\GG| \ge 2\tilde{\mathbf{b}}_i$, and that for $|\GG| = |\tilde{\mathbf{b}}_i|$ they are at most $1/3$ of the value for $|\GG|=0$. 

\subsection{Further Application of General Perturbation Framework to TBG}\label{approximationframework1}

While Eqs.~\ref{recurrencemacdonald} to~\ref{eq:M_recurrance} represent the general perturbation theory of Hamiltonians with a linear (growing) ramping term for  almost zero energy bands, we need further simplifications to practically apply them  to the TBG problem. However, the form of the $(\kk-\QQ)\cdot \sigma+ H_{An, An}$, which is not nicely invertible (although it can be inverted), and the form of $H_{Bn-1, An}$ (see App.~\ref{Appendix1} for the notation of these matrix elements), which is not diagonal, makes the matrix manipulations difficult, and un-feasible analytically for more than $2$ shells. Hence further approximations are necessary in order to make analytic progress.

First, we want to estimate the order of magnitudes of $P_{L-n}$ and $R_{L - n}$ terms in  Eqs.~\ref{eq:P_recurrance} and~\ref{eq:M_recurrance}. Recall that our energy is measured in units of $v_F k_\theta$, which for angle of $1^\circ$ is around $180\si{\milli\electronvolt}$. We note the following facts:
\begin{itemize}

\item The diagonal terms $H_{\mathbf{k}An}$ are of order $|n-|\kk||$, while the $H_{\mathbf{k}Bn}$ are of order $|n+1-|\kk||$ with $\mathbf{k}$ in the first Brillouin zone ($|\kk|<1$). Therefore, $H_{\mathbf{k} B1} \ge 1$, $H_{\mathbf{k} A2} > 1$, and all the other $H_{\mathbf{k}An}, H_{\mathbf{k}Bn}$ are considerably larger. This shows that $M_{n+1}$ in Eq. (\ref{perturbationHamiltonian1}) is of order $n$, due to the dominance of the momentum term in relation to the hopping terms.

\item $H_{An Bn}$ and $H_{Bn-1 An}$ are proportional to $T_j$, so they are of order $\alpha = w_{1}/(v_F k_\theta)$. Near the first magic angle ($\theta \approx 1^\circ$, or $w_{1}\approx 1/\sqrt{3}$ in units of $v_F k_\theta$), $\alpha \approx 0.6 / \theta$ with the angle in degrees (hence \emph{smaller angles have larger $\alpha$}). By Eq. (\ref{perturbationHamiltonian1}), this means the matrices $N_n\sim H_{Bn An+1}$ are of order $\alpha$.

\end{itemize}

These facts allow us to estimate $P_n$ in Eq. (\ref{eq:P_recurrance}):
%TODO: Check to make sure this equation is right it seems far too good. Messed up the factors somewhere
\begin{align}
    P_{n} &\propto |N_{n}|^2 |M_{n+1}|^{-2}|P_{n+1}| + 1 \nonumber \\
    &\propto (v_Fk_\theta)^2 \alpha^2 (v_F k_\theta n)^{-2}|P_{n+1}| + 1 \nonumber \\
    &= \alpha^2 n^{-2}|P_{n + 1}| + 1
\end{align}
For $n \geq 2$ therefore $P_n = 1$ up to a correction term no more than $\alpha^2 n^{-2} < 0.1$. Therefore we are justified (up to a 10\% error) of neglecting all $P_n$,  $n \geq 2$   terms. 
%This simplification arises due to the dominance of the momentum term in relation to the hopping terms.
Similarly, using these estimates and substituting into  $R_{ n}$ in Eq.~\ref{eq:M_recurrance}, we see that
\begin{align}
    |R_{n}| &\leq  \frac{\alpha^2}{(n+1)^{2}} |R_{n + 1}| + \frac{(v_F k_\theta)\alpha^2}{(n+1)} \nonumber \\
    &\leq 0.04 |R_{n + 1}| + 0.09 (v_F k_\theta)
\end{align}
when $n \geq 2$ at the first magic angle $\alpha \approx0.6$. Again this will allow us to neglect the $R_n$ term for $n \geq 2$. 

This means that shells after the first one can be neglected at the first magic angle. More generally, only the first $N$ shells will be needed for understanding the $N\textsuperscript{th}$ magic angle.

In order to see the validty of the above approximations more concretely, it is instructive to write down the 2-shell ($A1, B1, A2, B2$)  Hamiltonian explicitly, and estimate the contribution of the second shell. $A1$ and $B1$ are $12$-dimensional Hilbert spaces while $A2$ and $B2$ are $24$-dimensional Hilbert spaces, see App.~\ref{Appendix1}. Further shells are only a generalization of the ones below. We write the eigenvalue equation:
\begin{widetext}
\begin{eqnarray}
&(H_{kA1}+ H_{A1,A1} ) \psi_{A1} +  H_{A1,B1} \psi_{B1}= E\psi_{A1} \nonumber \\
& H_{A1,B1}^\dagger \psi_{A1} +H_{kB1} \psi_{B1} + H_{B1, A2} \psi_{A2} = E\psi_{B1} \nonumber \\
& H_{B1,A2}^\dagger \psi_{B1} + (H_{kA2}+ H_{A2,A2} ) \psi_{A2} + H_{A2,B2} \psi_{B2}= E\psi_{A2} \nonumber \\
& H_{A2,B2}  \psi_{A2}+H_{kB2} \psi_{B2}= E\psi_{B2} %\nonumber 
\end{eqnarray}
We integrate out from the outer shell to the first, to obtain the equations:
\begin{eqnarray}
&(H_{kA1}+ H_{A1,A1} ) \psi_{A1} +  H_{A1,B1} \psi_{B1}= E\psi_{A1}, \nonumber \\
& H_{A1,B1}^\dagger \psi_{A1} +( H_{kB1} + H_{B1, A2}( E-  (H_{kA2}+ H_{A2,A2} ) - H_{A2,B2}(E-H_{kB2} )^{-1}  H_{A2,B2}^\dagger   )^{-1}  H_{B1, A2}^\dagger  )\psi_{B1}  = E\psi_{B1}
\end{eqnarray} 
and to finally obtain:
\beq \label{eq:H-2shell-folding}
\begin{split}
E&\psi_{A1}=[H_{kA1}+ H_{A1,A1}  + \\
&H_{A1,B1}  (E- H_{kB1} - H_{B1, A2}( E-  (H_{kA2}+ H_{A2,A2} ) 
- H_{A2,B2}(E-H_{kB2} )^{-1}  H_{A2,B2}^\dagger   )^{-1} H_{B1, A2}^\dagger   )^{-1} H_{A1,B1}^\dagger ] \psi_{A1}.
\end{split}
\eeq
\end{widetext}
 Solving the above equation would give us the eigenstate energies, as well as the reduced eigenstate wavefunctions $\psi_{A1}$. However, even for two shells above, this is not analytically solvable, hence further approximations are necessary. We implement our approximations here.

\begin{itemize}

\item  First, focusing on the first magic angle of $1^\circ$, from numerical calculations we know that the energy of the active bands $|E| < 60\si{\milli\electronvolt}\approx 0.3 v_F k_\theta$. Hence $EH_{\mathbf{k} B1}^{-1} < 0.3$ and furthermore $E H_{\mathbf{k} Bn}^{-1}, E H_{\mathbf{k} An}^{-1} < 0.3n^{-1}$ for $n \geq 2$. 
%The fact that $|E H_{\mathbf{k} Bn}^{-1}|,\ |E H_{\mathbf{k} An}^{-1}| < 0.3n^{-1}$ for $n \geq 2$ 
This justifies  the approximation around the first magic angle:
\begin{eqnarray}\label{eq:E-Happ1}
&(E- H_{kB1})^{-1}=  - H_{kB1}^{-1}-  E H_{kB1}^{-2},
\end{eqnarray}
and
\begin{eqnarray}\label{eq:E-Happ2}
(E- H_{k(A,B)n})^{-1}=  - H_{k(A,B)n}^{-1}-  E H_{k(A,B)n}^{-2}
\end{eqnarray}  
for $n\ge2$. \emph{Region of validity of this approximation:} this approximation is independent on $w_{0},w_1$, the inter-layer tunneling. It however, depends on $\theta$ as well as on the energy range of the bands we are trying to approximate. For example, for $\theta=0.3^\circ$, an energy range $|E|\le60$meV would mean that  $|E/v_F k_\theta| \le  1$. This gives $|E H_{\mathbf{k} Bn}^{-1}|, |E H_{\mathbf{k} An}^{-1}| < n^{-1}$ and hence we would only be able to neglect shells larger than $n=3$.  In particular, in order to obtain convergence for bands of energy $E$ at angle $\theta$, we can neglect the shells at distance $n=2+ [E/ v_F k_\theta]$ (where $x$ means the integer part of $x$). Hence, as the twist angle is decreased, and if we are interested in obtaining convergent results for bands at a fixed energy, we will need to increase our shell cutoff to obtain a faithful representation of the energy bands. If we keep the number of shells fixed, we will obtain faithful (meaning in good agreement with the infinite cutoff limit) energies only for bands in a smaller energy window as we decrease the twist angle. Notice that this approximation does not depend on $w_{0}, w_1$ and hence  it is \emph{not} an approximation in the inter-layer coupling. 

\item The second approximation is regarding $w_{0},w_1$: because $\alpha = w_1/v_F k_\theta  \approx 0.6 $ at the first magic angle, we can do a perturbation expansion in the powers of  $\alpha$. We remark that $H_{\kk,Bn},H_{\kk,An}\sim n\gg\alpha$ for $n\ge 2$ and $\theta =1^\circ$. We also remark that  $H_{\kk,B1}^{-1}\alpha \le 0.6$ for all $\kk$ in the first $BZ$ (the largest value, $H_{K_M,B1}^{-1}\alpha = 0.6$ is reached for $\kk$ at the $K_M$ corner of the moir\'e BZ). As such, we find terms of the following form scale as 
%can make the following approximations:
\begin{eqnarray}\label{eq:E-Happ3}
H_{AnBn} H_{k{A,B}n}^{-1} H_{AnBn}^\dagger\sim \alpha^2 n^{-1},\  (n\ge 2) \nonumber \\
H_{Bn-1An} H_{k{A,B}n}^{-1} H_{Bn-1An}^\dagger\sim \alpha^2 n^{-1},\  (n\ge 2) \nonumber \\
H_{B1A1} H_{k{B}1}^{-1} H_{B1A1}^\dagger\sim \alpha^2\ .\qquad\qquad\qquad
\end{eqnarray}
%\begin{widetext}
%\begin{eqnarray}
%&(1- H_{AnBn} H_{k{A,B}n}^{-2} H_{AnBn}^\dagger)^{-1}= 1+ H_{AnBn} H_{k{A,B}n}^{-2} H_{AnBn}^\dagger,\;\; (n\ge 2) \nonumber \\ 
%&(1- H_{BnAn+1} H_{k{A,B}n}^{-2} H_{BnAn+1}^\dagger)^{-1}= 1+ H_{AnBn} H_{k{A,B}n}^{-2} H_{AnBn}^\dagger,\;\; (n\ge 2) \nonumber \\ 
%&(1- H_{B1A1} H_{k{B}1}^{-2} H_{B1A1}^\dagger)^{-1}= 1+ H_{A1B1} H_{k{B}1}^{-2} H_{A1B1}^\dagger.
%\end{eqnarray}
%\end{widetext}

\end{itemize}

With Eqs.~(\ref{eq:E-Happ1})-(\ref{eq:E-Happ3}), one can see that in Eq. (\ref{eq:H-2shell-folding}), the leading order contributions of the terms involving the second shell ($A2,B2$) are roughly $\sim |H_{A1,B1}|^2 |H_{kB1}|^{-2} |H_{B1, A2}|^2|H_{kA2}|^{-1} \sim \alpha^4/2 \sim 0.05$. It is hence a relatively good approximation to neglect shells higher than $n=1$ for angle $\theta=1^\circ$. For example, at the $K_M$ point, neglecting the $n=2$ shell will induce a less than $10\%$ percent error .
%(the order of magnitude of  %$H_{B1A2} H_{k{A}2}^{-2} H_{B1A2}^\dagger \approx 0.6^2/3$.$\alpha^4$). 
\emph{Region of validity of this approximation:} Notice that as the twist  angle is decreased, $\alpha$ increases. In general, the relative error of the $n$-th shell is roughly $H_{Bn-1An} H_{k{A}n}^{-2} H_{Bn-1An}^\dagger \sim \alpha^2/n^2$, so we can neglect the shells for which $n\gg\alpha$ where $\gg$ should be considered twice the value of $\alpha$. Hence, for an angle of $0.5^\circ$ ($\alpha =1.2$) we can neglect  all shells greater than $3$, etc. For angle $1/n$ of the first magic angle we can neglect all shells above $n+1$.

All the above remarks, which were made for the $\Gamma_M$ centered model, can also be extended to the $K_M$-centered model in Fig.~\ref{fig:MagicManifold}b. In particular, the Tripod model in  Fig.~\ref{MagicManifold2}b, containing only the $A1, A2$ shells, is a good approximation to the infinite model around the Dirac point, giving the correct first magic angle. 

\subsection{Further Approximation of the $1$-Shell ($A1, B1$) Hamiltonian in TBG}\label{approximationframework2}
 
In the previous section we claimed that, remarkably,  a relatively good approximation of the low energy BM model can be obtained by taking a  cutoff of 1 shell, where we only consider the \emph{first} $A$ sub-shell and the first $B$ sub-shell. The eigenvalue equations are 
\begin{eqnarray}
    &(H_{\mathbf{k}A1}+ H_{A1,A1} ) \psi_{A1} +  H_{A1,B1} \psi_{B1} = E\psi_{A1} \nonumber \\
    &H_{A1,B1}^\dagger \psi_{A1} +H_{\mathbf{k}B1} \psi_{B1} = E\psi_{B1} \label{2shell},
\end{eqnarray}
which can be solved for $\psi_{B1}$ to obtain
\begin{equation}
    \psi_{B1} = (E - H_{\mathbf{k} B1})^{-1}H^\dagger_{A1, B1} \psi_{A1}.
\end{equation}
%Plugging this back into the first Eq.~\ref{2shell} 
Eliminating $\psi_{B1}$ we find the eigenvalue equation for the first $A$ shell (which includes the coupling to the first $B$ shell):
\begin{equation}
    (H_{\mathbf{k} A1}+ H_{A1,A1} + H_{A1,B1}  (E- H_{\mathbf{k} B1})^{-1}H_{A1,B1}^\dagger ) \psi_{A1} = E\psi_{A1}.
\end{equation}
This is a $12\times 12$ non-linear eigenvalue equation in $E$. At this point we will make a few assumptions in order to simplify the eigenvalue equation. In particular, we would like to make this a linear matrix eigenvalue equation. Since we are interested close to $E = 0$ we may assume that $E \ll H_{\mathbf{k} B1}$. This allows us to treat the $B$ shell perturbatively, obtaining
\begin{equation}
    (H_{\mathbf{k} A1}+ H_{A1,A1} - H_{A1,B1}  H_{\mathbf{k} B1}^{-1}H_{A1,B1}^\dagger) \psi_{A1} = E\psi_{A1}.
\end{equation}
Our approximation Hamiltonian is
\begin{equation} \label{Happrox1}
    H_{\text{Approx} 1}(\kk) = H_{\mathbf{k} A1}+ H_{A1,A1} - H_{A1,B1}  H_{\mathbf{k} B1}^{-1}H_{A1,B1}^\dagger.
\end{equation} We note that  $  H_{\text{Approx} 1}(\kk)$ is a further perturbative Hamiltonian for the $n=1$ shell ($A1, B1$). For $\kk$ small, around the $\Gamma_M$-point, we expect this to be an excellent approximation of the $n=1$ shell Hamiltonian (and since the $n=1$ shell is a good approximation of the infinite shell, then  $ H_{\text{Approx} 1}(\kk) $ is expected to be an excellent approximation of the full $BM$ Hamiltonian close to the $\Gamma_M$ point). The good approximation is expected to deteriorate as $\kk$ gets closer to the boundary of the MBZ, since $H_{A1,B1}  H_{\mathbf{k} B1}^{-1}H_{A1,B1}^\dagger$ increases as $\kk$ approaches the MBZ boundary. This is because $  H_{\mathbf{k} B1}^{-1}$  has larger  terms as $\kk$ approaches the MBZ boundary. However, we expect still moderate qualitative agreement with the BM Hamiltonian.  We also predict that taking $2$ shells ($A1,B1,A2,B2$) would give an extremely good approximation to the infinite shell BM model.

\subsection{Numerical Confirmation of Our Perturbation Scheme}

\begin{figure}
\begin{centering}
\includegraphics[width=1.0\linewidth]{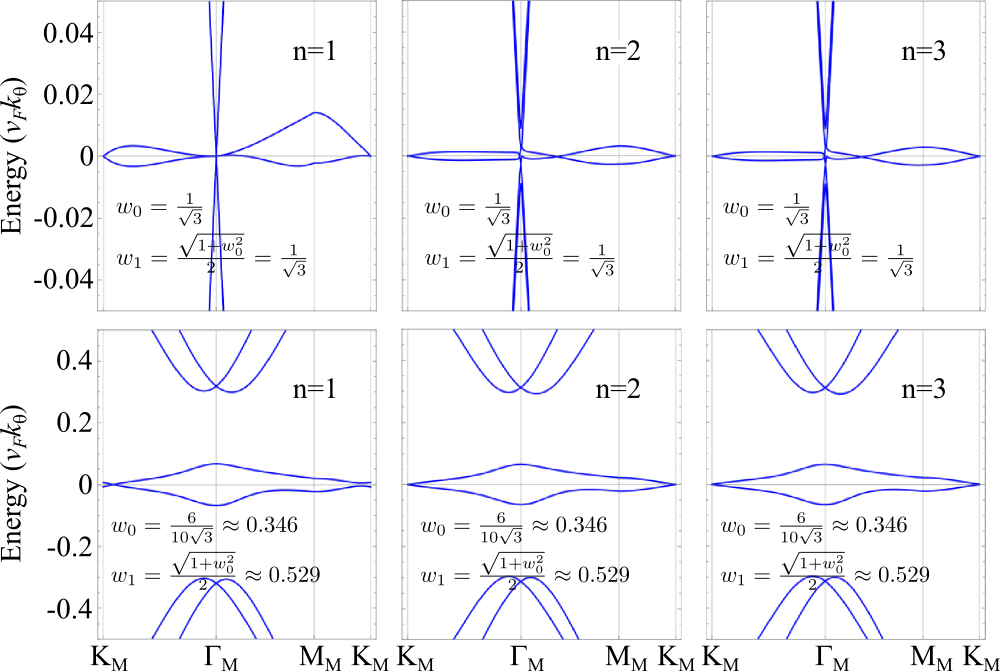}
\par\end{centering}

\protect\caption{\label{NumericalConfirmation1} Comparison of the different cutoff shells of the BM model in Eq.~\ref{moireham1}, for two values of $w_0, w_1$.  (more data available in App.~\ref{Appendix2}). We clearly see that $n=2$ has reached the infinite cutoff limit (the band structure does not change from $n=2$ and $n=3$, while $n=1$ (only one shell, $A1, B1$ subshells) shows excellent agreement around  the $\Gamma_M$  point, and good agreement even away from the $\Gamma_M$ -point (for example see the second row). }
\end{figure}

The  series of approximations performed in Secs.~\ref{approximationframework1} and~\ref{approximationframework2} are thoroughly numerically verified  at length in App. \ref{Appendix2}. We here present only a small part of the  highlights. In Fig.~\ref{NumericalConfirmation1} we present the $n=1, 2,3$ shells (one shell is made out of $A,B$ sub=shells) results of the BM Hamiltonian in Eq.~\ref{moireham1}, for two values of $w_0, w_1$. We virtually see no change between $2$ and $3$ shells (see also App~.\ref{Appendix2}) we verify this for higher shells and for many more values of $w_0, w_1$, around - and away from, within some manifolds $(w_0,w_1)$ explained in Sec.~\ref{StoryOfTwoLattices} - the magic angle. Hence our perturbation framework works well, and confirms the irrelevance of the $n>2$ shells. The $n=1$ shell band structure in Fig.~\ref{NumericalConfirmation1}, while in excellent agreement to the $n=2$ shells around the $\Gamma_M$ point, contains some quantitative differences from the $n=2$ shell (equal to the infinite cutoff) away from the $\Gamma_M$ point. However, the generic aspects of the band structure - low bandwidth, almost exact degeneracy (at $n=1$, becoming exact with machine precision in the $n>2$) at the $K_M$ point are still present even in the $n=1$ case, as our perturbative framework predicts in Secs.~\ref{approximationframework1} and~\ref{approximationframework2}.

\begin{figure}
\begin{centering}
\includegraphics[width=1\linewidth]{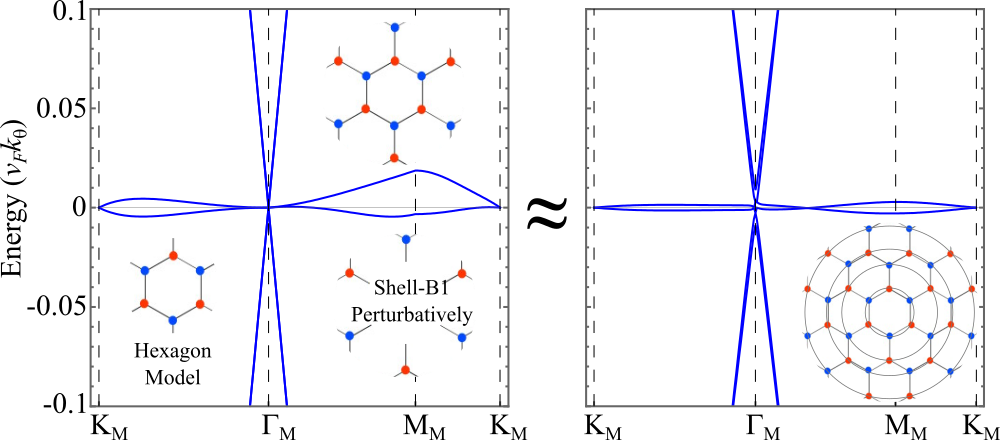}
\par\end{centering}

\protect\caption{\label{fig:HexagonalModelApproximation}  Band structure of the approximation $H_{\text{Approx} 1}(\kk)$ to the $1$-shell Hamiltonian, versus the infinite limit approximation,  for the $w_0=w_1=1/\sqrt{3}$ magic point. The $n=1$-shell Hamiltonian band structure is undistinguishable from   $H_{\text{Approx} 1}(\kk)$, and is plotted in the App.~\ref{Appendix2}. }

\end{figure}

\begin{figure}
\begin{centering}
\includegraphics[width=1.0\linewidth]{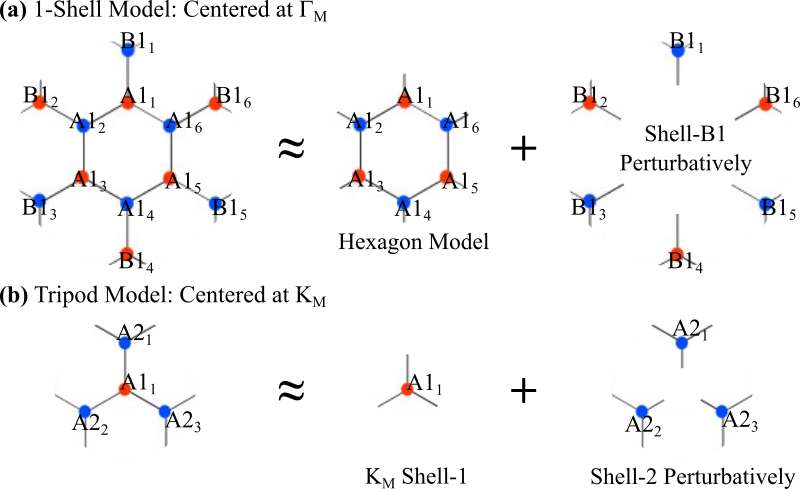}
\par\end{centering}
\protect\caption{\label{MagicManifold2}  The two types of approximate models used for analytics. (a) The  shell 1 (A1, B1) model which we have theoretically argued and numerically substantiated to represent a good approximation for values $w_0, w_1 \le 1/\sqrt{3}$. Analytically, we will first solve it by perturbation theory around the Hexagon model, which involves  only the $A1$ sites. The shell $B1$ will be added perturbatively to obtain $H_{\text{Approx} 1}(\kk)$ in  Eq.~\ref{Happrox1}. (A second way to solve for this Hamiltonian will be presented later)  (b) The Tripod model, which involves the two shells $A1$ (also known as the $K_M$ point) and $A2$. Due to the same considerations as for the $\Gamma_M$-centered model, this should be a good approximation for the infinite shell model for $w_0, w_1 \le 1/\sqrt{3}$. This is the same model as solved by Bistritzer and MacDonald \cite{bistritzer_moire_2011}. We find that the ``magic angle'' at which the Dirac velocity vanishes at the $K_M$ point is given by $w_1=1/\sqrt{3},\;\;  \forall w_0$.  }
\end{figure}

 Our approximations of the $n=1$ shell Hamiltonain in Sec.~\ref{approximationframework2} have brought us to the perturbative   $H_{\text{Approx} 1}(\kk) $ in Eq.~\ref{Happrox1}. Around the first magic angle, we claim that this  Hamiltonian is a good approximation to the band structure of the $n=1$ shell, especially away from MBZ boundary. The $n=1$ shell is only a 15\% difference on the $n=2$ shell and that the $n=1$ shell is within $5\%$ of the thermodynamic limit, we then make the approximation that $H_{\text{Approx} 1}$ explains the band structure of TBG within about 20\%.  The approximations are visually presented in  Fig.~\ref{MagicManifold2}a, and the band structure of the approximation $H_{\text{Approx} 1}$ to the $1$-shell Hamiltonian is presented in Fig.~\ref{fig:HexagonalModelApproximation}. We see that around the $\Gamma_M$-point, the Hamiltonian $H_{\text{Approx} 1}(\kk) $ in Eq.~\ref{Happrox1} has very good  match to the BM Hamiltonian  Eq.~\ref{moireham1}, while away from the $\Gamma_M$ point the qualitative agreement - small bandwidth, crossing at (close to) $K_M$ (the crossing is at $K_M$ for the infinite shell cutoff by symmetry, but can deviate slightly from $K_M$ for finite cutoff).

In App.~\ref{Appendix2} we present many different tests which confirm all aspects of our perturbative framework, different twist angles and $AA$, $AB$ coupling. We test the $n=1,2,3,4,\ldots$ shells, and also further test the validity of the approximation  $H_{\text{Approx} 1}(\kk)$ to the $n=1$ shell Hamiltonian in Sec.~\ref{approximationframework2}.

\section{Analytic Calculations on the BM Model: Story of Two Lattices}\label{StoryOfTwoLattices}

We will now analytically study the approximate Hamiltonian in Eq.~\ref{Happrox1}. While in Secs.~\ref{approximationframework1} and~\ref{approximationframework2} we have focused on the $\Gamma_M$-centered lattice, the same approximations can be made in the $K_M$ centered lattice, where the $H_{\text{Approx} 1}(\kk)$ changes to $ H_{\text{Approx} 1}(\kk) = H_{\mathbf{k} A1} + H_{A1,A2}  H_{\mathbf{k} A2}^{-1}H_{A1,A2}^\dagger$. The two types of approximations are schematically in Fig.~\ref{MagicManifold2} in the $\Gamma_M$  and $K_M$-centered lattice. First, we start with the Tripod model Fig.~\ref{MagicManifold2}b, to extend the Bistritzer-Macdonald calculation of the magic angle in the isotropic limit and find a ``first magic manifold'', where the Dirac velocity vanishes in the Tripod model (and is very close to vanishing in the infinite shell BM model. We then solve the 1-shell $\Gamma_M$ -centered model Fig.~\ref{MagicManifold2}a, defined by Eq.~\ref{Happrox1} which is supposed to faithfully describe TBG at and above the magic angle, as proved in Sec.~\ref{perturbationframework1}. This is a $12\times 12$ Hamiltonian, with no known analytic solutions, formed by shell 1: $A1, B1$, where the $B$ part of the first shell, $B1$ is taken into account perturbatively, as  $H_{A1,B1}  H_{kB1}^{-1}H_{A1,B1}^\dagger $.

\subsection{The $K_M$-centered ``Tripod Model'' and the First Magic Manifold}\label{tripodmodelfirstmagicmanifold}

For completeness we solve for the magic angle in the model in the  $K_M$-centered Model of Fig.~\ref{fig:MagicManifold} by taking only $4$-sites, one in shell $A1$ and $3$ in shell $A2$. We call this approximation, depicted in Fig.~\ref{MagicManifold2}b,  the Tripod model. This model is identical to the one solved by Bistritzer and Macdonald in the  isotropic limit. However, we will solve for the Dirac velocity away from the isotropic limit, to find a manifold $w_1(w_0)$ where the Dirac velocity vanishes. The tripod Hamiltonian $H_{\text{Tri}} (\mathbf{k}, w_0, w_1)$, with $\kk$ measured from  the $K_M$ point, reads

\begin{eqnarray}
&H_{\text{Tri}}(\mathbf{k}, w_0, w_1)=  \\ &
\begin{pmatrix}
\kk\cdot \sigma &T_1(w_0, w_1) &T_2(w_0, w_1)&T_3(w_0, w_1) \\
T_1(w_0, w_1) & (\kk-\qq_1)\cdot \sigma  &0 & 0 \\
 T_2(w_0, w_1) &0    &(\kk-\qq_2)\cdot \sigma &0   \\
   T_3(w_0, w_1) & 0  &0 & (\kk-\qq_3)\cdot \sigma   \end{pmatrix}\nonumber
\end{eqnarray} 
The Schrodinger equation in the basis $(\psi_{A1_1}, \psi_{A2_1},\psi_{A2_2},\psi_{A2_3})$ reads:
\begin{eqnarray}
    &\mathbf{k} \cdot \mathbf{\sigma} \psi_{A1_1} + \sum_{i=1,2,3} T_i(w_0,w_1) \psi_{A2_i} = E \psi_{A1_1} \\
    &T_i \psi_{A1_1}+ (\mathbf{k}-\mathbf{q}_i)\cdot \mathbf{\sigma} \psi_{A2_i} = E \psi_{A2_i},\;\; i=1,2,3.
\end{eqnarray}
from the second equation we find $\psi_{A2_i} = (E- (\kk-\qq_i)\cdot \sigma_i)^{-1} T_i \psi_{A1_1}$ and plug it into the first equation to obtain:
\begin{align}
&E \psi_{A1_1} = \mathbf{k}\cdot \mathbf{\sigma} \psi_{A1_1} + \sum_{i=1}^3 T_i \frac{E+ (\mathbf{k}-\mathbf{q}_i) \cdot \sigma}{E^2- (\mathbf{k}-\mathbf{q}_i)^2} T_i \psi_{A2_i}   \nonumber
\\&\approx \mathbf{k}\cdot \mathbf{\sigma} \psi_{A1_1} - \sum_{i=1}^3 T_i (E+ (\mathbf{k}-\mathbf{q}_i) \cdot \mathbf{\sigma})(1 + 2 \mathbf{k} \cdot \mathbf{q}_i) T_i  \psi_{A2_i}
\end{align}
where we neglect $E^2$ as small and expand the denominator to first order in $\mathbf{k}$ to focus on momenta near the $K_M$ Dirac point. Keeping only first order terms in $E, \mathbf{k}$ (not their product as they are both similarly small), and using that $|q_i|=1, \; \forall i=1,2,3$, we find
\begin{equation}
    (1-3 w_1^2) \kk \cdot \sigma  \psi_{A1_1} = (1+ 3(w_0^2+ w_1^2))E \psi_{A1_1}. \label{tripodeq2}
\end{equation}
and hence we find that the Dirac velocity vanishes on a  manifold of $w_0, w_1$ given by $w_1=\frac{1}{\sqrt{3}}$ and $\forall w_0$, which we call the first magic manifold. 
%\begin{eqnarray}
%&K_M-\text{centered Tripod model, }\nonumber \\ & \text{ Dirac velocity at $K_M$ vanishes} :\nonumber \\ &  w_1=\frac{1}{\sqrt{3}},\;\; \forall \; w_0 \end{eqnarray}
The angle for which the Dirac velocity vanishes at the $K_M$ point  is hence not a magic angle but a magic manifold. However, a further restriction needs to be imposed: $w_0$ cannot be too large, since from our approximation scheme in Secs.~\ref{approximationframework1} and~\ref{approximationframework2},  if $w_0 \gg 1/\sqrt{3}$, the Tripod model would not be a good approximation for the BM model with a  large number of shells; hence we restrict to $w_0 \le 1/\sqrt{3}$, and define 
\begin{eqnarray}
%& \text{ Dirac velocity at $K_M$ vanishes:} \nonumber \\ &\boxed{ 
\textbf{First Magic Manifold: } \quad w_0\le w_1=\frac{1}{\sqrt{3}}.
%, \;w_0 \le \frac{1}{\sqrt{3}}
\end{eqnarray}
The Tripod model, Fig.~\ref{fig:MagicManifold}b,  in which we found the first magic manifold, does not respect the \emph{exact} $C_{2x}$ symmetry of the lattice, although it becomes asymptotically accurate as the number of shells increases. The magic angle also does not explain analytically the flatness of bands, since it only considers the velocity vanishing at one point. However, the value obtained by BM for the magic angle is impressive; despite considering only four sites and the $K_M$ point, the bands do not change much after adding more shells, and they are flat throughout the whole Brillouin zone, not only around the $K_M$ point. 
%However, this calculation gives the physics around only one point, $K_M$ in the moire BZ where the band is flat (zero second derivative). 
Why is the entire band so flat at this value? We answer this question by examining the $\Gamma_M$-centered model below.

\subsection{The $\Gamma_M$-centered Hexagon Model and the Second Magic Manifold}\label{HexagonModel}

In Sec.~\ref{approximationframework2}, we introduced a yet unsolved approximate model  $H_{\text{Approx} 1}(\kk)$  in Eq.~\ref{Happrox1}, the $\Gamma_M$-centered model in Fig.~\ref{fig:MagicManifold}a. This model respects all the symmetries of TBG, and we have showed in App.~\ref{Appendix2} that it represents a good approximation to the infinite cutoff limit.  As we can see in Fig.~\ref{fig:TBLGBSShells2}, the band dispersions of the $n = 1$ shell model is very similar to that of $n = 2$. After $n = 2$ shells the difference to the infinite cutoff band structure is not visible by eye. 

 An analytic solution for the  $12\times 12$ Hamiltonian $H_{\text{Approx} 1}(\kk)$  in Eq.~\ref{Happrox1} is \emph{not possible} at every $\kk$. We hence separate the Hamiltonian into $H_{\text{Hex}}(\mathbf{k}, w_0, w_1) = H_{kA1}+ H_{A1,A1}$, then treat the smaller part $ H_{A1,B1}  H_{kB1}^{-1}H_{A1,B1}^\dagger$ perturbatively, for $w_0, w_1 \le \sqrt{3}$.  We will try to solve the first (largest) part of   $H_{\text{Approx} 1}(\kk)$: the $A1$ shell model- $H_{\text{Hex}}(\mathbf{k}, w_0, w_1) = H_{kA1}+ H_{A1,A1}$-  which we call the Hexagon model:

\begin{widetext}
\begin{equation}
H_{\text{Hex}}(\kk, w_0, w_1)= 
\begin{pmatrix}
(\kk-\qq_1)\cdot \sigma &T_2(w_0, w_1) &0  & 0 & 0 &T_3(w_0, w_1) \\
T_2(w_0, w_1) & (\kk+\qq_3)\cdot \sigma  & T_1(w_0, w_1)   & 0&  0&0 \\
  0 &T_1(w_0, w_1)    &(\kk-\qq_2)\cdot \sigma &T_3(w_0, w_1)  &0& 0  \\
   0 & 0  &T_3(w_0, w_1)& (\kk+\qq_1)\cdot \sigma &  T_2(w_0, w_1)  &0  \\
0 & 0  &0 &T_2(w_0, w_1)  & (\kk-\qq_3)\cdot \sigma &T_1(w_0, w_1)  \\
T_3(w_0, w_1) & 0  & 0 & 0 & T_1(w_0, w_1) &(\kk+\qq_2)\cdot \sigma 
\end{pmatrix} \label{HexagonModel1}
\end{equation} 
\end{widetext} This is still a $12\times 12$ Hamiltonian and its eigenstates cannot be analytically obtained at general $\kk$. In particular, it is also not illuminating to focus on a $12\times 12$ Hamiltonian when we want to focus on the physics of the  $2$ active bands and the low-energy physics of the dispersive passive bands. As such we make a series of approximations, which also elucidate some of the questions posed in Fig.~\ref{fig:Questions1}. We first analytically find a set of bands which can act as a perturbation theory treatment.

\subsubsection{Energies of the Hexagon Model at $\mathbf{k}=0$ for arbitrary $w_0$, $w_1$}

The only momentum where the Hexagon model $H_{\text{Hex}}(\kk, w_0, w_1)$ can be solved is the $\Gamma_M$ point. This is fortunate, as this point preserves all the symmetries of TBG, and is a good starting point for a perturbative theory. We find the 12 eigenenergies of $H_{\text{Hex}}(\kk=0, w_0, w_1)$ given in Tab. \ref{tab:6-fold-eigenvalue}. 

\begin{table}[htbp]
\centering
\begin{tabular}{c|c|c|c}
\hline
Band & Energy at $\mathbf{k=0}$ for any $w_0,w_1$ & $w_0=w_1=\frac{1}{\sqrt{3}}$ & dege.  \\
\hline
$E_1$ & $2 w_1 - \sqrt{1+ w_0^2}$ & $0$ & $1$  \\
\hline
$E_2$ & $-2 w_1 + \sqrt{1+ w_0^2}$ & $0$ & $1$  \\
\hline
$E_{3,4}$ & $-\frac{1}{2} (\sqrt{4+ w_0^2} - \sqrt{9 w_0^2+ 4w_1^2})$ & $0$ & $2$  \\
\hline
$E_{5,6}$ & $\frac{1}{2} (\sqrt{4+ w_0^2} - \sqrt{9 w_0^2+ 4w_1^2})$ & $0$ & $2$  \\
\hline
$E_{7,8}$ & $-\frac{1}{2} (\sqrt{4+ w_0^2} + \sqrt{9 w_0^2+ 4w_1^2})$ & $-\sqrt{13/3}$ & $2$  \\
\hline
$E_{9,10}$ & $\frac{1}{2} (\sqrt{4+ w_0^2} + \sqrt{9 w_0^2+ 4w_1^2})$ & $\sqrt{13/3}$ & $2$  \\
\hline
$E_{11}$ & $-2 w_1 - \sqrt{1+ w_0^2}$ & $-4/\sqrt{3}$ & $1$  \\
\hline
$E_{12}$ & $2 w_1 + \sqrt{1+ w_0^2}$ & $4/\sqrt{3}$ & $1$  \\
\hline
\end{tabular}
\caption{Eigenvalues of the Hexagon model in Eq. (\ref{HexagonModel1}) at $\Gamma_M$ point ($\kk=\mathbf{0}$). The values for general $w_0,w_1$ and for $w_0=w_1=\frac{1}{\sqrt{3}}$ are given, and dege. is short for degeneracy.}\label{tab:6-fold-eigenvalue}
\end{table}

%\begin{widetext}
%\begin{eqnarray}
%&E_1(\kk= 0, w_0, w_1) = 2 w_1 - \sqrt{1+ w_0^2}, \;\;\; \text{ 1-fold degenerate} \nonumber \\
%&E_2(\kk= 0, w_0, w_1) =-  2 w_1 + \sqrt{1+ w_0^2}, \;\;\; \text{ 1-fold degenerate} \nonumber \\
%& E_{3,4}(\kk= 0, w_0, w_1) =  -\frac{1}{2} (\sqrt{4+ w_0^2} - \sqrt{9 w_0^2+ 4w_1^2}), \;\;\; \text{ 2-fold degenerate} \nonumber \\
%&E_{5,6}(\kk= 0, w_0, w_1) = \frac{1}{2} (\sqrt{4+ w_0^2} - \sqrt{9 w_0^2+ 4w_1^2}), \;\;\; \text{ 2-fold degenerate} \nonumber \\
%&E_{7,8}(\kk= 0, w_0, w_1) =  -\frac{1}{2} (\sqrt{4+ w_0^2} + \sqrt{9 w_0^2+ 4w_1^2}), \;\;\; \text{ 2-fold degenerate} \nonumber \\&
%E_{9,10}(\kk= 0, w_0, w_1) =  \frac{1}{2} (\sqrt{4+ w_0^2} + \sqrt{9 w_0^2+ 4w_1^2}), \;\;\; \text{ 2-fold degenerate} \nonumber \\
%&E_{11}(\kk= 0, w_0, w_1) = -2 w_1-\sqrt{1+ w_0^2}, \;\;\; \text{ 1-fold degenerate} \nonumber \\
%&E_{12}(\kk= 0, w_0, w_1) = 2 w_1+ \sqrt{1+ w_0^2}, \;\;\; \text{ 1-fold degenerate} 
%\end{eqnarray} 
%\end{widetext}

By analyzing these energies as a function of $w_0, w_1$, we can answer  the question (1) in  Fig.~\ref{fig:Questions1} and give arguments for question (3) in  Fig.~\ref{fig:Questions1}. Numerically, at (and around) the first magic angle - which as per the Tripod model is defined as $w_1= 1/\sqrt{3}$ - and in the isotropic limit $w_0=w_1$, the system exhibits very flat active $2$-bands, not only around the $K_M$ point but everywhere in the MBZ. It also exhibits a very small gap (sometimes non-existent) between the active bands and the passive bands, around the values $w_0 = w_1=1/\sqrt{3}$. The Hexagon model  $H_{\text{Hex}}(\kk, w_0, w_1)$ explains both these observations. We find that the eigenenergies of $H_{\text{Hex}}(\mathbf{k}=0,  w_0= 1/\sqrt{3} , w_1= 1/\sqrt{3} )$ - in the isotropic limit -  are given in the third column of Tab. \ref{tab:6-fold-eigenvalue}. 
%\begin{eqnarray}
%&E_1(\mathbf{k=0}, w_0= \frac{1}{\sqrt{3}} , w_1= \frac{1}{\sqrt{3}} ) =0, \;\;\; \text{ 1-fold } \nonumber \\
%&E_2(\mathbf{k=0},  w_0= \frac{1}{\sqrt{3}} , w_1= \frac{1}{\sqrt{3}} ) =0, \;\;\; \text{ 1-fold }\nonumber \\
%&E_{3,4}(\mathbf{k=0},  w_0= \frac{1}{\sqrt{3}} , w_1= \frac{1}{\sqrt{3}}) = 0, \;\;\; \text{ 2-fold } \nonumber \\
%&E_{5,6}(\mathbf{k=0},  w_0= \frac{1}{\sqrt{3}} , w_1= \frac{1}{\sqrt{3}})  =  0, \;\;\; \text{ 2-fold } \nonumber \\
%&E_{7,8}(\mathbf{k=0},  w_0= \frac{1}{\sqrt{3}} , w_1= \frac{1}{\sqrt{3}} )  =  -\sqrt{13/3}, \;\;\; \text{ 2-fold } \nonumber \\&
%E_{9,10}(\mathbf{k=0},   w_0= \frac{1}{\sqrt{3}} , w_1= \frac{1}{\sqrt{3}} )  =  \sqrt{13/3}, \;\;\; \text{ 2-fold } \nonumber \\
%&E_{11}(\mathbf{k=0}, w_0= \frac{1}{\sqrt{3}} , w_1= \frac{1}{\sqrt{3}}  )  = -4/\sqrt{3}, \;\;\; \text{ 1-fold} \nonumber \\
%&E_{12}(\mathbf{k=0},  w_0= \frac{1}{\sqrt{3}} , w_1= \frac{1}{\sqrt{3}} )  = 4/\sqrt{3} \;\;\; \text{ 1-fold }  \label{EnergiesAtMagicPoint}
%\end{eqnarray}
Remarkably, in the isotropic limit $w_0=w_1$, and at the first magic angle $w_1=1/\sqrt{3}$, the bands at the $\Gamma_M$ point are $6$-fold degenerate at energy $0$. The two active bands are degenerate with the two passive bands above them and the two passive bands below them. This degeneracy is fine-tuned, but the degeneracy breaking terms in the next shells (sub-shells $B1, A2, B2,\ldots $ etc) are perturbative. Hence the gap between the  active and the passive bands will remain small in the isotropic limit, answering  question (1) in Fig.~\ref{fig:Questions1}.

From the Tripod model, the two active bands have energy zero at the $K_M$ point, and vanishing velocity at $w_1= \frac{1}{\sqrt{3}}$. Moreover, they also have energy zero at the $\Gamma_M$ point in the Hexagon model (a good approximation for the infinite case at the $\Gamma_M$ point). This now gives us $\emph{two}$ points ($\Gamma_M, K_M$) in the MBZ where the bands have zero energy; at one of those points, the band velocity vanishes. This gives us more analytic arguments that the band structure remains flat than just the $K_M$ point velocity, i.e. point (3) in Fig.~\ref{fig:Questions1}. We further try to establish band properties away from the $\Gamma_M, K_M$ points by performing a further perturbative treatment of $H_{\text{Hex}}(\kk, w_0, w_1)$ using the eigenstates at $\Gamma_M$.

\subsubsection{  $\mathbf{k} \ne 0$  Six-Band approximation of the Hexagon Model In the Isotropic Limit}

In the isotropic limit at $w_0=w_1=1/\sqrt{3}$, the $6$-fold degeneracy point of the Hexagon model $H_{\text{Hex}}(\kk, w_0, w_1)$ at $\Gamma_M$ prevents the development of a Hamiltonian for the two active bands. However, since the gap ($= \sqrt{13/3}$) between the $6$ zero modes  $E_{1\ldots 6}(\mathbf{k}= 0, w_0= \frac{1}{\sqrt{3}} , w_1= \frac{1}{\sqrt{3}} ) $ in Tab.~\ref{tab:6-fold-eigenvalue} and the rest of the bands $E_{7\ldots 12}(\mathbf{k}= 0, w_0= \frac{1}{\sqrt{3}} , w_1= \frac{1}{\sqrt{3}} )$  is large at $\Gamma_M$, we can build a $6$-band $k \cdot p$ Hamiltonian away from the $\Gamma_M$ point:
\begin{widetext} 
 \begin{eqnarray} \label{6foldperturbationofHexagonmodel}
&H^{\text{6 band}}_{ij} (\mathbf{k}) = \langle \psi_{E_i} | H_{\text{Hex}}(\mathbf{k},  w_0= w_1=  \frac{1}{\sqrt{3}}) -H_{\text{Hex}}(\mathbf{k}=0,  w_0= w_1=  \frac{1}{\sqrt{3}} ) |\psi_{E_j} \rangle
=\langle \psi_{E_i} |I_{6\times 6} \otimes \mathbf{k} \cdot \vec{\sigma} |\psi_{E_j} \rangle\ ,
%\nonumber \\ &\langle \psi_i(\mathbf{k}=0,  w_0= w_1=  \frac{1}{\sqrt{3}} ) | H_{\text{Hex}}(\mathbf{k},  w_0= w_1=  \frac{1}{\sqrt{3}}) -H_{\text{Hex}}(\mathbf{k}=0,  w_0= w_1=  \frac{1}{\sqrt{3}} ) | \psi_j(\mathbf{k}=0,  w_0=w_1= \frac{1}{\sqrt{3}}) \rangle = \nonumber \\ &= \langle \psi_i(\mathbf{k}=0,  w_0= w_1=  \frac{1}{\sqrt{3}} ) | I_{6\times 6} \otimes \mathbf{k} \cdot \vec{\sigma} | \psi_j(\mathbf{k}=0,  w_0=w_1= \frac{1}{\sqrt{3}}) \rangle  
\end{eqnarray}  \end{widetext}
where $ | \psi_{E_j} \rangle $ with $j=1,\ldots 6$ are the zero energy eigenstates of $H_{\text{Hex}}(\mathbf{k}=0,  w_0= w_1=  \frac{1}{\sqrt{3}} )$. We find these eigenstates in App.~\ref{Appendix3}, where we place them in $C_3, C_{2x}$ eigenvalue multiplets. The $6\times 6$ Hamiltonian is the smallest effective Hamiltonian at the isotropic point, due to the $6$-fold degeneracy of bands at $\Gamma_M$.

The  explicit form of the Hamiltonain $H^{\text{6 band}} (\mathbf{k})$ is given in App.~\ref{Appendix3}, Eq.~\ref{6foldperturbationofHexagonmodelHamiltonian}. Due to the large gap between the $6$-bands (degenerate at $\Gamma_M$) and the rest of the bands, it should present a good approximation of the Hexagon model at finite $\kk$ for  $w_0 = w_1= \sqrt{3}$. The approximate $H^{\text{6 band}} (\mathbf{k}) $ is still not generically diagonalizable (solvable) analytically. However, we can obtain several important properties analytically. First, the characteristic polynomial
\begin{eqnarray}
&Det[E- H^{\text{6 band}}(\mathbf{k}) ]=0 \implies \nonumber \\ &(13 E^2- 12(k_x^2+k_y^2)E+ k_x(k_x^2-3 k_y^2))^2=0
\end{eqnarray} Or, parametrizing $(k_x, k_y)= k(\cos\theta, \sin\theta)$, where $|\kk| =k$, we have
\begin{eqnarray}
 (13 E^3- 12k^2E+ k^3\cos(3\theta) )^2=0\ .
\end{eqnarray} The characteristic polynomial reveals several properties of the $6$-band approximation to the Hexagon model:
\begin{itemize}

\item The exponent of $2$ in the characteristic polynomial reveals that all bands of this approximation to the Hexagon model are exactly doubly degenerate. This explains the almost degeneracy of the flat bands (point (3) in Fig.~\ref{fig:Questions1}), but furthermore it explains why the passive bands, even though highly dispersive, are almost degenerate for  a large momentum range around the $\Gamma_M$-point in the full model (see Fig.~\ref{fig:TBLGBSShells1}): they are exactly degenerate in the $6$-band approximation to Hexagon model; corrections to this approximation come from the remaining $6$ bands of the Hexagon model, which reside extremely far   (energy $\sqrt{13/3}$), or from the $B_1$ shell, which we established is at most $20\%$ in the MBZ - and smaller around the $\Gamma_M$ point. Thus, the almost double degeneracy of the passive bands pointed out in  (2) of Fig.~\ref{fig:Questions1} is explained.

\item Along the $\Gamma_M-K_M$ line we have $k_x=0,k_y=k$ and hence the characteristic polynomial becomes 
\beq
\Gamma_M-K_M:\;\; (13 E^3- 12k_y^2E)^2=0
\eeq This implies two further properties: \textbf{(1)} The ``active'' bands of the approximation of the Hexagon mode are exactly flat at $E=0$ for the whole  $\Gamma_M-K_M$ line, thereby explaining their flatness for a range of momenta; notice that our prior derivations found that the active bands have zero energy at $K_M$, $\Gamma_M$ and vanishing Dirac velocity at $K_M$ for $w_0 = w_1= \sqrt{3}$; our current derivation shows that the approximately flat bands along the whole $\Gamma_M-K_M$ line originate from the doubly degenerate zero energy bands of the Hexagon model.  \textbf{(2)} The dispersive (doubly degenerate) passive bands, for $w_0 = w_1= \sqrt{3}$, have a linear dispersion 
\begin{equation}
E= \pm \sqrt{12/13} k
\end{equation}
along $\Gamma_M-K_M$, with velocity $2 \sqrt{3/13}=0.960769$, close to the Dirac velocity. This explains property (2) in Fig.~\ref{fig:Questions1}. Note that the velocity is equal to $2/(E_{9,10}(\mathbf{k}= 0, w_0= 1/\sqrt{3} , w_1= 1/\sqrt{3} ))$  or two over the gap to the first excited state.  This approximation is visually shown in Fig.~\ref{fig:HexagonalModelApproximation1}.

\item Remarkably, the eigenstates along along the $\Gamma_M-K_M$ line can also be obtained (See App.~\ref{Appendix4}). Along this line, the eigenstates of all bands of the $H^{\text{6 band}}$ Hamiltonian approximation to the Hexagon model are $k_y$ independent! (see App.~\ref{Appendix4})

\item Along the $\Gamma_M-M_M$ line ($k_x=k, k_y=0$)  the characteristic polynomial becomes
\beq
\Gamma_M-M_M:\; (k+E)^2(k^2-13k E+13 E^2)^2=0.
\eeq  Hence the energies are:  $E= -k$, a highly dispersive (doubly degenerate) hole branch passive band of velocity $-1$; $E=\frac{1}{2}(1+\frac{3}{\sqrt{13}})k$ ($\approx 0.916025 k$), another highly dispersive doubly degenerate electron branch passive band. This explains property (2) in Fig.~\ref{fig:Questions1}. Notice that this velocity is $\frac{1}{2}(1+\frac{1}{E_{9,10}(\mathbf{k}= 0, w_0= 1/\sqrt{3} , w_1= 1/\sqrt{3} )})$. The third dispersion is $E=\frac{1}{2}(1-\frac{3}{\sqrt{13}})k$ ($\approx 0.0839749 k$), a weakly dispersive doubly degenerate active bands. This explains the very weak, but nonzero dispersion of the bands on $\Gamma_M-M_M$.  The eigenstates along this line can also be obtained (See App.~\ref{Appendix4}). The approximation is visually shown in Fig.~\ref{fig:HexagonalModelApproximation1}.

\item Along the $\Gamma_M-M_M$, the eigenstates of all bands of the $H^{\text{6 band}}$ Hamiltonian approximation to the Hexagon model are $k_x$ independent! (see App.~\ref{Appendix4})

\item In the $6$-band model, eigenstates are independent of $\kk$ on the manifold $k_x= k_y$. %{\color{red} BL: where is this proved? Or is it a typo? $k_x=k_y$ does not look like a special direction.} 

\begin{figure}
\begin{centering}
\includegraphics[width=1.0\linewidth]{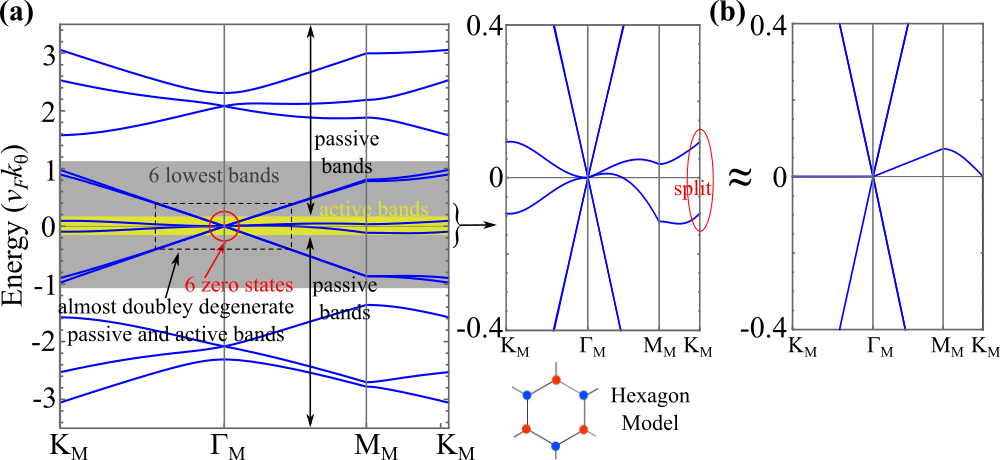}
\par\end{centering}

\protect\caption{\label{fig:HexagonalModelApproximation1}  Band structure of the $6$-band  approximation$H^{\text{6 band}}$ to the Hexagon model for the $w_0=w_1=1/\sqrt{3}$ magic point.
(a) The 6 zero energy eigenstates at $\Gamma_M$ marked by the red circle are used to obtain a perturbative Hamiltonian for the 6 lowest bands across all the BZ. As the 6 bands are very well separated from the other 6, we expect a good approximation over a large part of the BZ.
The active and passive bands in the dashed square are almost doubly degenerate.
In the right panel, the 6 lowest bands of the hexagon model, for a smaller energy range, are shown.
Notice the passive bands are undistinguishably 2-fold degenerate by eye (not an exact degeneracy, they split close to $K_M$ see left plot)
Note the Dirac feature of the passive bands. 
The active bands split at $K_M$ in the hexagon model, but the B1 shell addition makes them degenerate.
(b) The first order approximation to the Hexagon model using the 6 zero energy bands at the $\Gamma_M$ point gives exactly doubly degenerate bands over the whole BZ. It gives the correct velocity of the Dirac Nodes, zero dispersion of active bands on $\Gamma_M-K_M$ and a small dispersion of active bands on $\Gamma_M-M_M$, with known velocities. 
Along these lines, all eigenstates are k-independent.
}
\end{figure}
\end{itemize}

\subsubsection{Energies of the Hexagon Model at $\mathbf{k}=0$ Away From the Isotropic Limit and the Second Magic Manifold. }

In the isotropic limit (which coincides with the magic angle of the Tripod model),  $w_0=w_1=1/\sqrt{3}$, due to the 6-fold degeneracy of the $\Gamma_M$ point, it is impossible to obtain an approximate Hamiltonian that is less than a $6\times 6$ matrix. Moving away from the isotropic limit, and staying in the range of approximations  $w_0, w_1 \le \frac{1}{\sqrt{3}}$, the Hexagon model is a good starting point for a perturbative expansion. We now ask what values of $w_1, w_0$ might have a ``simple''expression for their energies. 

We see that if $w_1= \frac{\sqrt{1+ w_0^2}}{2}$, the $6$-fold degeneracy at the $\Gamma_M$ point at zero energy for $w_1=1/\sqrt{3}$ splits into a $2 (\text{enforced})+2(\text{accidental})+2(\text{enforced})$-fold degeneracy. There is  an \emph{accidental} $2$-fold degeneracy of the active bands at zero energy, and a gap to the passive bands which have an \emph{symmetry enforced} degeneracy. The $2$-fold accidental degeneracy at zero energy along  $w_1= \frac{\sqrt{1+ w_0^2}}{2}$ is the important property  of this manifold in parameter space. The eigenvalues of the Hexagon model in this case are given in Tab. \ref{tab:EnergiesAtMagicManifold1}.
%$\forall\; w_0\le 1/\sqrt{3}$:

\begin{table}[htbp]
\centering
\begin{tabular}{c|c|c}
\hline
Band & Energy at $\mathbf{k=0}$ at $w_1=\frac{\sqrt{1+w_0^2}}{2}$ & dege.  \\
\hline
$E_{1,2}$  & $0$ & $2$  \\
\hline
$E_{3,4}$  & $\frac{\sqrt{10 w_0^2+1}-\sqrt{w_0^2+4}}{2}$ & $2$  \\
\hline
$E_{5,6}$  & $-\frac{\sqrt{10 w_0^2+1}-\sqrt{w_0^2+4}}{2}$ & $2$  \\
\hline
$E_{7,8}$  & $-\frac{\sqrt{10 w_0^2+1}+\sqrt{w_0^2+4}}{2}$ & $2$  \\
\hline
$E_{9,10}$ & $\frac{\sqrt{10 w_0^2+1}+\sqrt{w_0^2+4}}{2}$ & $2$  \\
\hline
$E_{11}$  & $-2\sqrt{1+ w_0^2}$ & $1$  \\
\hline
$E_{12}$ & $2\sqrt{1+ w_0^2}$ & $1$  \\
\hline
\end{tabular}
\caption{Eigenvalues of the Hexagon model in Eq. (\ref{HexagonModel1}) at $\Gamma_M$ point ($\kk=\mathbf{0}$) at the Second Magic Manifold $w_1=\frac{\sqrt{1+w_0^2}}{2}$. The notation dege. is short for degeneracy.}\label{tab:EnergiesAtMagicManifold1}
\end{table}

%\begin{eqnarray}
%&E_{1,2}(\mathbf{k}= 0, w_0, w_1= \frac{\sqrt{1+ w_0^2}}{2}) = 0 \nonumber \\
%&E_{3,4}(\mathbf{k}= 0, w_0, w_1= \frac{\sqrt{1+ w_0^2}}{2}) =  \frac{\left(\sqrt{10 w_0^2+1}-\sqrt{w_0^2+4}\right)}{2}  \nonumber \\
%&E_{5,6}(\mathbf{k}= 0, w_0, w_1= \frac{\sqrt{1+ w_0^2}}{2}) = -\frac{\left(\sqrt{10 w_0^2+1}-\sqrt{w_0^2+4}\right)}{2}  \nonumber \\
%&E_{7,8}(\mathbf{k}= 0, w_0, w_1= \frac{\sqrt{1+ w_0^2}}{2}) =  -\frac{ \left(\sqrt{w_0^2+4}+\sqrt{10 w_0^2+1}\right) }{2}\nonumber \\&
%E_{9,10}(\mathbf{k}= 0, w_0, w_1= \frac{\sqrt{1+ w_0^2}}{2}) =  \frac{ \left(\sqrt{w_0^2+4}+\sqrt{10 w_0^2+1}\right)}{2} \nonumber \\
%&E_{11}(\mathbf{k}= 0, w_0, w_1= \frac{\sqrt{1+ w_0^2}}{2}) = -2\sqrt{1+ w_0^2}\nonumber \\
%&E_{12}(\mathbf{k}= 0, w_0, w_1= \frac{\sqrt{1+ w_0^2}}{2}) = 2 \sqrt{1+ w_0^2}\label{EnergiesAtMagicManifold1}
%\end{eqnarray}  

Although the perturbative addition of the $B1$ shell will split the $\Gamma_M$ point $E_{1,2}(\mathbf{k}= 0, w_0, w_1= \frac{\sqrt{1+ w_0^2}}{2}) = 0$  degeneracy, we find that this zero energy doublet of the Hexagon model is particularly useful to calculate a $k \cdot p$ perturbation theory of the active bands, as many perturbative terms cancel. In particular, we see that the gap between the active band zero energy doublet and the passive bands ($E_{3,4}(\mathbf{k}= 0, w_0, w_1= \frac{\sqrt{1+ w_0^2}}{2})$) of the Hexagon model becomes large in the chiral limit($E_{3,4}(\mathbf{k}= 0, w_0=0, w_1= \frac{\sqrt{1+ w_0^2}}{2} =1/2) =-1/2$). We note that this explains property (4) of Fig.~\ref{fig:Questions1}: from the Hexagon model, the gap between the active and the passive bands  is, in effect, proportional to $w_1-w_0$. Since the bandwidth of the TBG model is known to be smaller than this gap, we will use the $\Gamma_M$ point doublet of states $E_{1,2}(\mathbf{k}= 0, w_0, w_1= \frac{\sqrt{1+ w_0^2}}{2}) = 0$ to perform a perturbative expansion. We define this paramter manifold as the ``Second Magic Manifold'':
\begin{eqnarray}
%& \text{ Hexagon model $\Gamma_M$ doublet of $E=0$ active bands:}  \\ 
&\textbf{Second Magic Manifold: }  w_1=\frac{\sqrt{1+w_0^2}}{2}, \;w_0 \le 1/\sqrt{3}. \nonumber \end{eqnarray}

\section{Two-Band Approximations on The Magic Manifolds}\label{MagicManifolds}

\subsubsection{Differences between the First and Second Magic Manifolds}

We have defined two manifolds in parameter space where the two active bands of the Hexagon model are separated from the passive bands. Hence, we can do a perturbative expansion in the inverse of the gap from the passive to the active bands.  We first briefly review the differences between the two Magic Manifolds

$ \textbf{First Magic Manifold: } w_0\le w_1=1/\sqrt{3}:$

\begin{itemize}

\item For these values of $w_0, w_1$, the Dirac velocity at $K_M$ vanishes in the Tripod model, which is a good approximation to the infinite cutoff model. Hence the velocity at the $K_M$ point in the infinite model should be small. The Dirac node is at $E=0$.
\item One end of the first magic manifold, the isotropic point $w_0=w_1=1/\sqrt{3}$ is also the end-point of the second magic manifold, and exhibits the $6$-fold degeneracy at $E=0$ at the $\Gamma_M$ point in the Hexagon model. 
\item Away from the isotropic point, on the first magic manifold, a gap opens everywhere between  the $6$ states of the Hexagon model. At the $\Gamma_M$ point, the $6$-fold degenerate bands at the isotropic limit split when going away from this limit, into a 2 (symmetry enfroced) -1-1-2 (symmetry enforced) degeneracy configuration. Hence the two active bands, in the Hexagon model, split from each other in the first magic manifold.
\item The splitting of the active bands in the Hexagon model in the first magic manifold is corrected by the addition of the $B1$ shell  as the term $ H_{A1,B1}  H_{\mathbf{k} B1}^{-1}H_{A1,B1}^\dagger$ in Eq.~\ref{Happrox1}. 

\item The active bands, when computed with the full Hamiltonian without approximation, are very flat on the first magic manifold (much flatter than on the second magic manifold), and there is a full, large gap to the passive bands (see Fig.~\ref{fig:MagicManifold2Data}).

\begin{figure}
\begin{centering}
\includegraphics[width=1\linewidth]{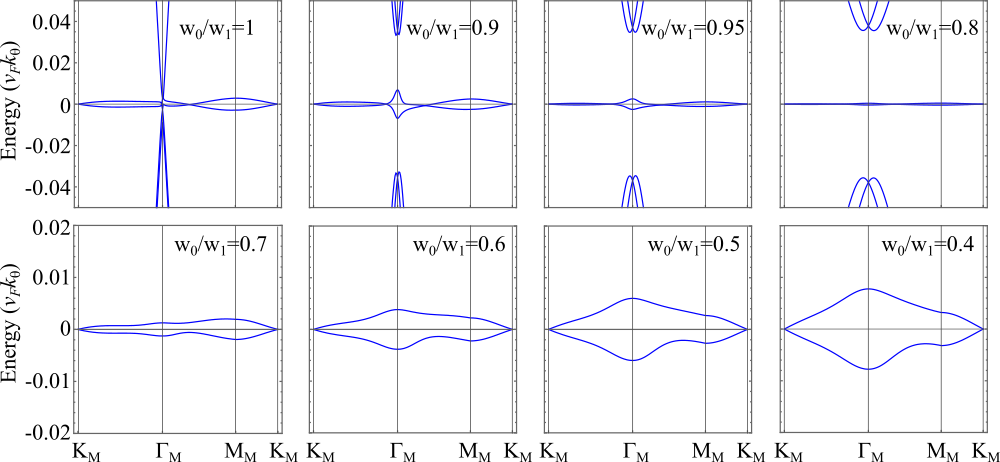}
\par\end{centering}

\protect\caption{\label{fig:MagicManifold2Data} Plots of the active bands band structure on the first magic manifold, $w_1=1/\sqrt{3}$, $w_0\le\sqrt{3}$, for a large number of shells.
In the second row, the gap to the passive bands is large and outside the range.
The Dirac velocity is small for all values of $w_0/w_1$ (it vanishes in the Tripod model, but has a finite value once further shells are included), and the bands are extremely flat. The ratio of active bands bandwidth to the active-passive band gap decreases upon decreasing $w_0/w_1$. }

\end{figure}

\end{itemize}

$\textbf{Second Magic Manifold: } w_1=\frac{\sqrt{1+w_0^2}}{2}, \;w_0 \le 1/\sqrt{3}:$

\begin{itemize}

\item The Hexagon model exhibits a doublet of zero energy active bands at $\Gamma_M$ along the entire second magic manifold.

\item  One end of the second magic manifold, the isotropic point $w_1=w_0 = 1/\sqrt{3}$ is also the end-point of the first magic manifold, and exhibits a $6$-fold degeneracy at $E=0$ at the $\Gamma_M$ point in the Hexagon model and a vanishing Dirac velocity in the Tripod model.

\item Away from the isotropic point, on this manifold, the bands do not have a vanishing velocity at the Dirac point. 

\item The eigenstates of the active bands are simple (simpler than on the first magic manifold) on this manifold, with simple matrix elements (as proved below). A perturbation theory can be performed away from the $\Gamma_M$ point and away from this manifold to obtain a general Hamiltonian for $\kk, w_0, w_1$. The $B1$ shell can then also be included perturbatively as the term $ H_{A1,B1}  H_{\mathbf{k} B1}^{-1}H_{A1,B1}^\dagger$ in Eq.~\ref{Happrox1}. 

\item The active bands are not the flattest on this manifold. They are much less flat than on the first magic manifold, due to the fact that the Dirac velocity does not vanish (is not small) at the $K_M$ point on the second magic manifold.

\end{itemize}

\subsection{Two-Band Approximation for the Active Bands of the Hexagon Model on the Second Magic Manifold}\label{twobandsecondmagicmanifoldhexagon}

We now try to obtain a $2$-band model on the manifold $ w_1= {\sqrt{1+w_0^2}}/{{2}}, \;\; \forall w_0\le 1/\sqrt{3} $, for which we use the $\Gamma_M$-point  $H_{\text{Hex}}(\mathbf{k}=0, w_0,  w_1=\frac{ \sqrt{1+w_0^2}}{2})$ as zeroth order Hamiltonian and perform a $\kk \cdot \mathbf{p}$ expansion away from the $\Gamma_M$ point.  

Fig.~\ref{fig:MagicManifold2Data} shows that away from the isotropic limit, the gap  that opens at the  $\Gamma_M$ point between the formerly $6$-fold degenerate bands can be much larger than the bandwidth of the active bands even for modest deviations from the isotropic limit. We have explained this from the behavior of the $6$-band approximation to the Hexagon model, and from knowing the analytic form of the $\Gamma_M$-point energy levels in the Hexagon model.  We have also obtained the eigenstates of all the $\Gamma_M$-energy levels in App.~\ref{Eigenstatesolutiononthesecondmagicmanifold}. It is then sufficiently accurate to treat the manifold of the \emph{two $\Gamma_M$-point zero energy states at $ w_1= \frac{\sqrt{1+w0^2}}{{2}},\forall w_0\le 1/\sqrt{3} $} as the bases of the perturbation theory.

To perform a $2$-band model approximation to the Hexagon model, we take the unperturbed Hamiltonian to be $H_{\text{Hex}}(\mathbf{k}=0, w_0, w_1 = \sqrt{1+ w_0^2}/2)$ (the Hexagon model on the second magic manifold) in Eq.~\ref{HexagonModel1}. For this Hamiltonian we are able to obtain \emph{all the eigenstates analytically} in App.~\ref{Eigenstatesolutiononthesecondmagicmanifold}. The perturbation Hamiltonian, on the second magic maifold is 
\begin{eqnarray} \label{eq-Hpert_k_w0}
&H_{\text{perturb}}(\mathbf{k}, w_0) = H_{\text{Hex}}(\mathbf{k}, w_0, w_1 = \frac{ \sqrt{1+w_0^2}}{2})- \nonumber \\ &  - H_{\text{Hex}}(\mathbf{k}=0, w_0,  w_1=\frac{ \sqrt{1+w_0^2}}{2})= I_{6\times 6} \otimes \mathbf{k} \cdot \vec{\sigma}. 
\end{eqnarray}
The manifold of states which are kept as ``important'' are the two zero energy eigenstates of $H_{\text{Hex}}(\mathbf{k}=0, w_0, w_1 = \sqrt{1+ w_0^2}/2)$, given in Eq.~\ref{zeroenergyeigenstatesonsecondmanifold}. This manifold will be denoted as $\psi$ with a band index $m\in\{1,2\}$.
%$m, m', m'',m''', \ldots \in 1,2$. 
The manifold of ``excited'' states, that will be integrated out, is 
made up of the eigenstates Eqs.~\ref{energyeigenstatesonsecondmanifold1}, \ref{energyeigenstatesonsecondmanifold2}, \ref{energyeigenstatesonsecondmanifold3} and~\ref{energyeigenstatesonsecondmanifold4}, each doubly degenerate and Eqs.~\ref{energyeigenstatesonsecondmanifold5} and~\ref{energyeigenstatesonsecondmanifold6}, each nondegenerate. This manifold will be denoted as $\psi$ with a band index $l\in\{3,4,\cdots,12\}$.
%$l, l', l'', l''', \ldots \in 3,4,\ldots, 12$. 
We now give the expressions for the perturbation theory up to the fifth order. We here give only the final results, for the expression of the matrix elements computed in perturbation theory, see App.~\ref{HamiltonianMatrixElements1}.

We first note that the first order (linear in $\kk$)  perturbation term is $H^{(1)}_{m m'}  (\mathbf{k}, w_0) =\langle \psi_m | H_{\text{perturb}} (\mathbf{k}, w_0)  |\psi_{m'} \rangle =0$. This is  a particular feature of  the second magic manifold and renders the perturbation theory  simple. Furthermore, it implies that, on the second magic manifold, the active bands of the Hexagon model have a quadratic touching at the $\Gamma_M$ point, as confirmed numerically.  Due to the vanishing of these matrix elements, one can perform quite a large order perturbative  expansion.

It can be shown that the $n$-th order perturbation is proportional to $1/(3w_0^2-1)^{n-1}$, with symmetry-preserving functions of $\kk$ (see App.~\ref{HamiltonianMatrixElements1}). Up to the 5-th order, the full $2$-band approximation to the Hexagon Hamiltonian can be expressed as:
\beq
\begin{split}
&H_{\text{2band}}^{\text{Hex}}(\mathbf{k}, w_0, w_1= \frac{\sqrt{1+w_0^2}}{2})=\nonumber \\ &= d_0(\mathbf{k} ,w_0) \sigma_0+ d_1(\mathbf{k} ,w_0)(\sigma_y+\sqrt{3} \sigma_x), \label{2bandhexagonprojected2}
\end{split}
\eeq
where
\beq
\begin{split}
&d_0(\mathbf{k} ,w_0) =\frac{4  w_0}{9\sqrt{w_0^2+1}  \left(1-3 w_0^2\right)^2 }\Big[  \left(w_0^2-3\right)    - \\ &\frac{4  \left(29 w_0^6-223 w_0^4-357 w_0^2-9\right) }{9 \left(1-3 w_0^2\right)^2 \left(w_0^2+1\right)} \left(k_x^2+k_y^2\right)\Big]k_x \left(k_x^2-3 k_y^2\right),
\end{split}
\eeq and
\beq
\begin{split}
&d_1(\mathbf{k} ,w_0)=\frac{4 w_0^2}{3\sqrt{w_0^2+1} \left(3 w_0^2-1\right)} \\ &\times\Big[-1+\frac{2(35 w_0^4+68 w_0^2+9 )\left(k_x^2+k_y^2\right)}{9 \left(w_0^2+1\right) \left(3 w_0^2-1\right)^2} \Big]  \left(k_x^2+k_y^2\right),
\end{split}
\eeq
while the Pauli matrices $\sigma_{j}$ here are in the basis defined in App. \ref{Eigenstatesolutiononthesecondmagicmanifold-a} (rather than the basis of graphene sublattice). In particular, we note that the eigenstates of the $k\cdot p$ model $H_{\text{2band}}^{\text{Hex}}(\mathbf{k}, w_0, w_1= \frac{\sqrt{1+w_0^2}}{2})$ are independent of $\kk$ up to the fifth order perturbation within the hexagon model.

\subsection{ Away from the Second Magic Manifold: 2-Band Active Bands Approximation of the Hexagon Model  }

We now want to perform calculations away from the second magic manifold, and possibly connect the perturbation theory with the first magic manifold.  There are two ways of doing this, while still using the $\Gamma_M$-point wave-functions as a basis (we cannot solve the Hexagon model exactly at any other $\kk$-point). One way is to solve for the wave-functions at the $\Gamma_M$-point for all $w_0, w_1$, and use these states to build a perturbation theory that way. However, away from the special first and second magicmanifolds, the expression of the ground-states is complicated. The second way, is to use the eigenstates already obtained for the second magic manifold $w_1=\frac{\sqrt{1+ w_0^2}}{2}$ and obtain a perturbation away from the second magic manifold. In this section we choose the latter.

We take the unperturbed Hamiltonian to be $H_{\text{Hex}}(\kk=0, w_0, w_1 = \sqrt{1+ w_0^2}/2)$ (the Hexagon model on the second magic manifold) in Eq.~\ref{HexagonModel1}. For this Hamiltonian we are able to obtain \emph{all the eigenstates analytically} in App.~\ref{Eigenstatesolutiononthesecondmagicmanifold}. The perturbation Hamiltonian, away the second magic manifold is 
\begin{eqnarray} &H_{\text{perturb}}(\kk, w_0, w_1) =H_{\text{Hex}}(\kk, w_0, w_1) - \nonumber \\ &-H_{\text{Hex}}(\kk=0, w_0,  w_1=\frac{ \sqrt{1+w_0^2}}{2})= \nonumber \\ &= I_{6\times 6} \otimes \kk \cdot \vec{\sigma} + H_{\text{Hex}}(\kk=0, 0,  w_1-\frac{ \sqrt{1+w_0^2}}{2})
\end{eqnarray}
We now give the expressions for the perturbation theory up to the fourth order. We here give only the final results, for the expression of the matrix elements computed in perturbation theory, see Apps.~\ref{HamiltonianMatrixElements1} and~\ref{HamiltonianMatrixElements2}.

%\subsubsection{Results up to the Fourth Order}
We first note that the first order Hamiltonian is
\beq \label{hexagonmodelfirstorderawayfromsecondmagicmanifold}
H^{(1)}_{m m'}  (\kk, w_0,w_1)=( \frac{\sqrt{w_0^2+1}}{2}-w_1 )(\sigma_y+ \sqrt{3}\sigma_x)
\eeq
Hence we find there is now a linear order term in the Hamiltonian - as it should since the two states degenerate at $\Gamma_M$ on the second magic manifold are no longer degenerate away from it. Because of this, many other terms in the further degree perturbation theory become nonzero, and the perturbation theory has a more complicated form. We present all details in App.~\ref{HamiltonianMatrixElements2} and here show only the final result, up to fourth order. We can label the two band Hamiltonian as:
\beq
\begin{split}
H_{\text{2band}}^{\text{Hex}}(\kk,& w_0, w_1)= d_0(\kk ,w_0,w_1) \sigma_0\\
& + d_1(\kk ,w_0,w_1)(\sigma_y+\sqrt{3} \sigma_x)\ , \label{hexagon2bandprojection11}
\end{split}
\eeq
where the expressions of $d_0(\kk, w_0,w_1)$ and $d_1(\kk, w_0,w_1)$ are given in Eqs. (\ref{hexagon2bandprojection1-d0}) and (\ref{hexagon2bandprojection1-d1}) in App. \ref{HamiltonianMatrixElements2}.
% \begin{widetext}
%\begin{eqnarray}
%& d_0(\kk, w_0,w_1)=\frac{4 k_x w_0 \left(w_0^2-3\right) \left(k_x^2-3 k_y^2\right)}{9 \left(1-3 w_0^2\right)^2 \sqrt{w_0^2+1}}-\frac{8 k_x w_0 \left(w_0^2+21\right) \left(k_x^2-3 k_y^2\right) \left(\sqrt{w_0^2+1}-2 w_1\right)}{9 \left(3 w_0^2-1\right)^3}\sigma_0  
%\end{eqnarray}
%and 
%\begin{eqnarray}
%& d_1(\kk, w_0,w_1)=( \frac{\sqrt{w_0^2+1}}{2}-w_1 )-\frac{4 w_0^2 \left(k_x^2+k_y^2\right)}{3 \sqrt{w_0^2+1} \left(3 w_0^2-1\right)}-\frac{2 \left(7 w_0^2+3\right) \left(k_x^2+k_y^2\right) \left(\sqrt{w_0^2+1}-2 w_1\right)}{3 \left(1-3 w_0^2\right)^2}+\nonumber \\ &  +\frac{4 \left(k_x^2+k_y^2\right) \left(2 w_0^2 \left(35 w_0^4+68 w_0^2+9\right) \left(k_x^2+k_y^2\right)-9 \left(w_0^2+1\right) \left(10 w_0^4+9 w_0^2+3\right) \left(2 w_1-\sqrt{w_0^2+1}\right)^2\right)}{27 \left(w_0^2+1\right)^{3/2} \left(3 w_0^2-1\right)^3} 
%\end{eqnarray}
%\end{widetext}
The perturbation is made on the zero energy eigenstates of $H_{\text{Hex}}(\kk=0, w_0,  w_1=\frac{ \sqrt{1+w_0^2}}{2})$. If $w_1=\frac{\sqrt{1+w_0^2}}{2}$, then the expressions reduce to our previous Hamiltonian Eq.~\ref{2bandhexagonprojected2}. Notice that so far, remarkably the eigenstates are not $\kk$-dependent, they are just the eigenstates of $(\sigma_y+ \sqrt{3}\sigma_x)$.

\subsection{2-active Bands Approximation of the $n=1$ Shell Model $H_{\text{Approx}1}(\kk)$ on the Second Magic Manifold }

In Sec.~\ref{twobandsecondmagicmanifoldhexagon} we have obtained an effective model for the  two active bands of the Hexagon model  on the second magic manifold $ w_1= \frac{\sqrt{1+w0^2}}{{2}}, \;\; \forall w_0\le 1/\sqrt{3} $ using the $\Gamma_M$-point  $H_{\text{Hex}}(\kk=0, w_0,  w_1=\frac{ \sqrt{1+w_0^2}}{2})$ as zeroth order Hamiltonian. We expect this to be valid around the $\Gamma_M$ point.  We know that a good approximation of the TBG involves at least $n=1$ shells: the $A1$ sub-shell, which is the Hexagon model, and the $B1$ sub-shell, which is taken into account perturbatively in $H_{\text{Approx}1}(\kk)$ of Eq.~\ref{Happrox1}. After detailed calculations given in App.~\ref{app-B1shellapprox}, we find the first order perturbation Hamiltonian given by
\begin{equation}
\begin{split}
H^{(B1)} (\kk, w_0, & w_1) =\frac{1}{ \prod_{i=1,2,3} |k-2q_i|^2 |k+2q_i|^2 } \\
& \times\sum_{\mu=0,x,y,z}\widetilde{d}_\mu(\kk, w_0, w_1) \sigma_\mu\ , \label{B1shellprojected1}
\end{split}
\end{equation}
where $\widetilde{d}_\mu(\kk, w_0, w_1)$ are given in Eqs. (\ref{eq-B1-d0})-(\ref{eq-B1-dy}) of App.~\ref{app-B1shellapprox}. This represents the first order $H_{\text{Approx}1}(\kk)$ projected into the zero energy bands of the Hexagon model on the second magic manifold. We note that the $B1$ shell perturbation expressions can only be obtained to first order. Second and higher orders are particularly tedious and not illuminating. Note that, to first order in perturbation theory on the second magic manifold, only the  term $ H_{A1,B1}  H_{kB1}^{-1}H_{A1,B1}^\dagger$ contributes to the approximate $2$-band Hamiltonian. Also we obtained the perturbation of $H_{A1,B1}  H_{kB1}^{-1}H_{A1,B1}^\dagger$ for \emph{generic $w_0, w_1$}  projected into the second magic manifold $\Gamma_M$ point bands of the Hexagon model.

\subsection{ 2-Band Approximation for the Active Bands of the $n=1$ Shell Model $H_{\text{Approx}1}(\kk)$  in Eq.~\ref{Happrox1} for any $w_0, w_1\le \frac{1}{\sqrt{3}}$ }

We are now in a position to describe the 2 active bands of the approximate Hamiltonain of the $1$-shell model in Eq.~\ref{Happrox1}, $H_{\text{Approx} 1}= H_{kA1}+ H_{A1,A1} - H_{A1,B1}  H_{kB1}^{-1}H_{A1,B1}^\dagger$ 
by adding $H^{(B1)} (\kk, w_0,w_1)$ of Eq.~\ref{B1shellprojected1} to $H_{\text{2band}}^{\text{Hex}}(\kk, w_0, w_1)$ of Eq.~\ref{hexagon2bandprojection11}.  We note that this is still perturbation theory performed by using the $\Gamma_M$-point  $H_{\text{Hex}}(\kk=0, w_0,  w_1=\frac{ \sqrt{1+w_0^2}}{2})$ as zeroth order Hamiltonian:
\beq \label{1shelltwobandmodelgeneralw0w1fromsecondmagicangle}
H_{\text{2band}}(\kk, w_0, w_1)= H_{\text{2band}}^{\text{Hex}}(\kk, w_0, w_1) +H^{(B1)} (\kk, w_0,w_1) 
\eeq
We now find some of the predictions of this Hamiltonian.

\begin{figure}
\begin{centering}
\includegraphics[width=1\linewidth]{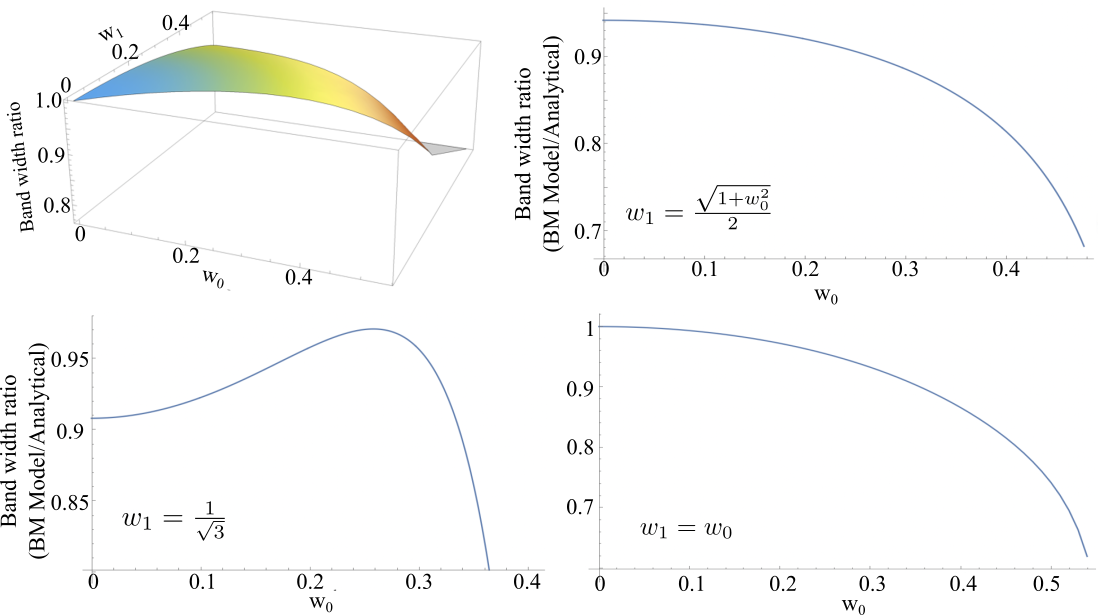}
\par\end{centering}
\protect\caption{\label{fig.BandwidthofFlatBand1} Plots of the ratio of the bandwidth of the active bands for the large number of shells to the analytic bandwidth $\Delta$ in Eq.~\ref{bandwidthtblg}, for different values of $w_0, w_1$, including the two magic manifolds. In the regime of validity of our approximations, we can see that this ratio is substantially above 90\%. }
\end{figure}

%\subsubsection{$\Gamma_M$  point energetics of the two band model}

The energies of the two bands of the  Eq.~\ref{1shelltwobandmodelgeneralw0w1fromsecondmagicangle} at $\Gamma_M$ point are

\beq
E_\pm(w_0,w_1)= \pm (  \frac{-4 \sqrt{w_0^2+1} w_1+w_0^2+w_1^2+2}{2 \sqrt{w_0^2+1}})
\eeq over the full range of $w_0, w_1 \le 1/\sqrt{3}$.
Remarkably, we find an amazing agreement between the energy of the bands at $\Gamma_M$ point and the numerics. We find that the bandwidth of the flat band at $\Gamma_M$ point is
\beq
\Delta(w_0, w_1)=2 |E_\pm(w_0,w_1)|\label{bandwidthtblg}
\eeq
This matches incredibly well with the actual values. In Fig.~\ref{fig.BandwidthofFlatBand1} we plot  the ratio of actual active bandwidth at $\Gamma_M$ point from the large number of shell model to $\Delta$ in Eq.~\ref{bandwidthtblg}, for values $w_0<  1/\sqrt{3}$, $w_0<w_1< 1/\sqrt{3}$. Note that even though we are sometimes going far from the second magic manifold values $w_0, w_1=\sqrt{1+ w_0^2}/{2}$ where the perturbation theory is valid, the ratio holds up well, and is actually never smaller than $0.8$ or larger than $1$. We are using $w_0<w_1$ because the perturbation theory is around the manifold  $w_0, w_1=\sqrt{1+ w_0^2}/{2} \le\frac{1}{\sqrt{3}}$ for which $w_0<w_1$. For $w_1<w_0$ the approximation becomes worse, but is outside of the validity regime.

For the two magic manifolds, also shown in Figs.~\ref{fig.BandwidthofFlatBand1} and~\ref{fig.BandwidthofFlatBand2}, the agreement is very good. We point out several consistency checks. First, remarkably, the set of approximations that led us to finding a $2$-band Hamiltonian becomes \emph{exact} at some points

\begin{itemize}

\item The $\Gamma_M$ point Bandwidth at $w_0=w_1=1/\sqrt{3}$ vanishes $\Delta(\frac{1}{\sqrt{3}},\frac{1}{\sqrt{3}}) =0$. This degeneracy reproduces the exact result, in the $1$-shell model  (See $n=1$ in Fig.~\ref{fig:TBLGBSShells1}, the $6$-fold degeneracy at the $\Gamma_M$ point). The approximate model of the $1$-shell, $H_{\text{Approx} 1}$ of Eq.~\ref{Happrox1} also has an exact $6$-fold degeneracy at the $\Gamma_M$ point at $w_0=w_1=1/\sqrt{3}$ (the two bands here being part of the $6$-fold manifold). It is remarkable that our $2$-band projection perturbation approximation reproduces this degeneracy exactly, especially since it is supposed \emph{not} to work close to  $w_0=w_1=1/\sqrt{3}$ - where the gap  to the active bands is $0$ and the $\Gamma_M$ point becomes $6$-fold degenerate.

\item At $w_0= w_1=0$, the bandwidth at $\Gamma_M$ is  $\Delta(0, 0)=2$. This is again, an exact result \emph{for the infinite shell model}. Indeed, at the $\Gamma_M$ point, the BM Hamiltonian with zero interlayer coupling has a gap $=2|q_1|=2$. 

\item We now ask: what is the $w_0, w_1$ manifold, under this approximation, for which the $\Gamma_M$ point bandwidth is zero? This is easily solved to give:

\begin{figure}
\begin{centering}
\includegraphics[width=1\linewidth]{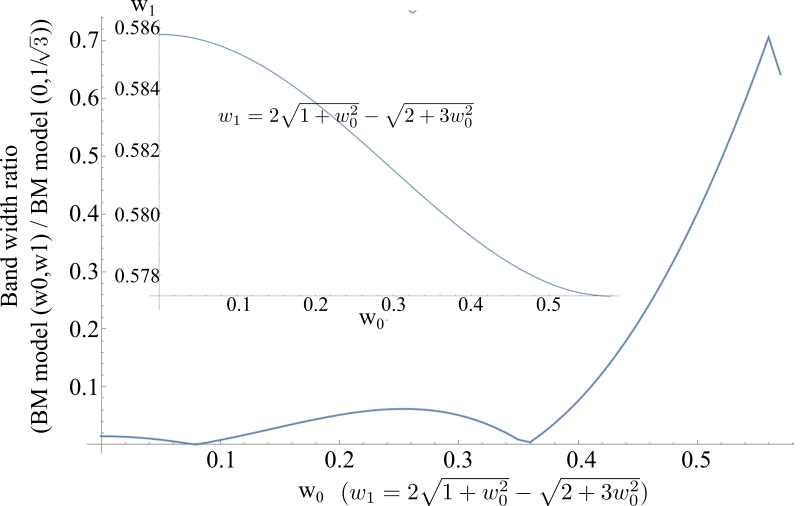}
\par\end{centering}
\protect\caption{\label{fig.BandwidthofFlatBand2}  $\Gamma_M$ point bandwidth of the active bands  (large number of shells) on the manifold $\Delta(w_0,w_1)=0$ ($w_1=2 \sqrt{1 + w_0^2} -\sqrt{2 + 3 w_0^2}$) of
zero analytic bandwidth (Eq.~\ref{bandwidthtblg}) divided by the bandwidth of the active bands in the chiral limit ($(w_0, w_1)= (0, \frac{1}{\sqrt{3}})$. Note that this number is extremely small away from $w_0=w_1= \frac{1}{\sqrt{3}}$, showing that our analytic manifold of smallest bandwidth ($\Delta(w_0,w_1)=0$) also exhibits small bandwidth in the large cell number.  Inset: the curve $w_1=2 \sqrt{1 + w_0^2} -\sqrt{2 + 3 w_0^2}$ for which  $\Delta(w_0,w_1)=0$ for $0\le w_0 \le \frac{1}{\sqrt{3}}$. Note that $w_1$ changes extremely little ~ $1\%$ (stays within $1\%$ of $\frac{1}{\sqrt{3}}$)  during the entire sweeping of $w_0$. }
\end{figure}

\beq
\begin{split}\label{eq-zerobandwidth}
&\qquad\text{\bf 2-band model with zero bandwidth at }\Gamma_M: \\ & \qquad w_1=2 \sqrt{w_0^2+1}-\sqrt{3 w_0^2+2},\;\;\; w_0 \in [0, \frac{1}{\sqrt{3}}].
\end{split}
\eeq
Fig.~\ref{fig.BandwidthofFlatBand2} plots the ratio of the bandwidth of the full BM model on this manifold to the bandwidth at at the chiral limit $w_0=0, w_1= \frac{1}{\sqrt{3}}$ (which is already really small!). We can see that, for most of the $w_0 \in (0,1/\sqrt{3})$ this ratio is below $0.1$, showing us that we have identified an extremely small bandwidth manifold.

\item What are the values of $w_1$ on this manifold? Remarkably, as can be seen in Fig.~\ref{fig.BandwidthofFlatBand2},  $w_1=2 \sqrt{w_0^2+1}-\sqrt{3 w_0^2+2}$ is an almost fully constant over the interval  $w_0 \in (0,1/\sqrt{3})$: it changes by around $1\%$ only. Moreover, its values (0.578-586) are very close to $1/\sqrt{3} \approx 0.57735$. Hence our approximation explains the flatness of the bands over the emph{ first magic manifold}, $0\le  w_0 \le \frac{1}{\sqrt{3}}, w_1 = \frac{1}{\sqrt{3}}$: This manifold is almost the same with the one for which our analytical approximate calculation gives zero gap. Hence property (6) of Fig.~\ref{fig:Questions1} is answered. 

\item At $w_0=0$, one has $w_1=2 \sqrt{w_0^2+1}-\sqrt{3 w_0^2+2}= 2- \sqrt{2}$ in Eq. (\ref{eq-zerobandwidth}), for which the bandwidth is $0$ in our perturbative model. As we show in App. \ref{app-1shellexact}, this value of $w_1$ coincides with the exact value for which the $\Gamma_M$ bandwidth is zero in the approximation Hamiltonian $H_{\text{Approx} 1}$ of  Eq.~\ref{Happrox1}. Furthermore, at $w_0=0$, the value $w_1= 2- \sqrt{2}$ also coincides with the exact value of zero $\Gamma_M$ bandwidth in the no-approximation Hamiltonian of the $n=1$ shell Hamiltonian (of $A1,B1$ subshells) (see App.~\ref{app-1shellexact}).

\item  At $w_0=0$, the value $w_1=2 \sqrt{w_0^2+1}-\sqrt{3 w_0^2+2}= 2- \sqrt{2}$ for which the bandwidth of our approximate $2$-band model is projected to be zero is numerically \emph{very} close to the value of $0.586$ quoted for the first magic angle in the chiral limit \cite{tarnopolsky_origin_2019}. In fact, at $w_0=0, w_1= 2- \sqrt{2}$ the bandwidth of the active bands is half of that at $w_1=0.586$. 

\end{itemize}

\subsection{Region of Validity of the 2-Band Model and Further Fine-Tuning}

The $2$-band approximation to the $n=1$ Shell Model has a radius of convergence in $\kk$ space in the first MBZ. This radius of convergence is easily estimated from the following argument. In Tab. \ref{tab:EnergiesAtMagicManifold1}, the (maximum) gap, at the $\Gamma_M$ point, between the active and the passive bands in the Hexagon model (and in the region $w_0 \le 1/\sqrt{3}$) is at $w_0=0$ and equals $1/2$. The distance, in the MBZ between $\Gamma_M$ and $K_M$ points equals to $1$. Hence we expect that our $2$-band model will work for $|\kk|\ll 1/2$, as our numerical results confirm.  The form factor matrices can be computed for this range of $\kk$ analytically, by using the full Hexagon Hamiltonian in Eq. \ref{hexagon2bandprojection11} plus the $B1$ shell perturbation in Eq. \ref{B1shellprojected1}. They will be presented in a future publication.

The $\kk = K_M$ point is outside the range of validity of  the $2$-band model, and hence this does not capture the gapless Dirac point for all values of $w_0,w_1$. However, with some physical intuition, we can obtain a $2$-band model that has a gap closing at the $K_M$ point.  In Fig. \ref{fig:HexagonalModelApproximation1} we see that the Hexagon model does not have a gap closing between the active bands at the $K_M$ point. However, in Figs. \ref{fig:TBLGBSShells5}, \ref{fig:TBLGBSShells6}, \ref{fig:TBLGBSShells7} we see that $ H_{\text{Approx} 1}(\kk)$ in Eq. \ref{Happrox1} has a gap closing close to, or almost at the $K_M$-point. This means that one of the main  roles of the $B1$ shell is to close the $K_M$ gap, leading to the Dirac point.

\begin{figure}
\begin{centering}
\includegraphics[width=1\linewidth]{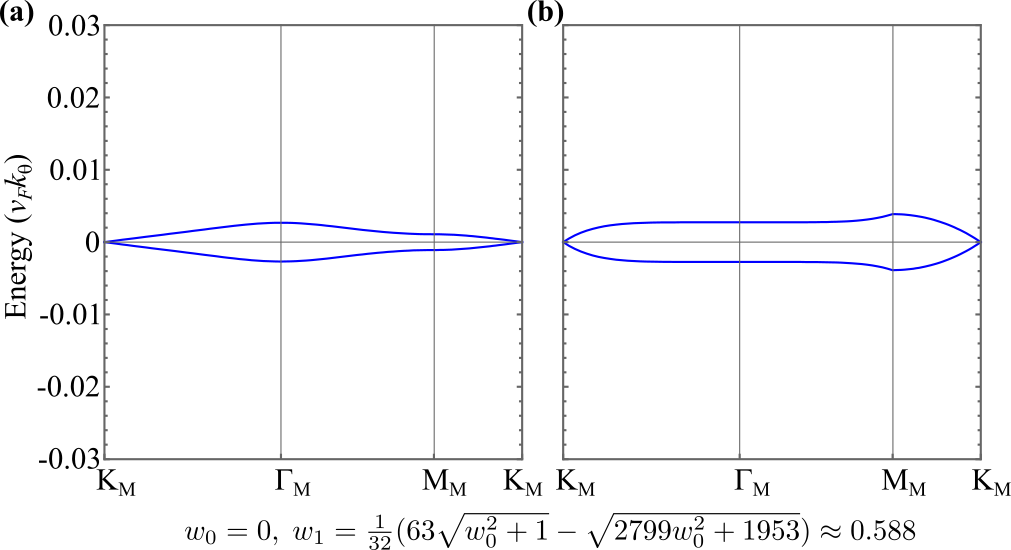}
\par\end{centering}
\protect\caption{\label{2BandModelFirstOrder}  Comparison between (a) the active bands of the BM model at the $w_0=0, w_1\approx0.588$ point and (b)the bands of the $2$-band first order approximation to $ H_{\text{Approx} 1}(\kk)$ in Eq. \ref{Happrox1}. Notice that the bandwith at the $\Gamma_M$ point is virtually identical, that the bands are flat, and that they close gap at the $K_M$ point.  }
\end{figure}

Hence we can use the $2$-band model of the first order approximation to the Hexagon model, Eq. \ref{hexagonmodelfirstorderawayfromsecondmagicmanifold}, $H^{(1)}_{m m'}  (\kk, w_0,w_1)=( \frac{\sqrt{w_0^2+1}}{2}-w_1 )(\sigma_y+ \sqrt{3}\sigma_x)$ along with the  $2$-band model first order approximation for the $B_1$-shell $H^{(B1)} (\kk, w_0,w_1)$ to obtain a first order $2-band$ approximation Hamiltonian: $H^{(1)} (\kk, w_0,w_1)+ H^{(B1)} (\kk, w_0,w_1)$. Note that $H^{(1)} (\kk, w_0,w_1)$, the $2$-band first order approximation to the Hexagon model, has two flat $\kk$ independent bands. We now impose the condition:  $H^{(1)} (\kk=K_M, w_0,w_1)+ H^{(B1)} (\kk=K_M, w_0,w_1)=0$ to find the manifold $(w_1, w_0)$ on which this condition happens. Notice that, a-priori, there is no guarantee that the result of this condition will give a manifold that is anywhere near the values of $w_1, w_0$ considered in this paper, for which our set of approximations is valid (i.e. $w_0,w_1$ not much larger than $1/\sqrt{3}$). We find:
\begin{eqnarray}\label{flatmanifold1}
&H^{(1)} (\kk=K_M, w_0,w_1)+ H^{(B1)} (\kk=K_M, w_0,w_1)=0 \nonumber \\ &\implies \nonumber \\ 
& \text{\bf 2-band model degenerate at } K_M :\qquad \qquad\qquad   \\
& w_1=\frac{1}{32} \left(63 \sqrt{w_0^2+1}-\sqrt{2977 w_0^2+1953}\right)\end{eqnarray}
Remarkably, we note that as $w_0$ is tuned from $1/\sqrt{3}$ to $0$, $w_1$ only changes from $(1/\sqrt{3}=0.57735$ and $\frac{3}{32} \left(21- \sqrt{217}\right)= 0.587726$!. Hence the isotropic point is included in this manifold, and $w_1$ changes by only about 2\% as $w_0$ is tuned from the isotropic point to the chiral limit. We hence propose this model as a first, heuristic $k\cdot p$ model for the active bands on the $w_1(w_0)$ manifold in Eq. \ref{flatmanifold1}. Importantly, this model will have (A) flat bands with small bandwidth; (B) identical gap between the active bands at the $\Gamma_M$ point with the TBG BM model; (C) gap closing at the $K_M$ point (Fig. \ref{2BandModelFirstOrder}).

\section{Conclusions}

In this paper we presented a series of analytically justified approximations to the physics of the BM model \cite{bistritzer_moire_2011}. These approximations allow for an analytic explanation of several properties of the BM model such as (1) the difficulty to stabilize the gap, in the isotropic limit from active to passive bands over a wide range of angles smaller than the first magic angle. (2) The \emph{almost} double degeneracy of the  passive bands in the isotropic limit, even away from the $\Gamma_M$-point, where no symmetry forces them to be. (3) The determination of the high group velocities of the passive bands. (4) The flatness of the active bands even away from the Dirac point, around the magic angle which has $w_1=1/\sqrt{3}$. (5) The large gap,  away from the isotropic limit, (with $w_1=1/\sqrt{3}$), between the active and passive bands, which increases immediately with decreasing $w_0$, while the bandwidth of the active bands does not increase. (6) The flatness of bands over the wide range of $w_0 \in [0, 1/\sqrt{3}]$, from chiral to the isotropic limit. Also, we provided a $2\times 2$ $k\cdot p$ Hamiltonian for the active bands, which allowed for an analytic manifold on which the bandwidth is extremely small: $w_1=2 \sqrt{w_0^2+1}-\sqrt{3 w_0^2+2},\;\;\; w_0 \in [0, \frac{1}{\sqrt{3}}]$.

However, the most important feature uncovered in this paper is the development of an analytic perturbation theory which justifies neglecting most of the matrix elements (form factors/overlap matrices, see Eq.~\ref{approximation2}), which will appear in the Coulomb interaction \cite{ourpaper3}. The exponential decay of these matrix elements with momentum will justify the use of the ``flat metric condition'' in Eq.~\ref{eqn-condition-at-nuMT} and allow for the determination of exact Coulomb interaction ground-states and excitations \cite{ourpaper3,ourpaper4,ourpaper5, ourpaper6}.

Future research in the BM model is likely to uncover many surprises. Despite the apparent complexity of the model and the need for numerical diagonalization, one cannot help but think that there is a $2\times 2$ $k\cdot p$ model valid over the whole area of the MBZ, for all $w_0, w_1$ around the first magic angle. Our $2$-band model is valid around the $\Gamma_M$ point - for a large interval but not for the entire MBZ, although we can fine tune to render the qualitative aspects valid at the $K_M$ point also. A future goal is to find an approximate summation, based on our perturbative expansion, where outer shells can be taken into account more carefully and possibly summed together in a closed-form series, thereby leading to a much more accurate $k\cdot p$ model. We leave this for future research.

\begin{acknowledgments}
We thank Aditya Cowsik and Fang Xie for valuable discussions. B.A.B thanks Michael Zaletel, Christophe Mora and Oskar Vafek for fruitful discussions. This work was supported by the DOE Grant No. DE-SC0016239, the Schmidt Fund for Innovative Research, Simons Investigator Grant No. 404513, and the Packard Foundation. Further support was provided by the NSF-EAGER No. DMR 1643312, NSF-MRSEC No. DMR-1420541 and DMR-2011750, ONR No. N00014-20-1-2303, Gordon and Betty Moore Foundation through Grant GBMF8685 towards the Princeton theory program, BSF Israel US foundation No. 2018226, and the Princeton Global Network Funds. B.L. acknowledge the support of Princeton Center for Theoretical Science at Princeton University in the early stage of this work.
\end{acknowledgments}

\appendix

\bibliography{TBLGHexalogy.bib,HexalogyInternalRefs.bib}

\newpage
\onecolumngrid
\tableofcontents

\section{Matrix Elements of the $\Gamma_M$-Centered Model}\label{Appendix1}

We introduce the shells in the $\Gamma_M$-centered model. The $An_j$ sites of the $n$'th $A$ shell (see Fig.~\ref{fig:MagicManifold}a) are situated at
\begin{eqnarray}
&Q_{An_j}= (n-1)(q_1-q_2) + (j-1) (q_2-q_3) + q_1, \;\; j=1\ldots n \nonumber \\
&Q_{An_{n+j}}= C_6 Q_{An_j}= (n-1)(q_1-q_3) + (j-1) (q_2-q_1) - q_3, \;\; j=1\ldots n \nonumber \\
&Q_{An_{2n+j}}= C_6^2 Q_{An_j}= (n-1)(q_2-q_3) + (j-1) (q_3-q_1) + q_2, \;\; j=1\ldots n \nonumber \\
&Q_{An_{3n+j}}= C_6^3 Q_{An_j}= (n-1)(q_2-q_1) + (j-1) (q_3-q_2) - q_1, \;\; j=1\ldots n \nonumber \\
&Q_{An_{4n+j}}= C_6^4 Q_{An_j}= (n-1)(q_3-q_1) + (j-1) (q_1-q_2) + q_3, \;\; j=1\ldots n \nonumber \\
&Q_{An_{5n+j}}= C_6^5 Q_{An_j}= (n-1)(q_3-q_2) + (j-1) (q_1-q_3) - q_2, \;\; j=1\ldots n
\end{eqnarray}
 There are $6n$ $A$ sites in the $n$'th shell. The $Bn_j$ sites of the $n$'th $B$ shell (see Fig.~\ref{fig:MagicManifold}a) are situated at

\begin{eqnarray}
&Q_{Bn_j}= Q_{An_j}+ q_1,  \;\; Q_{Bn_{n+j}}= Q_{An_{n+}j}- q_3,  \;\;  Q_{Bn_{2n+j}}= Q_{An_{2n+j}}+ q_2, \nonumber \\ & Q_{Bn_{3n+j}}= Q_{An_{3n+j}}- q_1, \;\;  Q_{Bn_{4n+j}}= Q_{An_{4n+j}}+ q_2,  \;\; Q_{Bn_{5n+j}}= Q_{An_{5n+j}}- q_2, \;\;\; j=1\ldots n
\end{eqnarray} There are $6n$  $B$ sites in the $n$'th shell. The basis we take for the BM Hamiltonian in Eq.~\ref{moireham1} is then
\begin{eqnarray}
&(A1,B1, A2, B2,\ldots AN, BN) \nonumber \\ &= (A1_1,A1_2,A1_3,A1_4,A1_5,A1_6, B1_1, B1_2 , B1_3, B1_4, B1_5, B1_6, A2_1,A2_2,  \ldots ) 
\end{eqnarray} where $N$ is the cutoff in the number of shells that we take. Each shell $n$ has $6n$ $A$ sites and $6n$ $B$ sites. 

The separation of shell $n= 1\ldots \infty $ into $A$ and $B$ is necessary in the $\Gamma_M$-centered model is necessary due to the structure of the matrix elements. Unlike in the $K_M$-centered model, where different shells hop from one to another but \emph{not} within a given shell, in the $\Gamma_M$-centered model, the $A$-shells hop between themselves too. Explicitly, the nonzero matrix elements within the $n$'th $A$ shell are called $H_{An,An}$: 
\begin{eqnarray}
&H_{An,An}= \nonumber \\ &An_n \leftrightarrow An_{n+1}: T_2; \; An_{2n} \leftrightarrow An_{2n+1}: T_1;\ An_{3n} \leftrightarrow An_{3n+1}: T_3; \nonumber \\ & An_{4n}\leftrightarrow An_{4n+1}: T_2;\; An_{5n}\leftrightarrow An_{5n+1} : T_1;\; An_{6n} \leftrightarrow An_{6n+1}: T_3
\end{eqnarray}  

 In the $B$ shell, there are no matrix elements between different $B$ sites, but there are matrix elements between the $A$ and $B$ sites in the same shell $n$. They are called $H_{An,Bn}$ and the nonzero elements are:
\begin{eqnarray}
&H_{An,Bn} = \nonumber\\ &An_j \leftrightarrow Bn_j: T_1;\;\; An_{n+j} \leftrightarrow Bn_{n+j}: T_3;\;\; An_{2n+j} \leftrightarrow Bn_{2n+j}: T_2;\;\;\nonumber \\ &An_{3n+j} \leftrightarrow Bn_{3n+j}: T_1;\;\; An_{4n+j} \leftrightarrow Bn_{4n+j}: T_3;\;\; An_{5n+j} \leftrightarrow Bn_{5n+j}: T_2;\;\;\nonumber \\ &j = 1\ldots n,\;\; n= 1 \ldots \infty 
\end{eqnarray} 

Last set of couplings are between the $n-1$'th $B$ shelll $Bn-1$ and the $n$'th shell $An$  are $H_{Bn-1,An}$ with nonzero matrix elements given by:
\begin{eqnarray}
&H_{Bn-1,An} =\nonumber \\
& Bn-1_j \leftrightarrow An_j: T_2, \;\; j= 1\ldots n-1; \;\;\; Bn-1_{j-1} \leftrightarrow An_j: T_3, \;\; j= 2 \ldots n; \nonumber \\ 
& Bn-1_{n+j} \leftrightarrow An_{n+j}: T_1, \;\; j= 1\ldots n-1; \;\;\; Bn-1_{n+j-1} \leftrightarrow An_{n+j}: T_2, \;\; j= 2 \ldots n; \nonumber \\ 
& Bn-1_{2n+j} \leftrightarrow An_{2n+j}: T_3, \;\; j= 1\ldots n-1; \;\;\; Bn-1_{2n+j-1} \leftrightarrow An_{2n+j}: T_1, \;\; j= 2 \ldots n; \nonumber \\ 
& Bn-1_{3n+j} \leftrightarrow An_{3n+j}: T_2, \;\; j= 1\ldots n-1; \;\;\; Bn-1_{3n+j-1} \leftrightarrow An_{3n+j}: T_3, \;\; j= 2 \ldots n; \nonumber \\ 
& Bn-1_{4n+j} \leftrightarrow An_{4n+j}: T_1, \;\; j= 1\ldots n-1; \;\;\; Bn-1_{4n+j-1} \leftrightarrow An_{4n+j}: T_2, \;\; j= 2 \ldots n; \nonumber \\ 
& Bn-1_{5n+j} \leftrightarrow An_{5n+j}: T_3, \;\; j= 1\ldots n-1; \;\;\; Bn-1_{5n+j-1} \leftrightarrow An_{5n+j}: T_1, \;\; j= 2 \ldots n; \nonumber \\ 
\end{eqnarray} The diagonal matrix elements are $(k- Q)\sigma \delta_{Q,Q'}$ where the $Q', Q$'s are given by the shell distance: We call these $H_{k An}$ or $H_{k Bn}$ depending on whether the $Q$ is on the $A$ or $B$ shell. Note that the Hamiltonian within the $B$ shell is  $H_{k Bn}$ while the Hamiltonian within the $A$ shell is  $H_{k An} + H_{An, An}$. We now have defined all the nonzero matrix elements of the Hamiltonian. In block-matrix form, it takes the expression
\begin{equation*}
H= 
\begin{pmatrix}
H_{kA1}+ H_{A1,A1} & H_{A1,B1} &0  & 0 & 0 &\cdots \\
 H_{A1,B1}^\dagger & H_{kB1}  & H_{B1, A2} & 0&  0&\cdots  \\
  0 &H_{B1, A2}^\dagger   &H_{kA2}+ H_{A2,A2} &H_{A2,B2}  &0& \cdots  \\
   0 & 0  &H_{A2,B2}^\dagger & H_{kB2}&  H_{B2, A3} &\cdots  \\
0 & 0  &0 &H_{B2,A3}^\dagger & H_{kA3}+ H_{A3, A3}& \cdots  \\
\vdots  & \vdots  & \ddots & \vdots  
\end{pmatrix}
\end{equation*}

\section{Numerical Confirmation of the Perturbative Framework}\label{Appendix2}

What our discussion in Secs.~\ref{approximationframework1} and~\ref{approximationframework2}  shows is that \textbf{(1)} For the first magic angle, we can neglect all shells greater than 2, while having a good approximation numerically. \textbf{(2)} For the next, smaller, magic angle, we need to keep more shells in order to obtain a good approximation.  We have tested that \emph{machine precision} convergence can be obtained for the active bands by choosing a cutoff of 5-6 shells. We test this next, along with other conclusions of Secs.~\ref{approximationframework1} and~\ref{approximationframework2}   In particular: 

\begin{itemize}
\item We first confirm our analytic conclusion that shells above $n>2$ do not change the spectrum for the first magic angle (and for larger angles than the first magic angle). Figs.~\ref{fig:TBLGBSShells1},\ref{fig:TBLGBSShells2} and~\ref{fig:TBLGBSShells3} show the spectrum for several values of $w_0, w_1$ around (or larger than) the first magic angle characterized by $w_0=1/\sqrt{3}$ for the $K_M$-centered model and by $w_0= w_1= 1/\sqrt{3}$ for the $\Gamma_M$-centered model model in Sec.~\ref{StoryOfTwoLattices}). For the $K_M$-centered model, the magic angle does not depend on $w_1$ but for the $\Gamma_M$-centered model it does, see Sec.~\ref{StoryOfTwoLattices}. For either $w_0$ or $w_1$ $\le 1/\sqrt{3}$, we see that the spectrum looks completely unchanged from $n=2$ to $n=4$ shells. From $n=2$ to $n=4$ shells, the largest change is smaller than $1\%$, and invisible to the naked eye. Above $n=4$ shells, the spectrum is numerically the same within machine precision. We confirm our first conclusion: \emph{To obtain a faithful model for TBG around the first magic angle, we can safely neglect all shells above $n=2$.} Keeping the $n=2$ shells gives us a Hamiltonian which contains the $A1, B1, A2, B2$ shells in Fig.~\ref{fig:MagicManifold}a, giving a Hamiltonian that is a $72\times 72$ matrix, too large for analytic tackling. Hence further approximations are necessary as per Secs.~\ref{approximationframework1} and~\ref{approximationframework2}, which we further numerically confirm.

\begin{figure}
\begin{centering}
\includegraphics[width=1\linewidth]{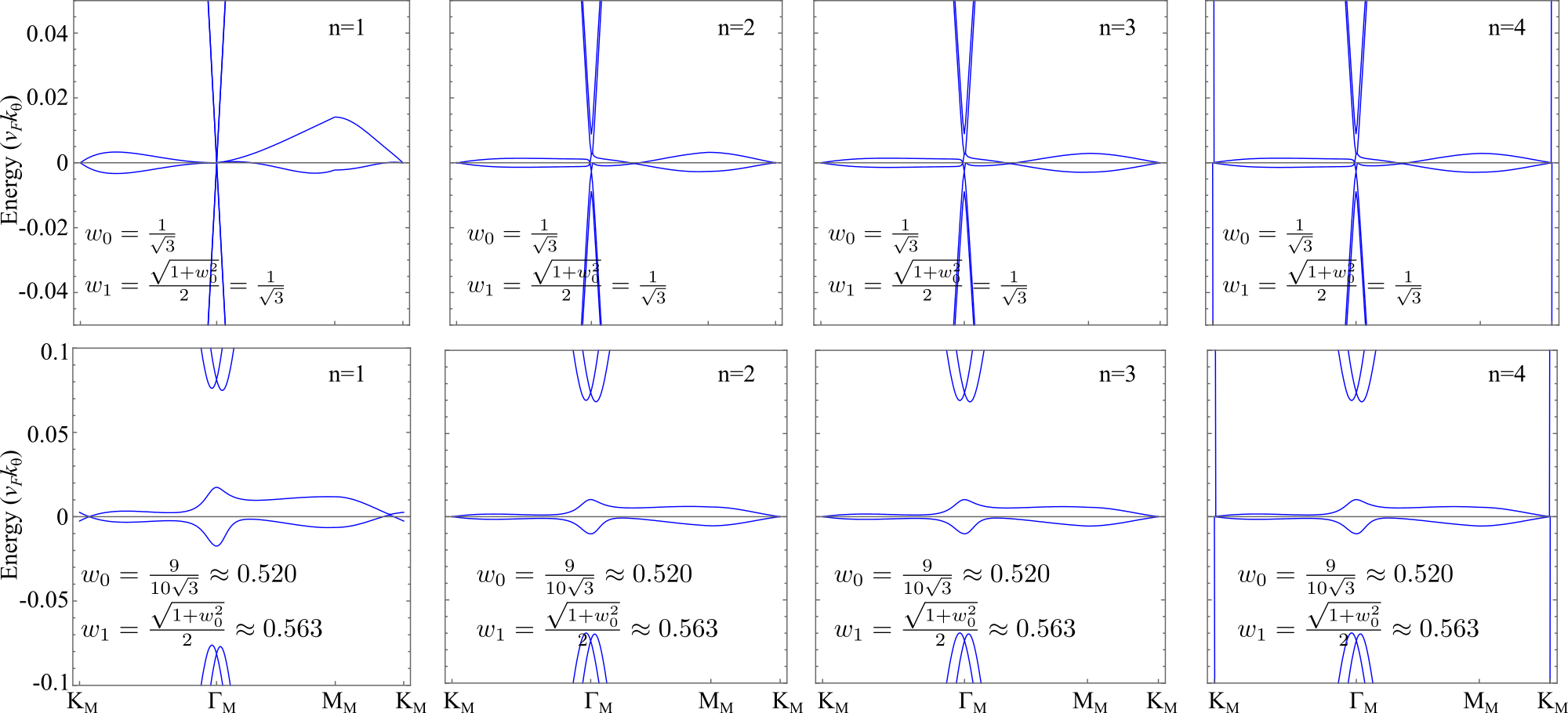}
\par\end{centering}

\protect\caption{\label{fig:TBLGBSShells1} Plots of the band structure for different parameters around the first magic angle, and for different ranges of the $y$-axis. Notice no change from $n=2$ to $n=4$, in agreement with the theoretical discussions}

\end{figure}

\begin{figure}
\begin{centering}
\includegraphics[width=1\linewidth]{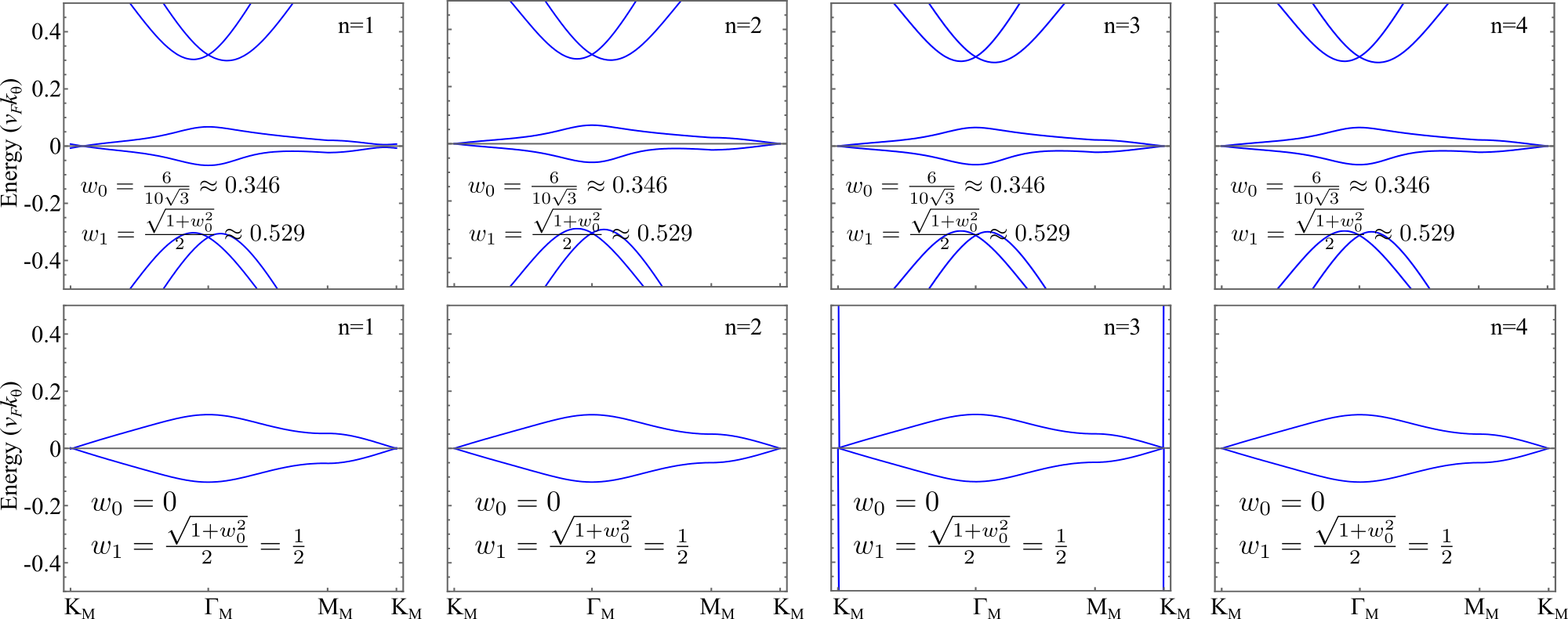}
\par\end{centering}

\protect\caption{\label{fig:TBLGBSShells2} Plots of the band structure for different parameters around the first magic angle, and for different ranges of the $y$-axis. Notice no change from $n=2$ to $n=4$, in agreement with the theoretical discussions}

\end{figure}

\begin{figure}
\begin{centering}
\includegraphics[width=1\linewidth]{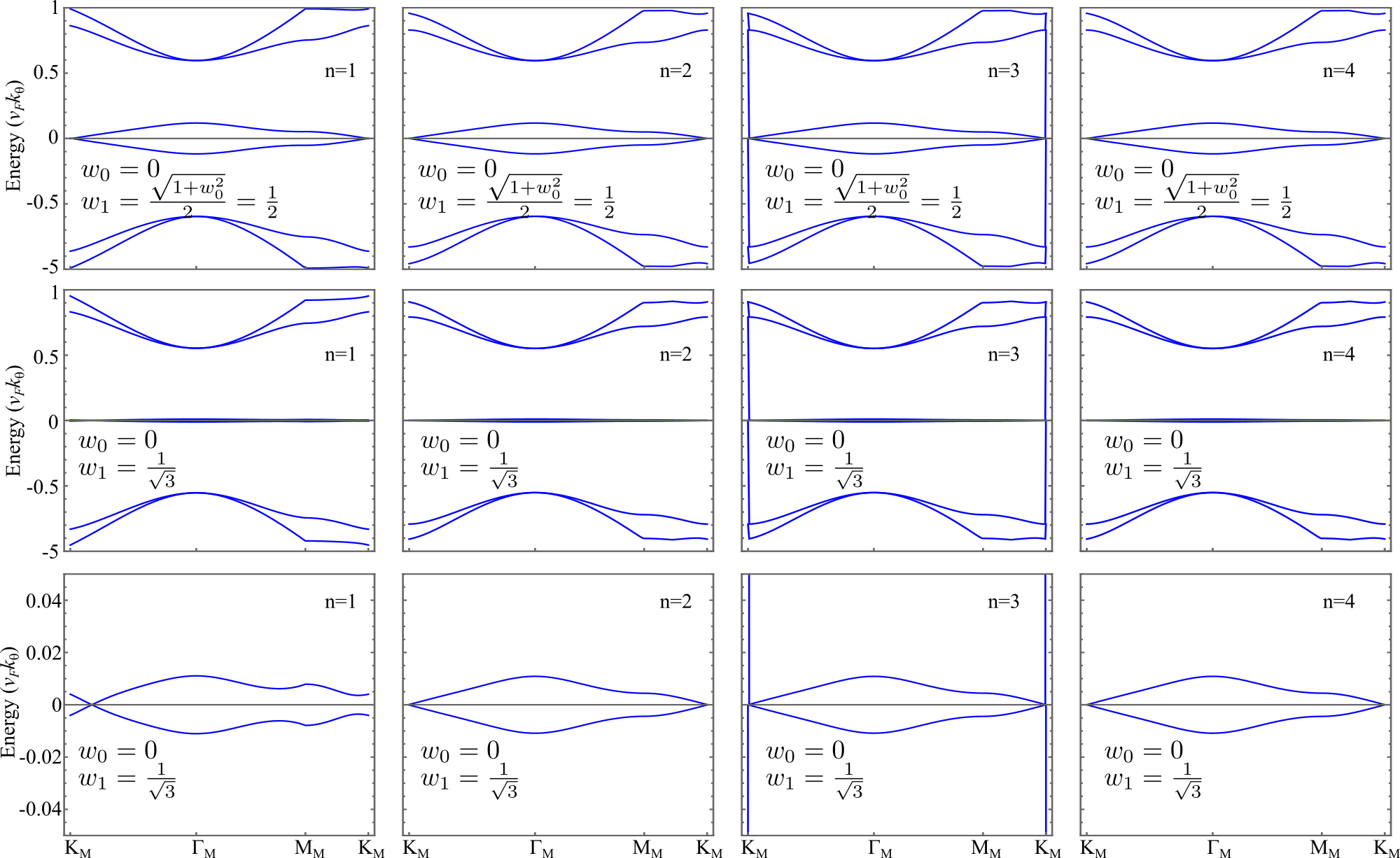}
\par\end{centering}

\protect\caption{\label{fig:TBLGBSShells3} Plots of the band structure for different parameters around the first magic angle, and for different ranges of the $y$-axis. Notice no change from $n=2$ to $n=4$, in agreement with the theoretical discussions}

\end{figure}

\item We confirmed our perturbation theory predictions of Secs.~\ref{approximationframework1} and~\ref{approximationframework2} for angles smaller than the first magic angle. In Fig.~\ref{fig:TBLGBSShells4} we confirm the analytic prediction that at angle $1/n$ times the first magic angle, we can neglect all the shells above $n+1$.

\item We confirmed  our perturbation theory predictions  Secs.~\ref{approximationframework1} and~\ref{approximationframework2} that - for the first magic angle and below, ($w_0, w_1 \le 1/\sqrt{3}$) -  keeping only the first shell induces only a $20\%$ error in the band structure. We have already established that keeping up to $n=2$ shells at the first magic angle gives the  correct band structure within less than $5\%$. Figs.~\ref{fig:TBLGBSShells1},\ref{fig:TBLGBSShells2} and~\ref{fig:TBLGBSShells3} also contain the $n=1$ shells band structure for a range of angles around and above the first magic angle $w_0, w_1\ge 1/\sqrt{3}$. We see that the band structures differ little to very little, while keeping the main characteristics, from  $n=1$ to $n=2$. In particular, in the chiral limit of $w_0=0$ and for $w_1=1/2$ (along what we call the \emph{Second Magic Manifold}, see Sec.~\ref{MagicManifolds}) the band structures do not visibly differ at all (see Fig~.\ref{fig:TBLGBSShells2}, lowest row) from $n=1$ to $n=2$. \emph{Hence for the first magic angle, to make analytic progress, we will consider only the $n=1$ shell, to a good approximation.} This gives a $24\times 24$ Hamiltonian, which is still analytically unsolvable. Hence further approximations are necessary, such as $H_{\text{Approx}1}(\kk)$ in Eq.~\ref{Happrox1}.

\begin{figure}
\begin{centering}
\includegraphics[width=1\linewidth]{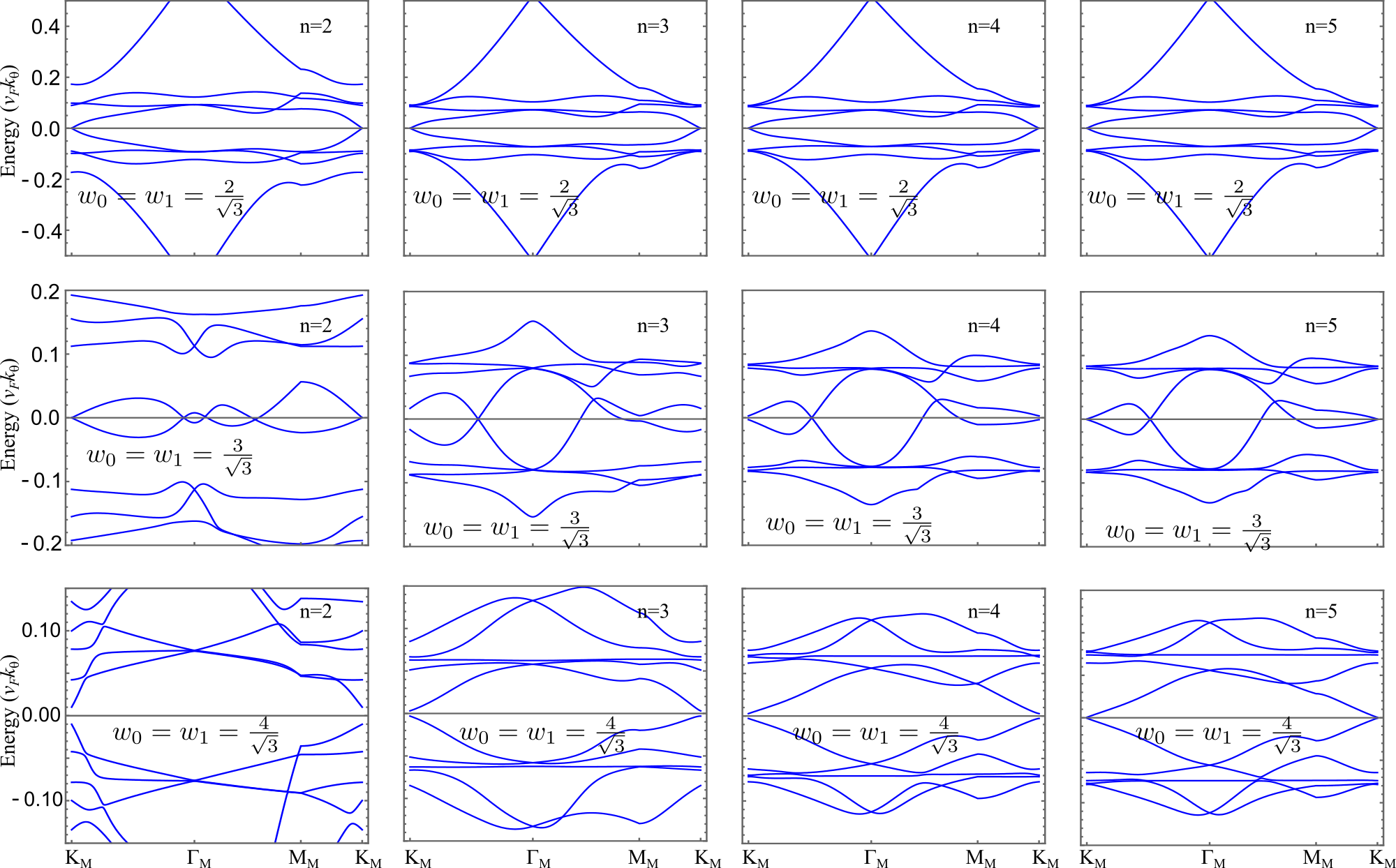}
\par\end{centering}

\protect\caption{\label{fig:TBLGBSShells4} Plots of the band structure for different parameters far away from the first magic angle: at half, a third and a fourth of the first magic angle. Notice that for and angle $1/n$ times the magic angle, we can neglect all shells above $n+1$, which confirms our perturbation theory result. For the first magic angle, above $n=2$ shells the band structure goes not change. For half the magic angle, the band structure above $n=3$ shells does not change (but the band structure at $n=2$ shells is changed compared to the $n=3$ band structure). For a third of the magic angle, the band structure above $n=4$ shells does not change (but the band structure at $n=2,3$ shells is changed compared to the $n=4$ band structure. For a quarter of the magic angle, the band structure above $n=5$ shells does not change (but the band structure at $n=2,3,4$ shells is changed  -- dramatically -- compared to the $n=6$ band structure. }
\end{figure}

\item We test the prediction that $H_{\text{Approx}1}(\kk)$ in Eq.~\ref{Happrox1} approximates well the band structure of TBG around (and for angles larger than) the magic angle for a series of values of $w_0, w_1 \le 1/\sqrt{3}$.  Figs.~\ref{fig:TBLGBSShells5},\ref{fig:TBLGBSShells6} and~\ref{fig:TBLGBSShells7}. We see remarkable agreement between $H_{\text{Approx}1}(\kk)$ and the $n=1$ Hamiltonian. We also see good agreement with the large shell limit. For values of the parameters $w_0=0, w_1= \frac{1}{2}$ in the \emph{Second Magic Manifold}, (see Sec.~\ref{MagicManifolds}), the $H_{\text{Approx}1}(\kk)$ and the $n=1,2,3\ldots$  shells give rise to bands \emph{undistinguishable by eye}(see Fig\ref{fig:TBLGBSShells6},  last row). \emph{\bf{ We will hence use $H_{\text{Approx}1}(\kk)$ as our TBG hamiltonian.}} This is a $12\times 12$ Hamiltonian that cannot be solved analytically. Hence further analytic approximations are necessary.

\begin{figure}
\begin{centering}
\includegraphics[width=1\linewidth]{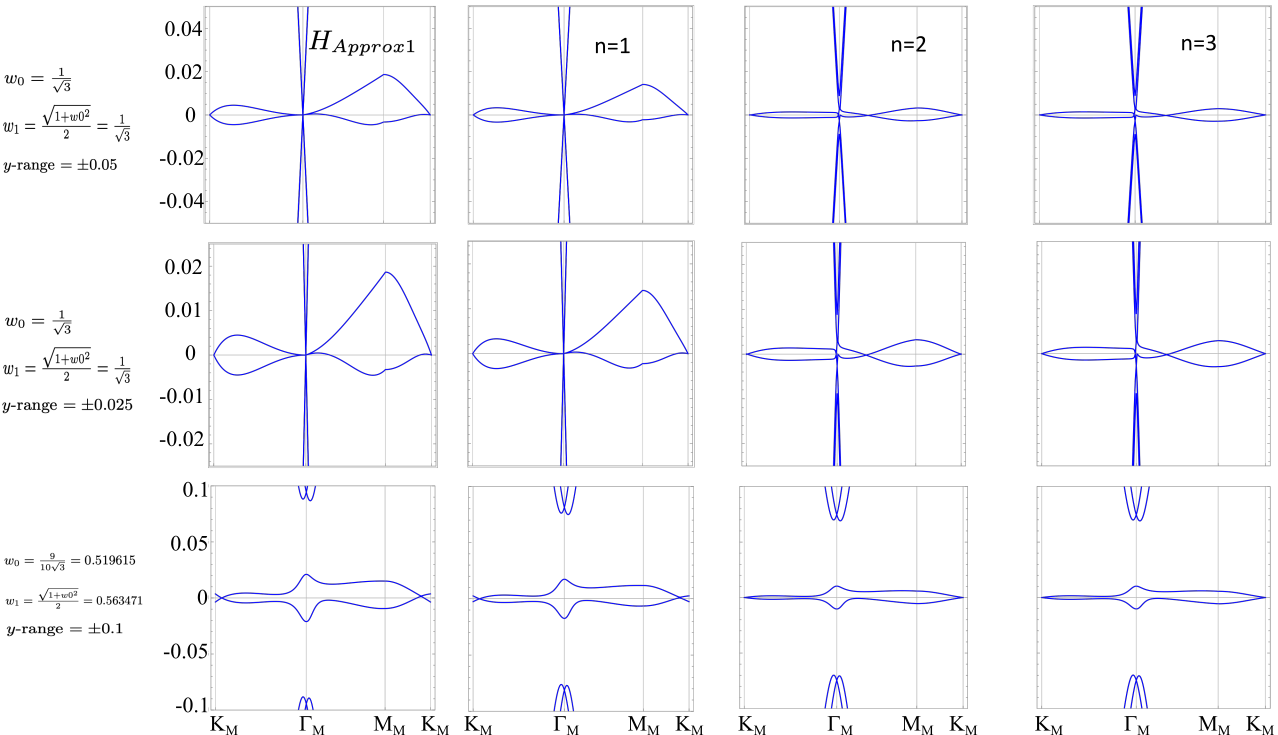}
\par\end{centering}

\protect\caption{\label{fig:TBLGBSShells5} Plots of the band structure of $H_{\text{Approx}1}$ for different parameters around the first magic angle, and for different ranges of the $y$-axis. For convenience, we also re-plot the $n=1,2,3$ shells band structure. Notice the good agreement of $H_{\text{Approx}1}$ with the $n=1$ shell Hamiltonian, and, further-on, the good approximation of the $n=2,3$ band structures by this Hamiltonian. For the chiral limit $w_0=9/10\sqrt{3}, w_1=\sqrt{1+w_0^2}/2$, the approximate $H_{\text{Approx}1}$ is a remarkably good approximation of the $n=1$ shell and a good approximation to the thermodynamic limit, albeit with the Dirac point slightly shifted.}

\end{figure}

\begin{figure}
\begin{centering}
\includegraphics[width=1\linewidth]{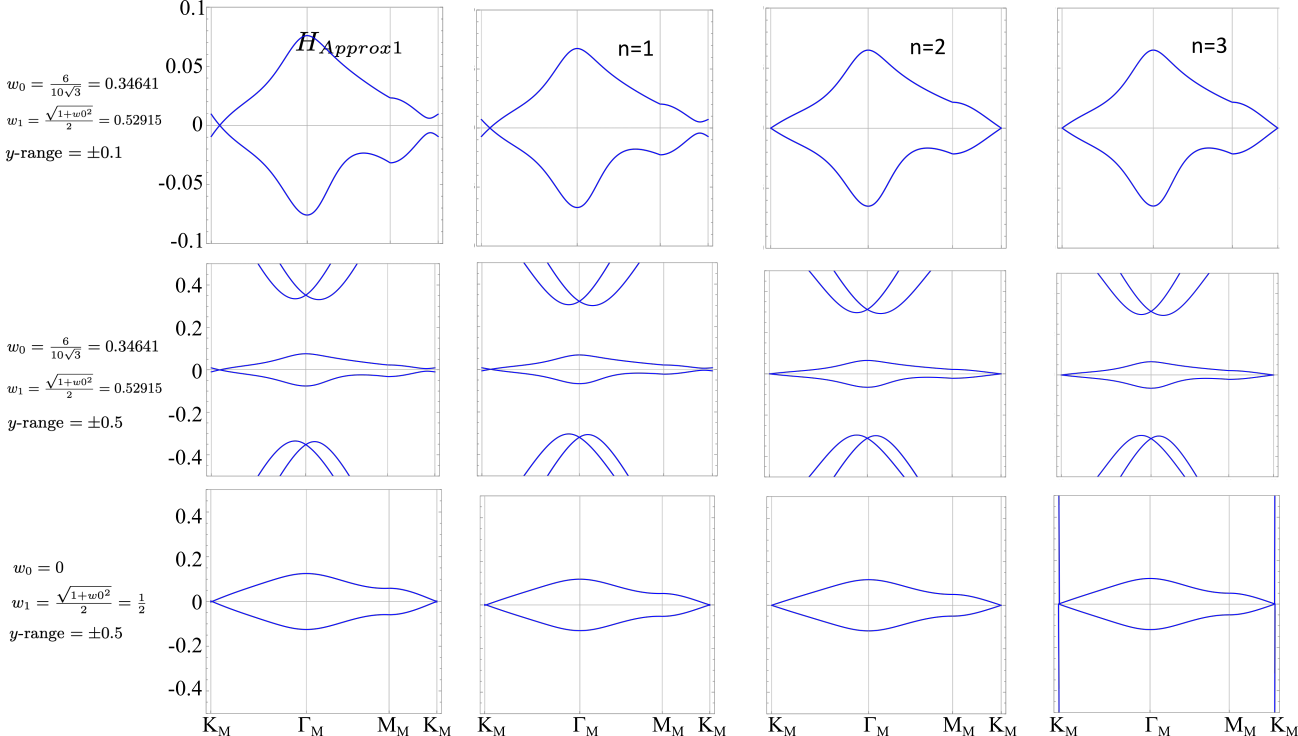}
\par\end{centering}

\protect\caption{\label{fig:TBLGBSShells6} Plots of the band structure of $H_{\text{Approx}1}$ for different parameters around the first magic angle, and for different ranges of the $y$-axis, which helps us focus on different bands. For convenience, we also re-plot the $n=1,2,3$ shells band structure. Notice the remarkable (almost undistinguishable by eye) agreement of $H_{\text{Approx}1}$ with the $n=1$ shell Hamiltonian, and the further-on good approximation of the $n=2,3$ band structures by this Hamiltonian. For the chiral limit $w_0=0, w_1=1/2$, the approximate $H_{\text{Approx}1}$ is a remarkably good approximation of the thermodynamic limit - undistinguishable by eye -, while for all other values it is a very good approximation. The Dirac point in the chiral limit $w_0=0, w_1= \sqrt{1+w_0^2}/2$ is at $K_M$ even for the $H_{\text{Approx}1}$.}

\end{figure}

\begin{figure}
\begin{centering}
\includegraphics[width=1\linewidth]{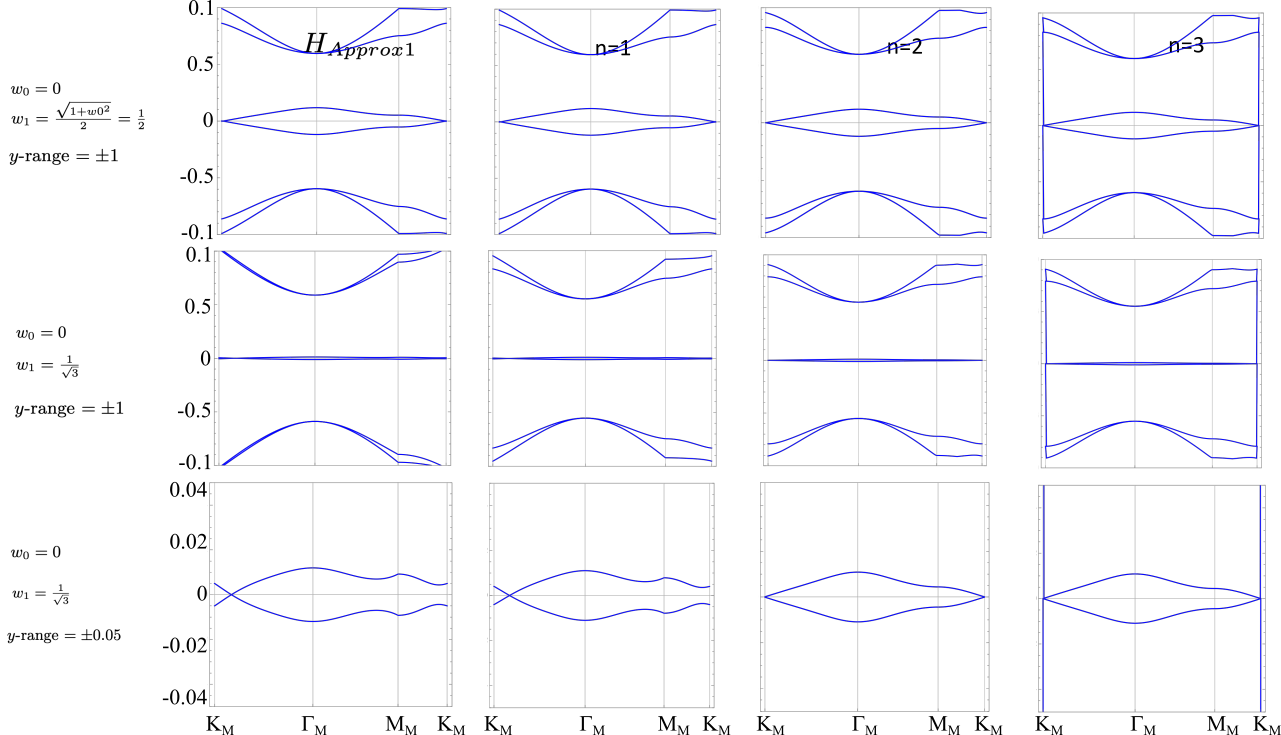}
\par\end{centering}

\protect\caption{\label{fig:TBLGBSShells7} Plots of the band structure of $H_{\text{Approx}1}$ for different parameters around the first magic angle, and for different ranges of the $y$-axis, which helps us focus on different bands. For convenience, we also re-plot the $n=1,2,3$ shells band structure. Notice the remarkable (almost undistinguishable by eye) agreement of $H_{\text{Approx}1}$ with the $n=1$ shell Hamiltonian, and the further-on good approximation of the $n=2,3$ band structures by this Hamiltonian. For the chiral limit $w_0=0, w_1=1/\sqrt{3}$, the approximate $H_{\text{Approx}1}$ is a remarkably good approximation of the $n=1$ Hamiltonian, and a good approximation to the thermodynamic limit. The Dirac point is  slightly moved away from the $K_M$ point.}

\end{figure}

\end{itemize}

\newpage
\section{Eigenstates of the  Hexagon Model at the $\Gamma_M$ point}\label{Appendix3}

We provide the explicit expressions for the six band model approximation for the Hexagon model at $w_0=w_1=1/\sqrt{3}$. The basis we choose is made of simultaneous eigenstates of $C_{3z}$ and $H$ for the states $ | \psi_j(\mathbf{k}=0,  w_0=w_1= \frac{1}{\sqrt{3}}) \rangle  = \psi_{E_j}$ $j=1\ldots 6$ in Eq.~\ref{6foldperturbationofHexagonmodel}:
\begin{equation}
\psi_{E_1}=\left(\begin{array}{c}
\zeta_1\\
e^{-i(2\pi/3)\sigma_z}\eta_1\\
e^{i(2\pi/3)\sigma_z}\zeta_1\\
\eta_1\\
e^{-i(2\pi/3)\sigma_z}\zeta_1\\
e^{i(2\pi/3)\sigma_z}\eta_1\\
\end{array}\right)\ ,
\qquad
\zeta_1=\frac{1}{2\sqrt{2}}\left(\begin{array}{c}
1\\
1
\end{array}\right)\ ,
\qquad
\eta_1=\frac{1}{\sqrt{3}}(-2i\sigma_z-\sigma_y)\zeta_1=\frac{1}{2\sqrt{6}}\left(\begin{array}{c}
-i\\
i
\end{array}\right)\ ,
\end{equation}

\begin{equation}
\psi_{E_2}=\left(\begin{array}{c}
\zeta_2\\
e^{-i(2\pi/3)\sigma_z}\eta_2\\
e^{i(2\pi/3)\sigma_z}\zeta_2\\
\eta_2\\
e^{-i(2\pi/3)\sigma_z}\zeta_2\\
e^{i(2\pi/3)\sigma_z}\eta_2\\
\end{array}\right)\ ,
\qquad
\zeta_2=\frac{1}{2\sqrt{6}}\left(\begin{array}{c}
1\\
-1
\end{array}\right)\ ,
\qquad
\eta_2=\frac{1}{\sqrt{3}}(-2i\sigma_z-\sigma_y)\zeta_2=\frac{1}{2\sqrt{2}}\left(\begin{array}{c}
-i\\
-i
\end{array}\right)\ ,
\end{equation}

\begin{equation}
\begin{split}
\psi_{E_3}=\left(\begin{array}{c}
\zeta_3\\
e^{-i(2\pi/3)(\sigma_z-\sigma_0)}\eta_3\\
e^{i(2\pi/3)(\sigma_z-\sigma_0)}\zeta_3\\
\eta_3\\
e^{-i(2\pi/3)(\sigma_z-\sigma_0)}\zeta_3\\
e^{i(2\pi/3)(\sigma_z-\sigma_0)}\eta_3\\
\end{array}\right)\ ,
&\qquad
\zeta_3=\frac{1}{\sqrt{26(5-\sqrt{13})}}\left(\begin{array}{c}
2\\
3-\sqrt{13}
\end{array}\right)\ , \\
&\eta_3=\frac{1}{\sqrt{3}}(\frac{\sigma_y}{2}+\frac{3i}{2}\sigma_x+i\sigma_z)\zeta_3=\frac{i}{\sqrt{78(5-\sqrt{13})}}\left(\begin{array}{c}
5-\sqrt{13}\\
1+\sqrt{13}
\end{array}\right)\ , \\\end{split}
\end{equation}

\begin{equation}
\begin{split}
\psi_{E_4}=\left(\begin{array}{c}
\zeta_4\\
e^{-i(2\pi/3)(\sigma_z-\sigma_0)}\eta_4\\
e^{i(2\pi/3)(\sigma_z-\sigma_0)}\zeta_4\\
\eta_4\\
e^{-i(2\pi/3)(\sigma_z-\sigma_0)}\zeta_4\\
e^{i(2\pi/3)(\sigma_z-\sigma_0)}\eta_4\\
\end{array}\right)\ ,
&\qquad
\zeta_4=\frac{1}{\sqrt{26(5+\sqrt{13})}}\left(\begin{array}{c}
2\\
3+\sqrt{13}
\end{array}\right)\ , \\
&\eta_4=\frac{1}{\sqrt{3}}(\frac{\sigma_y}{2}+\frac{3i}{2}\sigma_x+i\sigma_z)\zeta_4=\frac{i}{\sqrt{78(5+\sqrt{13})}}\left(\begin{array}{c}
5+\sqrt{13}\\
1-\sqrt{13}
\end{array}\right)\ , \\\end{split}
\end{equation}

\begin{equation}
\begin{split}
\psi_{E_5}=\left(\begin{array}{c}
\zeta_5\\
e^{-i(2\pi/3)(\sigma_z+\sigma_0)}\eta_5\\
e^{i(2\pi/3)(\sigma_z+\sigma_0)}\zeta_5\\
\eta_5\\
e^{-i(2\pi/3)(\sigma_z+\sigma_0)}\zeta_5\\
e^{i(2\pi/3)(\sigma_z+\sigma_0)}\eta_5\\
\end{array}\right)\ ,
&\qquad
\zeta_5=\frac{1}{\sqrt{26(5-\sqrt{13})}}\left(\begin{array}{c}
3-\sqrt{13}\\
2
\end{array}\right)\ , \\
&\eta_5=\frac{1}{\sqrt{3}}(\frac{\sigma_y}{2}-\frac{3i}{2}\sigma_x+i\sigma_z)\zeta_5=\frac{-i}{\sqrt{78(5-\sqrt{13})}}\left(\begin{array}{c}
1+\sqrt{13}\\
5-\sqrt{13}\\
\end{array}\right)\ , \\\end{split}
\end{equation}

\begin{equation}
\begin{split}
\psi_{E_6}=\left(\begin{array}{c}
\zeta_6\\
e^{-i(2\pi/3)(\sigma_z+\sigma_0)}\eta_6\\
e^{i(2\pi/3)(\sigma_z+\sigma_0)}\zeta_6\\
\eta_6\\
e^{-i(2\pi/3)(\sigma_z+\sigma_0)}\zeta_6\\
e^{i(2\pi/3)(\sigma_z+\sigma_0)}\eta_6\\
\end{array}\right)\ ,
&\qquad
\zeta_6=\frac{1}{\sqrt{26(5+\sqrt{13})}}\left(\begin{array}{c}
3+\sqrt{13}\\
2\\
\end{array}\right)\ , \\
&\eta_6=\frac{1}{\sqrt{3}}(\frac{\sigma_y}{2}+\frac{3i}{2}\sigma_x+i\sigma_z)\zeta_6=\frac{-i}{\sqrt{78(5+\sqrt{13})}}\left(\begin{array}{c}
1-\sqrt{13}\\
5+\sqrt{13}\\
\end{array}\right)\ . \\\end{split}
\end{equation}
The basis $\psi_{E_1},\psi_{E_2}$ has $C_{3z}=1$, the basis $\psi_{E_3},\psi_{E_4}$ has $C_{3z}=e^{i2\pi/3}$, and the basis $\psi_{E_5},\psi_{E_6}$ has $C_{3z}=e^{-i2\pi/3}$. 
The 6 by 6 Hamiltonian in Eq.~\ref{6foldperturbationofHexagonmodel} under these 6 basis takes the form
\begin{equation}\label{6foldperturbationofHexagonmodelHamiltonian}
\boxed{H^{\text{6 band}}_{ij} (\mathbf{k}, w_0=w_1= \frac{1}{\sqrt{3}})=\left(\begin{array}{ccc}
0_{2} & A_1 k_- & A_2^\dag k_+\\
A_1^\dag k_+ & 0_2 & A_3 k_-\\
A_2 k_- & A_3^\dag k_+ & 0_2\\
\end{array}\right)\ ,}
\end{equation}
where $k_\pm=k_x\pm i k_y$, $0_2$ is the 2 by 2 zero matrix, and
\begin{equation}
\begin{split}
&A_1=\left(\begin{array}{cc}
\frac{2 \sqrt{13}-13}{13 \sqrt{5-\sqrt{13}}} & \frac{\sqrt{6 \sqrt{13}+22}-1}{\sqrt{13 \left(\sqrt{13}+5\right)}} \\
\frac{1}{52} \left(\sqrt{13}-13\right) \sqrt{\sqrt{13}+5} & \sqrt{\frac{1}{26} \left(\sqrt{13}+4\right)}-\sqrt{\frac{3}{13 \left(\sqrt{13}+5\right)}}
\end{array}\right)\ ,\\
&A_2=\left(\begin{array}{cc}
\frac{2 \sqrt{13}-13}{13 \sqrt{5-\sqrt{13}}} & -\frac{1}{52} \left(\sqrt{13}-13\right) \sqrt{\sqrt{13}+5} \\
\frac{\sqrt{6 \sqrt{13}+22}-1}{\sqrt{13 \left(\sqrt{13}+5\right)}} & -\sqrt{\frac{1}{26} \left(\sqrt{13}+4\right)}+\sqrt{\frac{3}{13 \left(\sqrt{13}+5\right)}}
\end{array}\right)\ ,\\
&A_3=
\left(
\begin{array}{cc}
 \frac{1}{\sqrt{13}} & \frac{2 \sqrt{13}-5 \sqrt{6 \sqrt{13}+22}+\sqrt{78 \sqrt{13}+286}+2}{52 \sqrt{3}} \\
 \frac{2 \sqrt{13}-5 \sqrt{6 \sqrt{13}+22}+\sqrt{78 \sqrt{13}+286}+2}{52 \sqrt{3}} & -\frac{2 \left(\sqrt{13}+8\right)-\sqrt{6 \sqrt{13}+22}+\sqrt{78 \sqrt{13}+286}}{26 \left(\sqrt{13}+2\right)} \\
\end{array}
\right)\ .\\
\end{split}
\end{equation}
We note that $\psi_{E_1},\psi_{E_2}$ also serves as the Gamma point basis of the 2-band approximation at $w_1=\sqrt{1+w_0^2}/2$ in Sec.~\ref{MagicManifolds}.

\section{Eigenstates of  Along the $\Gamma_M -K_M$ line $k_x=0$ and on the $\Gamma_M-M_M$ line $k_y=0$}\label{Appendix4}

\subsection{Eigenstates of $H^{\text{6 band}}_{ij} (\mathbf{k} = (0,k_y) , w_0=w_1= \frac{1}{\sqrt{3}})$  }On the $\Gamma_M-K_M$ line, the energies (already mentioned in the main text) are

\begin{eqnarray}
E_{\text{6-band}} (\mathbf{k} = (0,k_y) , w_0=w_1= \frac{1}{\sqrt{3}})=  (-2 \sqrt{\frac{3}{13}} k_y,-2 \sqrt{\frac{3}{13}} k_y,2 \sqrt{\frac{3}{13}} k_y,2 \sqrt{\frac{3}{13}} k_y,0,0)
\end{eqnarray}
The energies have eigenstates (not orthonormalized yet)
\begin{eqnarray}
&\psi_{1; \text{6-band}} (\mathbf{k} = (0,k_y)  , w_0=w_1= \frac{1}{\sqrt{3}})=\nonumber\\
& (-\frac{1}{200} \sqrt{\frac{1}{221} \left(5570051 i \sqrt{3}-153112 \sqrt{13}+1077176 i \sqrt{39}+17078669\right)},\frac{191760161 i \sqrt{3}+166713618 \sqrt{13}-59265370 i \sqrt{39}-527508405}{200 \sqrt{2074} \left(13477 \sqrt{13}-45994\right)},\nonumber\\,& \frac{-2437915 i \sqrt{3}+698430 \sqrt{13}+569554 i \sqrt{39}-3303424}{100 \sqrt{22570} \left(49 \sqrt{13}-156\right)},\frac{23 i \left(26 i-1222 \sqrt{3}+86 i \sqrt{13}+221 \sqrt{39}\right)}{1300 \sqrt{370}},0,1) \nonumber \\
&\psi_{2; \text{6-band}} (\mathbf{k} = (0,k_y)  , w_0=w_1= \frac{1}{\sqrt{3}})=\nonumber\\
&( \frac{1}{200} (-23) \sqrt{\frac{1}{221} \left(37641 i \sqrt{3}+808 \sqrt{13}-2136 i \sqrt{39}-91159\right)},\frac{23}{100} \sqrt{\frac{705768 \sqrt{13}-8 i \sqrt{39 \left(886369537-160909896 \sqrt{13}\right)}+4606081}{26962}},\nonumber\\, &\frac{23 \left(-881719 i \sqrt{3}+56 \left(-687+3704 i \sqrt{3}\right) \sqrt{13}+52881\right)}{600 \sqrt{22570} \left(49 \sqrt{13}-156\right)},\frac{104 \left(775-596 i \sqrt{3}\right)+529 i \left(25 \sqrt{3}+23 i\right) \sqrt{13}}{2600 \sqrt{370}},1,0) \nonumber \\ 
&\psi_{3; \text{6-band}} (\mathbf{k} = (0,k_y)  , w_0=w_1= \frac{1}{\sqrt{3}})=\nonumber\\&
(\frac{1}{200} \sqrt{\frac{1}{221} \left(5570051 i \sqrt{3}+8 \left(19139-134647 i \sqrt{3}\right) \sqrt{13}+17078669\right)},\frac{-191760161 i \sqrt{3}+166713618 \sqrt{13}-59265370 i \sqrt{39}+527508405}{200 \sqrt{2074} \left(13477 \sqrt{13}+45994\right)},\nonumber\\,&\frac{2437915 i \sqrt{3}+698430 \sqrt{13}+569554 i \sqrt{39}+3303424}{100 \sqrt{22570} \left(49 \sqrt{13}+156\right)},\frac{23 \left(-1222 i \sqrt{3}+86 \sqrt{13}-221 i \sqrt{39}-26\right)}{1300 \sqrt{370}},0,1)
\nonumber \\ 
&\psi_{4; \text{6-band}} (\mathbf{k} = (0,k_y)  , w_0=w_1= \frac{1}{\sqrt{3}})=\nonumber\\&
(\frac{23}{200} \sqrt{\frac{1}{221} i \left(91159 i+37641 \sqrt{3}+808 i \sqrt{13}+2136 \sqrt{39}\right)},\frac{23}{100} \sqrt{\frac{-705768 \sqrt{13}+8 i \sqrt{39 \left(160909896 \sqrt{13}+886369537\right)}+4606081}{26962}},\nonumber\\,&\frac{23 i \left(52881 i+881719 \sqrt{3}+56 \sqrt{13} \left(3704 \sqrt{3}+687 i\right)\right)}{600 \sqrt{22570} \left(49 \sqrt{13}+156\right)},\frac{104 \left(775-596 i \sqrt{3}\right)+529 \left(23-25 i \sqrt{3}\right) \sqrt{13}}{2600 \sqrt{370}},1,0)
\nonumber \\ 
&\psi_{5; \text{6-band}} (\mathbf{k} = (0,k_y) , w_0=w_1= \frac{1}{\sqrt{3}})=\nonumber\\&
(\frac{1}{529} \sqrt{\frac{2}{51}} \left(710-19 i \sqrt{3}\right),\frac{2}{529} \sqrt{\frac{2}{1037}} \left(-2732+659 i \sqrt{3}\right),-\frac{1}{529} \sqrt{\frac{185}{61} \left(2483+5763 i \sqrt{3}\right)},0,\frac{1}{46} \left(47-19 i \sqrt{3}\right),1)
\nonumber \\ 
&\psi_{6; \text{6-band}} (\mathbf{k} =(0,k_y)  , w_0=w_1= \frac{1}{\sqrt{3}})=\nonumber\\&
(\frac{1}{46} \sqrt{\frac{185}{17}} \left(5 \sqrt{3}+11 i\right),\frac{1}{46} \sqrt{\frac{185}{1037}} \left(-57-71 i \sqrt{3}\right),\frac{3 \left(31-46 i \sqrt{3}\right)}{23 \sqrt{61}},1,0,0)
\end{eqnarray}
Fundamentally, what we notice is that the bands are $k_y$ independent!

\subsection{Eigenstates of $H^{\text{6 band}}_{ij} (\mathbf{k} = (k_x,0) , w_0=w_1= \frac{1}{\sqrt{3}})$  } On the $\Gamma_M-M_M$ line, the energies (already mentioned in the main text) are
\begin{eqnarray}
&E_{\text{6-band}} (\mathbf{k} = (k_x,0) , w_0=w_1= \frac{1}{\sqrt{3}})=  \nonumber \\ &(-k_x,-k_x,\frac{1}{26} \left(3 \sqrt{13}+13\right) k_x,\frac{1}{26} \left(3 \sqrt{13}+13\right) k_x,-\frac{1}{26} \left(3 \sqrt{13}-13\right) k_x,-\frac{1}{26} \left(3 \sqrt{13}-13\right) k_x)
\end{eqnarray}
The energies have eigenstates (not orthonormalized yet)
\begin{eqnarray}
&\psi_{1; \text{6-band}} (\mathbf{k} = (k_x,0) , w_0=w_1= \frac{1}{\sqrt{3}})=\nonumber\\
&(-\frac{219 \sqrt{3}+115 i}{52 \sqrt{34}},\frac{1609-63 i \sqrt{3}}{52 \sqrt{2074}},\frac{3 \left(1253+41 i \sqrt{3}\right)}{52 \sqrt{22570}},\frac{69 \left(-5-3 i \sqrt{3}\right)}{52 \sqrt{370}},0,1) \nonumber \\
&\psi_{2; \text{6-band}} (\mathbf{k} = (k_x,0) , w_0=w_1= \frac{1}{\sqrt{3}})=\nonumber\\
&(\frac{69 \sqrt{\frac{3}{34}}}{26},\frac{69 \left(9-i \sqrt{3}\right)}{52 \sqrt{2074}},-\frac{23 i \left(\sqrt{3}-151 i\right)}{52 \sqrt{22570}},\frac{277-112 i \sqrt{3}}{26 \sqrt{370}},1,0) \nonumber \\ 
&\psi_{3; \text{6-band}} (\mathbf{k} = (k_x,0) , w_0=w_1= \frac{1}{\sqrt{3}})=\nonumber\\&
(\frac{7 \left(-10569 i \sqrt{3}+17434 \sqrt{13}-2949 i \sqrt{39}+62876\right)}{\sqrt{34} \left(3 \sqrt{3}-i\right) \left(323 \sqrt{13}-65\right)},\frac{481425 i \sqrt{3}+307265 \sqrt{13}+145119 i \sqrt{39}+1454167}{4 \sqrt{2074} \left(323 \sqrt{13}-65\right)},\nonumber\\,& \frac{9 i \left(10385 i+10526 \sqrt{3}+4333 i \sqrt{13}+736 \sqrt{39}\right)}{2 \sqrt{22570} \left(61 \sqrt{13}-247\right)},\frac{69 \left(169 i \sqrt{3}+8 \sqrt{13}-45 i \sqrt{39}+26\right)}{52 \sqrt{370} \left(8 \sqrt{13}-29\right)},0,1) 
\nonumber \\ 
&\psi_{4; \text{6-band}} (\mathbf{k} = (k_x,0) , w_0=w_1= \frac{1}{\sqrt{3}})=\nonumber\\&
(\frac{69 \left(-1679 i \sqrt{3}+5303 \sqrt{13}-457 i \sqrt{39}+19129\right)}{2 \sqrt{34} \left(3 \sqrt{3}-i\right) \left(323 \sqrt{13}-65\right)},\frac{69 \left(6479 i \sqrt{3}+3374 \sqrt{13}+1939 i \sqrt{39}+12004\right)}{2 \sqrt{2074} \left(323 \sqrt{13}-65\right)}, \nonumber\\,& \frac{23 i \left(16877 i+3295 \sqrt{3}+4843 i \sqrt{13}+2705 \sqrt{39}\right)}{4 \sqrt{22570} \left(61 \sqrt{13}-247\right)},\frac{-36205 i \sqrt{3}-14941 \sqrt{13}+10699 i \sqrt{39}+64675}{104 \sqrt{370} \left(8 \sqrt{13}-29\right)},1,0)
\nonumber \\ 
&\psi_{5; \text{6-band}} (\mathbf{k} = (k_x,0) , w_0=w_1= \frac{1}{\sqrt{3}})=\nonumber\\&
(\frac{69 \left(-1679 i \sqrt{3}+5303 \sqrt{13}-457 i \sqrt{39}+19129\right)}{2 \sqrt{34} \left(3 \sqrt{3}-i\right) \left(323 \sqrt{13}-65\right)},\frac{69 \left(6479 i \sqrt{3}+3374 \sqrt{13}+1939 i \sqrt{39}+12004\right)}{2 \sqrt{2074} \left(323 \sqrt{13}-65\right)},\nonumber\\,& \frac{23 i \left(16877 i+3295 \sqrt{3}+4843 i \sqrt{13}+2705 \sqrt{39}\right)}{4 \sqrt{22570} \left(61 \sqrt{13}-247\right)},\frac{-36205 i \sqrt{3}-14941 \sqrt{13}+10699 i \sqrt{39}+64675}{104 \sqrt{370} \left(8 \sqrt{13}-29\right)},1,0)
\nonumber \\ 
&\psi_{6; \text{6-band}} (\mathbf{k} = (k_x,0) , w_0=w_1= \frac{1}{\sqrt{3}})=\nonumber\\&
(\frac{69 \left(1679 i \sqrt{3}+5303 \sqrt{13}-457 i \sqrt{39}-19129\right)}{2 \sqrt{34} \left(3 \sqrt{3}-i\right) \left(323 \sqrt{13}+65\right)},\frac{69 \left(-6479 i \sqrt{3}+3374 \sqrt{13}+1939 i \sqrt{39}-12004\right)}{2 \sqrt{2074} \left(323 \sqrt{13}+65\right)},\nonumber\\,& \frac{23 \left(-3295 i \sqrt{3}-4843 \sqrt{13}+2705 i \sqrt{39}+16877\right)}{4 \sqrt{22570} \left(61 \sqrt{13}+247\right)},\frac{i \left(64675 i+36205 \sqrt{3}+14941 i \sqrt{13}+10699 \sqrt{39}\right)}{104 \sqrt{370} \left(8 \sqrt{13}+29\right)},1,0)
\end{eqnarray}
Fundamentally, what we notice is that the bands are $k_x$ independent!

\section{Solutions of Eigenstates for the Hexagon Model}\label{Appendix5}

We now solve the eigenvalue equation
\beq
H_{\text{Hex}}(\mathbf{k}, w_0, w_1) \psi  = E \psi
\eeq  for the Hexagon model in Eq.~\ref{HexagonModel1} in the basis $\psi(\mathbf{k}, w_0, w_1) =( {\psi_{A1_1}, \psi_{A1_2},\psi_{A1_3},\psi_{A1_4},\psi_{A1_5},\psi_{A1_6}})(\mathbf{k}, w_0, w_1)$ where each $\psi_{A1_i}(\mathbf{k}, w_0, w_1)$ is a $2$-component spinor of Fig.~\ref{MagicManifold2}, for different values of $\mathbf{k}, w_0, w_1$.

\subsection{Eigenstate solution at $\mathbf{k}=0$ for Arbitrary $w_0, w_1$} 

The eigenvalue equation cannot be solved for general $\kk, w_0, w_1$ and we hence concentrate on several cases. First, we only can solve only the $\kk=0$ point. Using $|\vec{q}_i\cdot \vec{\sigma}| =1$, we find:

\begin{eqnarray}
&\psi_6 = \frac{E+ q_2 \cdot \sigma}{ E^2-1} (T_1 \psi_5+ T_3 \psi_1) \nonumber \\ 
&\psi_4 = \frac{E+ q_1 \cdot \sigma}{ E^2-1} (T_3 \psi_3+ T_2 \psi_5) \nonumber \\ 
&\psi_2 = \frac{E+ q_3 \cdot \sigma}{ E^2-1} (T_2 \psi_1+ T_1 \psi_3) \nonumber \\ 
&[(E+ q_3 \cdot \sigma)(E^2-1) - E(T_2^2+ T_1^2) - T_2 q_1\cdot \sigma T_2 - T_1 q_2 \cdot \sigma T_1]\psi_5 = T_2(E+ q_1\cdot \sigma) T_3 \psi_3  +T_1(E+ q_2 \cdot \sigma) T_3\psi_1 \nonumber \\ 
&[(E+ q_2 \cdot \sigma)(E^2-1) - E(T_1^2+ T_3^2) - T_1 q_3\cdot \sigma T_1 - T_3 q_1 \cdot \sigma T_3]\psi_3 = T_1(E+ q_3\cdot \sigma) T_2 \psi_1  +T_3(E+ q_1 \cdot \sigma) T_2\psi_5 \nonumber \\ 
&[(E+ q_1 \cdot \sigma)(E^2-1) - E(T_2^2+ T_3^2) - T_2 q_3\cdot \sigma T_2 - T_3 q_2 \cdot \sigma T_3]\psi_1 = T_2(E+ q_3\cdot \sigma) T_1 \psi_3  +T_3(E+ q_2 \cdot \sigma) T_1\psi_5 \label{eigenvaluesystem1}
\end{eqnarray} 
where  shorthand notation $T_i = T_i(w_0, w_1)$, $\psi_i= \psi_{A1_i}(\mathbf{k}=0, w_0, w_1)$. Using the expressions of $T_i$ from Eq.~\ref{interlayermatrixelements1}, we re-write the last $3$ equations above as:

\begin{eqnarray} 
&[E(E^2-1)\sigma_0 + q_3 \cdot \sigma(E^2-1 + w_0^2+ 2 w_1^2) - E(2(w_0^2+ w_1^2) \sigma_0 + w_0 w_1(\sigma_x+ \sqrt{3} \sigma_y))] \psi_5 \nonumber \\ &=\{  E[(w_0^2- \frac{w_1^2}{2}) \sigma_0 - w_0 w_1 \sigma_x + \frac{i \sqrt{3}}{2} w_1^2 \sigma_z] + (w_0^2- w_1^2) q_1\cdot \sigma\} \psi_3   \nonumber \\ &+ \{  E[(w_0^2- \frac{w_1^2}{2}) \sigma_0 + w_0 w_1 \frac{1}{2}(\sigma_x- \sqrt{3} \sigma_y)  - \frac{i \sqrt{3}}{2} w_1^2 \sigma_z] + (w_0^2- w_1^2) q_2\cdot \sigma\} \psi_1, \nonumber \\ &[E(E^2-1)\sigma_0 + q_2 \cdot \sigma(E^2-1 + w_0^2+ 2 w_1^2) - E(2(w_0^2+ w_1^2) \sigma_0 + w_0 w_1(\sigma_x- \sqrt{3} \sigma_y))] \psi_3 \nonumber \\ &=\{  E[(w_0^2- \frac{w_1^2}{2}) \sigma_0 + w_0 w_1\frac{1}{2}( \sigma_x+ \sqrt{3} \sigma_y) + \frac{i \sqrt{3}}{2} w_1^2 \sigma_z] + (w_0^2- w_1^2) q_3\cdot \sigma\} \psi_1   \nonumber \\ &+ \{  E[(w_0^2- \frac{w_1^2}{2}) \sigma_0 - w_0 w_1 \sigma_x  - \frac{i \sqrt{3}}{2} w_1^2 \sigma_z] + (w_0^2- w_1^2) q_1\cdot \sigma\} \psi_5, \nonumber \\ &[E(E^2-1)\sigma_0 + q_1 \cdot \sigma(E^2-1 + w_0^2+ 2 w_1^2) - E(2(w_0^2+ w_1^2) \sigma_0 -2 w_0 w_1\sigma_x)] \psi_1 \nonumber \\ &=\{  E[(w_0^2- \frac{w_1^2}{2}) \sigma_0 + w_0 w_1 \frac{1}{2}(\sigma_x+ \sqrt{3} \sigma_y)  - \frac{i \sqrt{3}}{2} w_1^2 \sigma_z] + (w_0^2- w_1^2) q_3\cdot \sigma\} \psi_3   \nonumber \\ &+ \{  E[(w_0^2- \frac{w_1^2}{2}) \sigma_0 + w_0 w_1 \frac{1}{2}(\sigma_x- \sqrt{3} \sigma_y)  + \frac{i \sqrt{3}}{2} w_1^2 \sigma_z] + (w_0^2- w_1^2) q_2\cdot \sigma\} \psi_5. \label{eigenvaluesystem2}
\end{eqnarray} 
Plugging in the expressions for the energy $E$, we can obtain the relations between $\psi_i$. However, these are messy, and we choose to find the eigenstates on several, simpler, manifolds in the $w_0, w_1$ parameter space.

\subsection{Eigenstate solution at $\mathbf{k}=0$ for on the second magic manifold $w_1= \sqrt{1+ w_0^2}/2$} \label{Eigenstatesolutiononthesecondmagicmanifold}

We first solve for the two zero eigenstates $E_{1,2}(\mathbf{k}= 0, w_0 , w_1= \frac{\sqrt{1+w_0^2}}{2}  ) =0$ of Tab.~\ref{tab:6-fold-eigenvalue}. Eq.~\ref{eigenvaluesystem1} become:
\begin{eqnarray}
&(3w_0^2-1) q_3\cdot \sigma \psi_5 = \frac{(3w_0^2-1) }{2} (q_1 \cdot \sigma \psi_3 + q_2 \cdot \sigma \psi_1)\nonumber \\&(3w_0^2-1) q_2\cdot \sigma \psi_3 = \frac{(3w_0^2-1) }{2} (q_3 \cdot \sigma \psi_1 + q_1  \cdot \sigma \psi_5)\nonumber \\&(3w_0^2-1) q_1\cdot \sigma \psi_1 = \frac{(3w_0^2-1) }{2} (q_3 \cdot \sigma \psi_3 + q_2 \cdot \sigma \psi_5)\label{eigenvaluesystem3}
\end{eqnarray}
We now have two cases

\subsubsection{Zero Energy eigenstate solution at $\mathbf{k}=0$ for on the second magic manifold $w_1= \sqrt{1+ w_0^2}/2$, $w_0\ne1/\sqrt{3}$ }\label{Eigenstatesolutiononthesecondmagicmanifold-a}
In this case $3w_0^2 -1\ne 0$ and Eq.~\ref{eigenvaluesystem3} become:
\beq
q_3\cdot \sigma \psi_5 = \frac{1 }{2} (q_1 \cdot \sigma \psi_3 + q_2 \cdot \sigma \psi_1);\;\;  q_2\cdot \sigma \psi_3 = \frac{1 }{2} (q_3 \cdot \sigma \psi_1 + q_1  \cdot \sigma \psi_5); \;\; q_1\cdot \sigma \psi_1 = \frac{1 }{2} (q_3 \cdot \sigma \psi_3 + q_2 \cdot \sigma \psi_5) \label{eigenvaluesystem4}
\eeq with solutions (for the two zero energy eigenstates)
\begin{eqnarray}
& \psi_1= (q_3 \cdot \sigma)(q_2 \cdot \sigma)\psi_3;\nonumber \\ 
&\psi_5= (q_2 \cdot \sigma)(q_3 \cdot \sigma) \psi_3;\nonumber \\ 
&\psi_4 = - q_1 \cdot \sigma(T_3+ T_2  (q_2 \cdot \sigma)(q_3 \cdot \sigma)) \psi_3;\nonumber \\ 
&\psi_2 = - q_3 \cdot \sigma(T_1+ T_2  (q_3 \cdot \sigma)(q_2 \cdot \sigma)) \psi_3;\nonumber \\ 
&\psi_6 = - q_2 \cdot \sigma(T_3  (q_3 \cdot \sigma)(q_2 \cdot \sigma)+ T_1  (q_2 \cdot \sigma)(q_3 \cdot \sigma)) \psi_3;
\end{eqnarray} The two independent zero energy eigenstates on the second magic manifold can be obtained by taking $\psi_3=(1,0)$ and $\psi_3=(0,1)$, respectively. However, they are not orthonormal and a further Gram-Schmidt must be performed to orthogonalize them. We obtain:
\begin{eqnarray}
&\psi_{E_1=0}(\mathbf{k}=0, w_0, w_1=\frac{\sqrt{1+w_0^2}}{2}  ) =\nonumber \\ & (-\frac{i \left(\sqrt{3}-i\right)}{2 \sqrt{6} \sqrt{w_0^2+1}},0,-\frac{\sqrt[6]{-1}}{\sqrt{6}},\frac{i w_0}{\sqrt{6} \sqrt{w_0^2+1}},\frac{1}{\sqrt{6} \sqrt{w_0^2+1}},0,-\frac{(-1)^{5/6}}{\sqrt{6}},-\frac{(-1)^{5/6} w_0}{\sqrt{6} \sqrt{w_0^2+1}},\frac{i \left(\sqrt{3}+i\right)}{2 \sqrt{6} \sqrt{w_0^2+1}},0,\frac{i}{\sqrt{6}},-\frac{\sqrt[6]{-1} w_0}{\sqrt{6} \sqrt{w_0^2+1}})\nonumber \\ 
&\psi_{E_2=0}(\mathbf{k}=0, w_0, w_1=\frac{\sqrt{1+w_0^2}}{2}  ) = \nonumber \\ &(\frac{i \left(\sqrt{3}+i\right) w_0}{2 \sqrt{6} \sqrt{w_0^2+1}},\frac{i \left(\sqrt{3}+i\right)}{2 \sqrt{6}},0,\frac{(-1)^{5/6}}{\sqrt{6} \sqrt{w_0^2+1}},-\frac{\sqrt[3]{-1} w_0}{\sqrt{6} \sqrt{w_0^2+1}},\frac{1}{\sqrt{6}},0,\frac{\sqrt[6]{-1}}{\sqrt{6} \sqrt{w_0^2+1}},\frac{w_0}{\sqrt{6} \sqrt{w_0^2+1}},-\frac{i \left(\sqrt{3}-i\right)}{2 \sqrt{6}},0,-\frac{i}{\sqrt{6} \sqrt{w_0^2+1}}) \label{zeroenergyeigenstatesonsecondmanifold}
\end{eqnarray}

\subsubsection{Non-zero Energy Eigenstate solutions at $\mathbf{k}=0$ for on the second magic manifold $w_1= \sqrt{1+ w_0^2}/2$, $w_0\ne1/\sqrt{3}$ } 

We can adopt the same strategy to build the other, non-zero energy  orthonormal eigenstates. It is tedious (analytic diagonalization programs such as Mathematica fail to provide a result, hence the algebra must be performed by hand) to write the details, but the final answer is, for the eigenstates of energies on the first magic manifold given in Tab.~\ref{tab:EnergiesAtMagicManifold1}:

\begin{eqnarray}
&\psi_{E_{3}}(\mathbf{k}=0, w_0, w_1=\frac{\sqrt{1+w_0^2}}{2}  ) = \frac{1}{4 \sqrt{6} \sqrt{\left(w_0^2+4\right) \left(10 w_0^2+1\right)}}  \nonumber \\&\times
(\left(\sqrt{3}+3 i\right) \left(\sqrt{3} w_0^2+i \sqrt{10 w_0^4+41 w_0^2+4}\right),\left(\sqrt{3}+3 i\right) \left(-2 \sqrt{10 w_0^2+1}+i w_0 \sqrt{w_0^2+1}\right),\nonumber \\&,-\left(\sqrt{3}-3 i\right) \left(2 \sqrt{w_0^2+1}+\sqrt{3} w_0 \sqrt{w_0^2+4}-i w_0 \sqrt{10 w_0^2+1}\right),-2 i \left(\sqrt{3} \sqrt{w_0^2+1} \sqrt{w_0^2+4}+6 w_0\right),12 w_0^2,0,\nonumber \\&,-\left(\sqrt{3}+3 i\right) \left(-2 \sqrt{w_0^2+1}+\sqrt{3} w_0 \sqrt{w_0^2+4}+i w_0 \sqrt{10 w_0^2+1}\right),\left(\sqrt{3}-i\right) \left(\sqrt{3} \sqrt{w_0^2+1} \sqrt{w_0^2+4}-6 w_0\right),\nonumber \\&,\left(\sqrt{3}-3 i\right) \left(\sqrt{3} w_0^2-i \sqrt{10 w_0^4+41 w_0^2+4}\right),-2 \sqrt{3} \left(2 \sqrt{10 w_0^2+1}-i w_0 \sqrt{w_0^2+1}\right),-12 w_0 \sqrt{w_0^2+4},-12 \left(\sqrt{3}+i\right) w_0)
\nonumber \\ 
&\psi_{E_{4}}(\mathbf{k}=0, w_0, w_1=\frac{\sqrt{1+w_0^2}}{2}  ) = \frac{1}{4 \sqrt{6} \sqrt{\left(w_0^2+4\right) \left(10 w_0^2+1\right)}} \nonumber \\&\times
(-\left(\sqrt{3}+i\right) \left(2 \sqrt{10 w_0^2+1}+i w_0 \sqrt{w_0^2+1}\right),-\left(\sqrt{3}+i\right) \left(3 \sqrt{3} w_0^2+i \sqrt{10 w_0^4+41 w_0^2+4}\right),\nonumber \\&, 2 \left(\sqrt{w_0^4+5 w_0^2+4}-6 \sqrt{3} w_0\right),\left(\sqrt{3}-i\right) \left(-2 \sqrt{w_0^2+1}+3 \sqrt{3} w_0 \sqrt{w_0^2+4}-i w_0 \sqrt{10 w_0^2+1}\right),\nonumber \\&,4 (-1)^{5/6} \left(2 \sqrt{10 w_0^2+1}+i w_0 \sqrt{w_0^2+1}\right),4 \sqrt{10 w_0^4+41 w_0^2+4},i \left(\sqrt{3}+i\right) \sqrt{w_0^4+5 w_0^2+4}-6 \left(\sqrt{3}-3 i\right) w_0,\nonumber \\&, \left(\sqrt{3}+i\right) \left(2 \sqrt{w_0^2+1}+3 \sqrt{3} w_0 \sqrt{w_0^2+4}+i w_0 \sqrt{10 w_0^2+1}\right),-2 w_0 \sqrt{w_0^2+1}+4 i \sqrt{10 w_0^2+1},\nonumber \\&,-\left(\sqrt{3}-i\right) \left(3 \sqrt{3} w_0^2-i \sqrt{10 w_0^4+41 w_0^2+4}\right),2 \left(1+i \sqrt{3}\right) \sqrt{w_0^4+5 w_0^2+4},-4 w_0 \sqrt{10 w_0^2+1}+8 i \sqrt{w_0^2+1}) \label{energyeigenstatesonsecondmanifold1}
\end{eqnarray}
\begin{eqnarray}
&\psi_{E_{5}}(\mathbf{k}=0, w_0, w_1=\frac{\sqrt{1+w_0^2}}{2}  ) = \frac{1}{4 \sqrt{6} \sqrt{\left(w_0^2+4\right) \left(10 w_0^2+1\right)}} \nonumber \\&\times
(\left(\sqrt{3}+3 i\right) \left(\sqrt{3} w_0^2+i \sqrt{10 w_0^4+41 w_0^2+4}\right),\left(\sqrt{3}+3 i\right) \left(2 \sqrt{10 w_0^2+1}+i w_0 \sqrt{w_0^2+1}\right),\nonumber \\&, \left(\sqrt{3}-3 i\right) \left(-2 \sqrt{w_0^2+1}+\sqrt{3} w_0 \sqrt{w_0^2+4}-i w_0 \sqrt{10 w_0^2+1}\right),2 i \left(\sqrt{3} \sqrt{w_0^2+1} \sqrt{w_0^2+4}-6 w_0\right),12 w_0^2,0,\nonumber \\&, \left(\sqrt{3}+3 i\right) \left(2 \sqrt{w_0^2+1}+\sqrt{3} w_0 \sqrt{w_0^2+4}+i w_0 \sqrt{10 w_0^2+1}\right),i \left(\sqrt{3}+3 i\right) \sqrt{w_0^2+1} \sqrt{w_0^2+4}-6 \left(\sqrt{3}-i\right) w_0,\nonumber \\&,\left(\sqrt{3}-3 i\right) \left(\sqrt{3} w_0^2-i \sqrt{10 w_0^4+41 w_0^2+4}\right),2 \sqrt{3} \left(2 \sqrt{10 w_0^2+1}+i w_0 \sqrt{w_0^2+1}\right),12 w_0 \sqrt{w_0^2+4},-12 \left(\sqrt{3}+i\right) w_0)
 \nonumber \\  
 &\psi_{E_{6}}(\mathbf{k}=0, w_0, w_1=\frac{\sqrt{1+w_0^2}}{2}  ) = \frac{1}{4 \sqrt{6} \sqrt{\left(w_0^2+4\right) \left(10 w_0^2+1\right)}} \nonumber \\&\times
 (2 \sqrt[6]{-1} \left(2 \sqrt{10 w_0^2+1}-i w_0 \sqrt{w_0^2+1}\right),-\left(\sqrt{3}+i\right) \left(3 \sqrt{3} w_0^2+i \sqrt{10 w_0^4+41 w_0^2+4}\right),\nonumber \\&, -2 \left(\sqrt{w_0^4+5 w_0^2+4}+6 \sqrt{3} w_0\right), -\left(\sqrt{3}-i\right) \left(2 \sqrt{w_0^2+1}+3 \sqrt{3} w_0 \sqrt{w_0^2+4}-i w_0 \sqrt{10 w_0^2+1}\right),\nonumber \\&, -4 (-1)^{5/6} \left(2 \sqrt{10 w_0^2+1}-i w_0 \sqrt{w_0^2+1}\right),4 \sqrt{10 w_0^4+41 w_0^2+4},\left(1-i \sqrt{3}\right) \sqrt{w_0^4+5 w_0^2+4}-6 \left(\sqrt{3}-3 i\right) w_0,\nonumber \\&,-\left(\sqrt{3}+i\right) \left(-2 \sqrt{w_0^2+1}+3 \sqrt{3} w_0 \sqrt{w_0^2+4}+i w_0 \sqrt{10 w_0^2+1}\right),-2 w_0 \sqrt{w_0^2+1}-4 i \sqrt{10 w_0^2+1},\nonumber \\&,-\left(\sqrt{3}-i\right) \left(3 \sqrt{3} w_0^2-i \sqrt{10 w_0^4+41 w_0^2+4}\right),-2 i \left(\sqrt{3}-i\right) \sqrt{w_0^4+5 w_0^2+4},4 w_0 \sqrt{10 w_0^2+1}+8 i \sqrt{w_0^2+1})\label{energyeigenstatesonsecondmanifold2}
\end{eqnarray}
\begin{eqnarray} &\psi_{E_{7}}(\mathbf{k}=0, w_0, w_1=\frac{\sqrt{1+w_0^2}}{2}  ) =\frac{1}{4 \sqrt{6} \sqrt{\left(w_0^2+4\right) \left(10 w_0^2+1\right)}} \nonumber \\&\times (\left(\sqrt{3}+3 i\right) \left(\sqrt{3} w_0^2-i \sqrt{10 w_0^4+41 w_0^2+4}\right),\left(\sqrt{3}+3 i\right) \left(2 \sqrt{10 w_0^2+1}+i w_0 \sqrt{w_0^2+1}\right),\nonumber \\&,-\left(\sqrt{3}-3 i\right) \left(2 \sqrt{w_0^2+1}+\sqrt{3} w_0 \sqrt{w_0^2+4}+i w_0 \sqrt{10 w_0^2+1}\right),-2 i \left(\sqrt{3} \sqrt{w_0^2+1} \sqrt{w_0^2+4}+6 w_0\right),12 w_0^2,0,\nonumber \\&, -\left(\sqrt{3}+3 i\right) \left(-2 \sqrt{w_0^2+1}+\sqrt{3} w_0 \sqrt{w_0^2+4}-i w_0 \sqrt{10 w_0^2+1}\right),\left(\sqrt{3}-i\right) \left(\sqrt{3} \sqrt{w_0^2+1} \sqrt{w_0^2+4}-6 w_0\right),\nonumber \\&,\left(\sqrt{3}-3 i\right) \left(\sqrt{3} w_0^2+i \sqrt{10 w_0^4+41 w_0^2+4}\right),2 \sqrt{3} \left(2 \sqrt{10 w_0^2+1}+i w_0 \sqrt{w_0^2+1}\right),-12 w_0 \sqrt{w_0^2+4},-12 \left(\sqrt{3}+i\right) w_0)
\nonumber \\ 
 &\psi_{E_{8}}(\mathbf{k}=0, w_0, w_1=\frac{\sqrt{1+w_0^2}}{2}  ) =\frac{1}{4 \sqrt{6} \sqrt{\left(w_0^2+4\right) \left(10 w_0^2+1\right)}} \nonumber \\&\times 
 (-\left(\sqrt{3}+i\right) \left(2 \sqrt{10 w_0^2+1}-i w_0 \sqrt{w_0^2+1}\right),\left(\sqrt{3}+i\right) \left(3 \sqrt{3} w_0^2-i \sqrt{10 w_0^4+41 w_0^2+4}\right),\nonumber \\&,-2 \left(\sqrt{w_0^4+5 w_0^2+4}-6 \sqrt{3} w_0\right),-\left(\sqrt{3}-i\right) \left(-2 \sqrt{w_0^2+1}+3 \sqrt{3} w_0 \sqrt{w_0^2+4}+i w_0 \sqrt{10 w_0^2+1}\right),\nonumber \\&,-2 \left(\sqrt{3}-i\right) \left(2 \sqrt{10 w_0^2+1}-i w_0 \sqrt{w_0^2+1}\right),4 \sqrt{10 w_0^4+41 w_0^2+4},\left(1-i \sqrt{3}\right) \sqrt{w_0^4+5 w_0^2+4}+6 \left(\sqrt{3}-3 i\right) w_0,\nonumber \\&,-\left(\sqrt{3}+i\right) \left(2 \sqrt{w_0^2+1}+3 \sqrt{3} w_0 \sqrt{w_0^2+4}-i w_0 \sqrt{10 w_0^2+1}\right),2 w_0 \sqrt{w_0^2+1}+4 i \sqrt{10 w_0^2+1},\nonumber \\&,\left(\sqrt{3}-i\right) \left(3 \sqrt{3} w_0^2+i \sqrt{10 w_0^4+41 w_0^2+4}\right),-2 i \left(\sqrt{3}-i\right) \sqrt{w_0^4+5 w_0^2+4},-4 w_0 \sqrt{10 w_0^2+1}-8 i \sqrt{w_0^2+1})
\label{energyeigenstatesonsecondmanifold3}
 \end{eqnarray}
 \begin{eqnarray}
 &\psi_{E_{9}}(\mathbf{k}=0, w_0, w_1=\frac{\sqrt{1+w_0^2}}{2}  ) =\frac{1}{4 \sqrt{6} \sqrt{\left(w_0^2+4\right) \left(10 w_0^2+1\right)}} \nonumber \\&\times
(\left(\sqrt{3}+3 i\right) \left(\sqrt{3} w_0^2-i \sqrt{10 w_0^4+41 w_0^2+4}\right),-\left(\sqrt{3}+3 i\right) \left(2 \sqrt{10 w_0^2+1}-i w_0 \sqrt{w_0^2+1}\right),\nonumber \\&,\left(\sqrt{3}-3 i\right) \left(-2 \sqrt{w_0^2+1}+\sqrt{3} w_0 \sqrt{w_0^2+4}+i w_0 \sqrt{10 w_0^2+1}\right),2 i \left(\sqrt{3} \sqrt{w_0^2+1} \sqrt{w_0^2+4}-6 w_0\right),12 w_0^2,0,\nonumber \\&, \left(\sqrt{3}+3 i\right) \left(2 \sqrt{w_0^2+1}+\sqrt{3} w_0 \sqrt{w_0^2+4}-i w_0 \sqrt{10 w_0^2+1}\right),-\left(\sqrt{3}-i\right) \left(\sqrt{3} \sqrt{w_0^2+1} \sqrt{w_0^2+4}+6 w_0\right),\nonumber \\&,\left(\sqrt{3}-3 i\right) \left(\sqrt{3} w_0^2+i \sqrt{10 w_0^4+41 w_0^2+4}\right),-2 \sqrt{3} \left(2 \sqrt{10 w_0^2+1}-i w_0 \sqrt{w_0^2+1}\right),12 w_0 \sqrt{w_0^2+4},-12 \left(\sqrt{3}+i\right) w_0)
 \nonumber \\ 
 &\psi_{E_{10}}(\mathbf{k}=0, w_0, w_1=\frac{\sqrt{1+w_0^2}}{2}  ) =\frac{1}{4 \sqrt{6} \sqrt{\left(w_0^2+4\right) \left(10 w_0^2+1\right)}} \nonumber \\&\times
(\left(\sqrt{3}+i\right) \left(2 \sqrt{10 w_0^2+1}+i w_0 \sqrt{w_0^2+1}\right),\left(\sqrt{3}+i\right) \left(3 \sqrt{3} w_0^2-i \sqrt{10 w_0^4+41 w_0^2+4}\right),\nonumber \\&,2 \left(\sqrt{w_0^4+5 w_0^2+4}+6 \sqrt{3} w_0\right),\left(\sqrt{3}-i\right) \left(2 \sqrt{w_0^2+1}+3 \sqrt{3} w_0 \sqrt{w_0^2+4}+i w_0 \sqrt{10 w_0^2+1}\right),\nonumber \\&,4 \sqrt[3]{-1} \left(w_0 \sqrt{w_0^2+1}-2 i \sqrt{10 w_0^2+1}\right),4 \sqrt{10 w_0^4+41 w_0^2+4},i \left(\sqrt{3}+i\right) \sqrt{w_0^4+5 w_0^2+4}+6 \left(\sqrt{3}-3 i\right) w_0,\nonumber \\&,\left(\sqrt{3}+i\right) \left(-2 \sqrt{w_0^2+1}+3 \sqrt{3} w_0 \sqrt{w_0^2+4}-i w_0 \sqrt{10 w_0^2+1}\right),2 w_0 \sqrt{w_0^2+1}-4 i \sqrt{10 w_0^2+1},\nonumber \\&,\left(\sqrt{3}-i\right) \left(3 \sqrt{3} w_0^2+i \sqrt{10 w_0^4+41 w_0^2+4}\right),2 \left(1+i \sqrt{3}\right) \sqrt{w_0^4+5 w_0^2+4},4 w_0 \sqrt{10 w_0^2+1}-8 i \sqrt{w_0^2+1})\label{energyeigenstatesonsecondmanifold4}
\end{eqnarray}

\begin{eqnarray}
&\psi_{E_{11}}(\mathbf{k}=0, w_0, w_1=\frac{\sqrt{1+w_0^2}}{2}  )=  \nonumber \\& (\frac{\left(\sqrt{3}-3 i\right) (w_0+i)}{12 \sqrt{w_0^2+1}},\frac{1}{12} \left(-\sqrt{3}+3 i\right),-\frac{1}{2 \sqrt{3}},-\frac{\sqrt[3]{-1} (w_0+i)}{2 \sqrt{3} \sqrt{w_0^2+1}},\frac{\left(\sqrt{3}+3 i\right) (w_0+i)}{12 \sqrt{w_0^2+1}},\frac{1}{2 \sqrt{3}}, \nonumber \\&, \frac{1}{12} \left(\sqrt{3}-3 i\right),-\frac{\left(\sqrt{3}-3 i\right) (w_0+i)}{12 \sqrt{w_0^2+1}},-\frac{w_0+i}{2 \sqrt{3} \sqrt{w_0^2+1}},\frac{1}{12} \left(-\sqrt{3}-3 i\right),\frac{1}{12} \left(\sqrt{3}+3 i\right),\frac{w_0+i}{2 \sqrt{3} \sqrt{w_0^2+1}}) \label{energyeigenstatesonsecondmanifold5}
\end{eqnarray}
\begin{eqnarray}
&\psi_{E_{12}}(\mathbf{k}=0, w_0, w_1=\frac{\sqrt{1+w_0^2}}{2}  )=\nonumber \\&(\frac{\left(\sqrt{3}-3 i\right) (w_0-i)}{12 \sqrt{w_0^2+1}},\frac{1}{12} \left(-\sqrt{3}+3 i\right),\frac{1}{2 \sqrt{3}},\frac{\left(\sqrt{3}+3 i\right) (w_0-i)}{12 \sqrt{w_0^2+1}},\frac{\left(\sqrt{3}+3 i\right) (w_0-i)}{12 \sqrt{w_0^2+1}},\frac{1}{2 \sqrt{3}},\nonumber \\&, \frac{1}{12} \left(-\sqrt{3}+3 i\right),\frac{\left(\sqrt{3}-3 i\right) (w_0-i)}{12 \sqrt{w_0^2+1}},-\frac{w_0-i}{2 \sqrt{3} \sqrt{w_0^2+1}},\frac{1}{12} \left(-\sqrt{3}-3 i\right),\frac{1}{12} \left(-\sqrt{3}-3 i\right),-\frac{w_0-i}{2 \sqrt{3} \sqrt{w_0^2+1}})\label{energyeigenstatesonsecondmanifold6}
\end{eqnarray}

\subsubsection{Zero Energy eigenstate solution at $\mathbf{k}=0$ for on the second magic manifold $w_1= \sqrt{1+ w_0^2}/2=w_0=1/\sqrt{3}$ } \label{6foldeigenstatessolution}
There are 6 zero Energies in Tab.~\ref{tab:6-fold-eigenvalue} at this point  $w_1= \sqrt{1+ w_0^2}/2=w_0=1/\sqrt{3}$.  They have already been given in App.~\ref{Appendix3}.

\section{Perturbation Theory for $H^{(1)}_{m m'}  (\kk, w_0) =0$, $E_m=0$ manifold}\label{fifthorderperturbation}

\subsection{Review Of Perturbation Theory}
We review the perturbation theory being performed in the main text. This formalism was first presented in Ref.~\cite{winklerbook}, but we go to higher order in current perturbation theory. We have a Hamiltonian $H^0$ whose eigenstates we know, and is hence purely diagonal in its eigenstate basis. We also have a perturbation Hamiltonian $H'$, with both diagonal and off-diagonal elements. Amongst the eigenstates of $H^0$ we have a set of eigenstates separated by a large gap from the others, which cannot be closed by the addition of $H'$,  and they represent the manifold we want to project in. These states are indexed by $m, m', m'', m''',\ldots$ while the rest of the eigenstates are indexed by $l, l', l'', l''', \dots$. These two form separate subspaces. We now want to find a Hamiltonian $H_{m m'}$ which incorporates the effects of $H'$ up to any desired order. We separate $H'$ into a diagonal part $H_1$ plus an off-diagonal part $H_2$ between these manifolds:
\begin{eqnarray}
&H'= H_1+ H_2 \nonumber \\ & (H_1)_{m m'}=\langle \psi_m |H'|\psi_{m'} \rangle;\;\; (H_1)_{l l'}=\langle \psi_l |H'|\psi_{l'} \rangle;\;\;  (H_2)_{m l}=\langle \psi_m |H'|\psi_{l} \rangle;\;\; (H_2)_{m m'}= (H_2)_{ll'}= (H_1)_{m l}=0
\end{eqnarray}
We also have
\beq
H|\psi_m\rangle= E_m| \psi_m\rangle,\;\; H|\psi_l\rangle= E_l| \psi_l\rangle
\eeq
 We look for a unitary transformation:
\beq
\tilde{H} = e^{-S} (H^0+ H') e^S
\eeq where $S(= - S^\dagger)$ has only matrix elements that are off-diagonal between the subspaces, i.e. $S_{ml}=0$. The unitary transformation is chosen such that the \emph{off-diagonal} part of $\tilde{H}$ is zero to the desired order ($H_{ml}=0$). Since we know $S, H_2$ are off diagonal and $H_1$ is diagonal, we find that $S$ can be obtained from the condition

\beq
\tilde{H}_{\text{``off-diagonal''}} = \sum_{j=0}^\infty \frac{1}{(2j+1)!}[H^0+H_1, S]^{2j+1} + \sum_{j=0}^\infty \frac{1}{(2j)!}[H_2, S]^{2j} =0
\eeq
(the off-diagonal Hamiltonian is zero). Once $S$ is found, the diagonal Hamiltonian is:
\beq
\tilde{H}_{\text{``diagonal''}} = \sum_{j=0}^\infty \frac{1}{(2j)!}[H^0+H_1, S]^{2j} + \sum_{j=0}^\infty \frac{1}{(2j+1)!}[H_2, S]^{2j+1} 
\eeq where $[A,B]^j=[[[[ [A,B],B],B],\ldots],B]$ where the number of $B$'s is equal to $j$.We then parametrize $S= S_1+ S_2+ S_3 + \ldots$, where $S_n$ is order $n$ in perturbation theory, i.e. in $H'$ (or equivalently, in $H_1$ or $H_2$.

The terms up to order $4$ are derived in Winkler's book \cite{winklerbook}, and for our simplified problem, they are presented in the main text. We have numerically checked their correctness. We here also present the fifth order term: this term is tedious, but we use a particularly nice property of our eigenstate space that $ (H_1)_{m m'}=\langle \psi_m |H'|\psi_{m'} \rangle=0$, $E_m=0$ for $m=1,2$ property is true only for $H'= I_{6\times 6} \otimes \kk\cdot \sigma $ and for the zero energy eigenstates $\psi_m, \;\; m=1,2$ of $H_0$=$H_{\text{Hex}}(\kk=0, w_0, w_1 = \sqrt{1+ w_0^2}/2)$.  To the desired order, we find:

\begin{eqnarray}
&(S_1)_{ml}= \frac{H'_{ml}}{E_l},\;\;\; (S_1)_{lm}= -\frac{H'_{lm}}{E_l}\nonumber \\ 
&(S_2)_{ml}= - \sum_{l'} \frac{H'_{ml'} H'_{l'l}}{E_l E_{l'}},\;\;\; (S_2)_{lm}=  \sum_{l'} \frac{H'_{ll'} H'_{l'm}}{E_l E_{l'}} \nonumber \\ 
&(S_3)_{ml}= \sum_{l', l''}\frac{H'_{ml'} H_{l'l''}H_{l'l}}{E_l E_{l'} E_{l''}} -\frac{1}{3} \sum_{l' m'} H'_{ml'} H_{l'm'} H_{m' l}(\frac{3}{E_l^2 E_{l'}} + \frac{1}{E_{l'}^2 E_l})\nonumber \\ 
&(S_3)_{lm}= -\sum_{l', l''}\frac{H'_{ll'} H_{l'l''}H_{l''m}}{E_l E_{l'} E_{l''}} +\frac{1}{3} \sum_{l' m'} H'_{lm'} H_{m'l'} H_{l'm}(\frac{3}{E_l^2 E_{l'}} + \frac{1}{E_{l'}^2 E_l})
\end{eqnarray}
Due to our property  $ (H_1)_{m m'}=\langle \psi_m |H'|\psi_{m'} \rangle=0$, $E_m=0$  on the second magic manifold, we find that the fourth order $S_4$ is not needed in order to obtain the $5$'th order diagonal Hamiltonian, as terms in the expression of the Hamiltonian that contain it cancel. We find that the $5$'th order Hamiltonian is:
\begin{eqnarray}
&\tilde{H}_{\text{``diagonal''}}^{(5)} =- S_2 H^0 S_3 - S_3 H^0 S_2- S_1 H_1 S_3- S_3 H_1 S_1- S_2 H_2 S_2 \nonumber \\ &-\frac{1}{6} (S_1 H^0 S_1 S_2 S_1+ S_1 H^0 S_2 S_1^2+ S_1 H^0 S_1^2 S_2+ S_2 H^0 S_1^3 + S_1 H_1 S_1^3 \nonumber \\ & +S_1 S_2 S_1 H^0 S_1+ S_2 S_1^2 H^0 S_1 + S_1^2 S_2 H^0 S_1 + S_1^3 H^0 S_2 + S_1^3 H_1 S_1)\nonumber \\ &+ \frac{1}{6}(H_2 S_2 S_1^2+ H_2 S_1^2 S_2 + H_2 S_1 S_2 S_1+ 3 (S_1 S_2 H_2 S_1 + S_2 S_1 H_2 S_1+ S_1^2 H_2S_2)  \nonumber \\ & -( S_2 S_1^2 H_2 + S_1^2 S_2 H_2 + S_1 S_2 S_1 H_2 )- 3( S_1 H_2 S_1 S_2+ S_1 H_2S_2 S_1 + S_2 H_2 S_1^2))
\end{eqnarray}
The matrix elements of these terms give:
\begin{eqnarray}
&\frac{1}{6}(H_2 S_2 S_1^2+ H_2 S_1^2 S_2 + H_2 S_1 S_2 S_1 -( S_2 S_1^2 H_2 + S_1^2 S_2 H_2 + S_1 S_2 S_1 H_2 ))_{mm'} \nonumber \\ &=-\frac{1}{6}  \sum_{l,l',l''} \sum_{m''}\frac{ (H'_{ml} H'_{ll'} H_{l'm''} H_{m'' l''} H_{l'' m'}   + H_{m'l''} H_{l'' m''} H_{m'' l'} H_{l'l} H_{lm}} {E_l E_{l'} E_{l''}}(\frac{1}{E_l}+ \frac{1}{E_{l'}} + \frac{1}{E_{l''}})
\end{eqnarray}

\begin{eqnarray}
&\frac{1}{6}(3 (S_1 S_2 H_2 S_1 + S_2 S_1 H_2 S_1+ S_1^2 H_2S_2) -3( S_1 H_2 S_1 S_2+ S_1 H_2S_2 S_1 + S_2 H_2 S_1^2))_{mm'} \nonumber \\ &=-\frac{1}{2}  \sum_{l,l',l''} \sum_{m''}\frac{ (H'_{ml} H'_{ll'} H_{l'm''} H_{m'' l''} H_{l'' m'}   + H_{m'l''} H_{l'' m''} H_{m'' l'} H_{l'l} H_{lm}} {E_l E_{l'} E_{l''}}(\frac{1}{E_l}+ \frac{1}{E_{l'}} + \frac{1}{E_{l''}})
\end{eqnarray}

\begin{eqnarray}
&-\frac{1}{6} (S_1 H^0 S_1 S_2 S_1+ S_1 H^0 S_2 S_1^2+ S_1 H^0 S_1^2 S_2+ S_2 H^0 S_1^3 + S_1 H_1 S_1^3+ \nonumber \\ &+S_1 S_2 S_1 H^0 S_1 + S_2 S_1^2 H^0 S_1 + S_1^2 S_2 H^0 S_1 + S_1^3 H^0 S_2 + S_1^3 H_1 S_1)  \nonumber \\ &=\frac{1}{6}  \sum_{l,l',l''} \sum_{m''}\frac{ (H'_{ml} H'_{ll'} H'_{l'm''} H'_{m'' l''} H'_{l'' m'}   + H'_{m'l''} H'_{l'' m''} H'_{m'' l'} H'_{l'l} H'_{lm}} {E_l E_{l'} E_{l''}}(\frac{1}{E_l}+ \frac{1}{E_{l'}} + \frac{1}{E_{l''}})
\nonumber 
\end{eqnarray}

\begin{eqnarray}
(- S_2 H^0 S_3 - S_3 H^0 S_2- S_1 H_1 S_3- S_3 H_1 S_1- S_2 H_2 S_2 )_{mm'} = \sum_{l,l',l'', l'''} \frac{H'_{ml}H'_{ll'}H'_{l'l''}H_{l''l'''} H_{l'''m'}}{E_l E_{l'} E_{l''} E_{l'''}}
\end{eqnarray}
Hence 
\begin{eqnarray}
&\tilde{H}_{\text{``diagonal''}}^{(5)} =   \sum_{l,l',l'', l'''} \frac{H'_{ml}H'_{ll'}H'_{l'l''}H_{l''l'''} H_{l'''m'}}{E_l E_{l'} E_{l''} E_{l'''}} \nonumber\\ &-\frac{1}{2}  \sum_{l,l',l''} \sum_{m''}\frac{ (H'_{ml} H'_{ll'} H_{l'm''} H_{m'' l''} H_{l'' m'}   + H_{m'l''} H_{l'' m''} H_{m'' l'} H_{l'l} H_{lm}} {E_l E_{l'} E_{l''}}(\frac{1}{E_l}+ \frac{1}{E_{l'}} + \frac{1}{E_{l''}}) 
\end{eqnarray}

\subsection{Calculations of the Hamiltonian Matrix Elements When First Order Vanishes}\label{HamiltonianMatrixElements1}

Here we calculate explicitly the perturbations of $H_{\text{perturb}} (\mathbf{k}, w_0)=I_{6\times 6} \otimes \mathbf{k} \cdot \vec{\sigma}$ in Eq. (\ref{eq-Hpert_k_w0}) up to the fifth order.

\subsubsection{First order}

The first order perturbation can be easily seen to be zero:
\beq
H^{(1)}_{m m'}  (\mathbf{k}, w_0) =\langle \psi_m | H_{\text{perturb}} (\mathbf{k}, w_0)  |\psi_{m'} \rangle =0\ .
\eeq

\subsubsection{Second Order}

\begin{eqnarray}
&H^{(2)}_{m m'}  (\mathbf{k}, w_0) =-\sum_{l=3\ldots 12}\frac{1}{E_l} \langle \psi_m | H_{\text{perturb}} (\mathbf{k}, w_0)  |\psi_{l} \rangle \langle \psi_l | H_{\text{perturb}} (\mathbf{k}, w_0)  |\psi_{m'} \rangle  \nonumber \\ &=\boxed{ -\frac{4 w_0^2 \left(k_x^2+k_y^2\right)}{3 \sqrt{w_0^2+1} \left(3 w_0^2-1\right)} (\sigma_y+ \sqrt{3}\sigma_x)}
\end{eqnarray}

\subsubsection{Third Order}

\begin{eqnarray}
&H^{(3)}_{m m'}  (\mathbf{k}, w_0) =\sum_{l,l' =3\ldots 12}\frac{1}{E_l E_{l'}} \langle \psi_m | H_{\text{perturb}} (\mathbf{k}, w_0)  |\psi_{l} \rangle \langle \psi_l | H_{\text{perturb}} (\mathbf{k}, w_0)  |\psi_{l'} \rangle \langle \psi_{l'}  | H_{\text{perturb}} (\mathbf{k}, w_0)  |\psi_{m'} \rangle  \nonumber \\ &=\boxed{  \frac{4 k_x w_0 \left(w_0^2-3\right) \left(k_x^2-3 k_y^2\right)}{9 \left(1-3 w_0^2\right)^2 \sqrt{w_0^2+1}} \sigma_0} \end{eqnarray}

\subsubsection{Fourth Order}

For the fourth order, there are two terms:

First, 
\begin{eqnarray}
&H^{(4_1)}_{m m'}  (\mathbf{k}, w_0) =\nonumber \\ &-\sum_{l,l',l'' =3\ldots 12}\frac{1}{E_l E_{l'}E_{l''}} \langle \psi_m | H_{\text{perturb}} (\mathbf{k}, w_0)  |\psi_{l} \rangle \langle \psi_l | H_{\text{perturb}} (\mathbf{k}, w_0)  |\psi_{l'} \rangle \langle \psi_{l'}  | H_{\text{perturb}} (\mathbf{k}, w_0)  |\psi_{l''} \rangle \langle \psi_{l''} | H_{\text{perturb}} (\mathbf{k}, w_0)  |\psi_{m'} \rangle  \nonumber \\ &= \boxed{ \frac{8 w_0^2 \left(w_0^4+16 w_0^2-9\right) \left(k_x^2+k_y^2\right)^2}{27 \left(w_0^2+1\right)^{3/2} \left(3 w_0^2-1\right)^3}    (\sigma_y+ \sqrt{3}\sigma_x) }\end{eqnarray}

Second,
\begin{eqnarray}
&H^{(4_2)}_{m m'}  (\mathbf{k}, w_0) =\nonumber \\ &\sum_{l,l' =3\ldots 12} \sum_{m''=1,2} \nonumber \\ &\frac{1}{E_l E_{l'}}(\frac{1}{E_l } +\frac{1}{ E_{l'}}) \langle \psi_m | H_{\text{perturb}} (\mathbf{k}, w_0)  |\psi_{l} \rangle \langle \psi_l | H_{\text{perturb}} (\mathbf{k}, w_0)  |\psi_{m''} \rangle \langle \psi_{m''}  | H_{\text{perturb}} (\mathbf{k}, w_0)  |\psi_{l'} \rangle \langle \psi_{l'} | H_{\text{perturb}} (\mathbf{k}, w_0)  |\psi_{m'} \rangle  \nonumber \\ &= \boxed{ \frac{16 w_0^2 \left(17 w_0^2+9\right) \left(k_x^2+k_y^2\right)^2}{27 \sqrt{w_0^2+1} \left(3 w_0^2-1\right)^3} (\sigma_y+ \sqrt{3}\sigma_x)}\end{eqnarray}

Notice that so far, the eigenstates are not $\kk$-dependent, they are just the eigenstates of $(\sigma_y+ \sqrt{3}\sigma_x)$.

\subsubsection{Fifth  Order}
The fifth order perturbation theory is not available in any book. Hence we derived it in App.~\ref{fifthorderperturbation}, for the special case for which the manifold $m$ of states we project in has the first order Hamiltonian $H^{(1)}_{m m'}  (\mathbf{k}, w_0) =0 $ and for which its energies are $E_m=0$.

The fifth order also has two terms, just like the $4$-th order (See App.~\ref{fifthorderperturbation}). We find:
\begin{eqnarray}
&  \sum_{l,l',l'', l'''} \frac{H'_{ml}H'_{ll'}H'_{l'l''}H_{l''l'''} H_{l'''m'}}{E_l E_{l'} E_{l''} E_{l'''}}  \nonumber \\ &=\boxed{\frac{32 k_x \left(w_0^2-3\right)^2 \left(2 w_0^2-1\right) w_0 \left(k_x^2-3 k_y^2\right) \left(k_x^2+k_y^2\right)}{81 \left(w_0^2+1\right)^{3/2} \left(3 w_0^2-1\right)^4} \sigma_0}
\end{eqnarray}

and
\begin{eqnarray}
&-\frac{1}{2}  \sum_{l,l',l''} \sum_{m''}\frac{ (H'_{ml} H'_{ll'} H_{l'm''} H_{m'' l''} H_{l'' m'}   + H_{m'l''} H_{l'' m''} H_{m'' l'} H_{l'l} H_{lm}} {E_l E_{l'} E_{l''}}(\frac{1}{E_l}+ \frac{1}{E_{l'}} + \frac{1}{E_{l''}})  \nonumber \\ & = \boxed{-\frac{16 k_x \left(11 w_0^4-94 w_0^2-9\right) w_0 \left(k_x^2-3 k_y^2\right) \left(k_x^2+k_y^2\right)}{27 \left(\sqrt{w_0^2+1} \left(3 w_0^2-1\right)^4\right)}\sigma_0}
\end{eqnarray}

We can clearly see the structure of the order $n$ Hamiltonian, as a perturbation in $1/(3w_0^2-1)^{n-1}$, with symmetry-preserving functions of $\kk$. The full $2$-band approximation to the Hexagon Hamiltonian is, up to $5$-th order, is
\begin{eqnarray}
&H_{\text{2band}}^{\text{Hex}}(\mathbf{k}, w_0, w_1= \frac{\sqrt{1+w_0^2}}{2})=\frac{4 w_0^2}{3\sqrt{w_0^2+1} \left(3 w_0^2-1\right)} \left[-1+\frac{2(35 w_0^4+68 w_0^2+9 )\left(k_x^2+k_y^2\right)}{9 \left(w_0^2+1\right) \left(3 w_0^2-1\right)^2} \right]  \left(k_x^2+k_y^2\right)(\sigma_y+ \sqrt{3}\sigma_x)\nonumber \\ &+
\frac{4  w_0}{9\sqrt{w_0^2+1}  \left(1-3 w_0^2\right)^2 }[  \left(w_0^2-3\right)    -\frac{4  \left(29 w_0^6-223 w_0^4-357 w_0^2-9\right) }{9 \left(1-3 w_0^2\right)^2 \left(w_0^2+1\right)} \left(k_x^2+k_y^2\right)]k_x \left(k_x^2-3 k_y^2\right)  \sigma_0
\end{eqnarray}
better expressed as:
\beq
\boxed{H_{\text{2band}}^{\text{Hex}}(\mathbf{k}, w_0, w_1= \frac{\sqrt{1+w_0^2}}{2})= d_0(\mathbf{k} ,w_0) \sigma_0+ d_1(\mathbf{k} ,w_0)(\sigma_y+\sqrt{3} \sigma_x) }\label{2bandhexagonprojected2}
\eeq where
\beq
d_0(\mathbf{k} ,w_0) =\frac{4  w_0}{9\sqrt{w_0^2+1}  \left(1-3 w_0^2\right)^2 }[  \left(w_0^2-3\right)    -\frac{4  \left(29 w_0^6-223 w_0^4-357 w_0^2-9\right) }{9 \left(1-3 w_0^2\right)^2 \left(w_0^2+1\right)} \left(k_x^2+k_y^2\right)]k_x \left(k_x^2-3 k_y^2\right) 
\eeq and
\beq
d_1(\mathbf{k} ,w_0)=\frac{\sqrt{1+w_0^2}}{2})=\frac{4 w_0^2}{3\sqrt{w_0^2+1} \left(3 w_0^2-1\right)} \left[-1+\frac{2(35 w_0^4+68 w_0^2+9 )\left(k_x^2+k_y^2\right)}{9 \left(w_0^2+1\right) \left(3 w_0^2-1\right)^2} \right]  \left(k_x^2+k_y^2\right)
\eeq

\subsection{Calculations of the Hamiltonian Matrix Elements When First Order Does Not Vanish}\label{HamiltonianMatrixElements2}

We take the unperturbed Hamiltonian to be $H_{\text{Hex}}(\kk=0, w_0, w_1 = \sqrt{1+ w_0^2}/2)$ (the Hexagon model on the second magic manifold) in Eq.~\ref{HexagonModel1}. For this Hamiltonian we are able to obtain \emph{all the eigenstates analytically} in App.~\ref{Eigenstatesolutiononthesecondmagicmanifold}. The perturbation Hamiltonian, away the second magic manifold is 
\begin{eqnarray} &H_{\text{perturb}}(\kk, w_0, w_1) = \nonumber \\ &  H_{\text{Hex}}(\kk, w_0, w_1) - H_{\text{Hex}}(\kk=0, w_0,  w_1=\frac{ \sqrt{1+w_0^2}}{2})= I_{6\times 6} \otimes \kk \cdot \vec{\sigma} + H_{\text{Hex}}(\kk=0, 0,  w_1-\frac{ \sqrt{1+w_0^2}}{2})
\end{eqnarray}

\subsubsection{First Order}

\beq
H^{(1)}_{m m'}  (\kk, w_0,w_1) =\langle \psi_m | H_{\text{perturb}} (\kk, w_0, w_1)  |\psi_{m'} \rangle =\boxed{( \frac{\sqrt{w_0^2+1}}{2}-w_1 )(\sigma_y+ \sqrt{3}\sigma_x)}
\eeq
Hence there is now a linear term in the Hamiltonian. Because of this, many other terms in the further degree perturbation theory become nonzero. 

\subsubsection{Second Order}

\begin{eqnarray}
&H^{(2)}_{m m'}  (\kk, w_0, w_1) =-\sum_{l=3\ldots 12}\frac{1}{E_l} \langle \psi_m | H_{\text{perturb}} (\kk, w_0)  |\psi_{l} \rangle \langle \psi_l | H_{\text{perturb}} (\kk, w_0)  |\psi_{m'} \rangle  \nonumber \\ &=\boxed{ -\frac{4 w_0^2 \left(k_x^2+k_y^2\right)}{3 \sqrt{w_0^2+1} \left(3 w_0^2-1\right)} (\sigma_y+ \sqrt{3}\sigma_x)}
\end{eqnarray}
The second order perturbation theory is unchanged!

\subsubsection{Third Order}

There are now two third order terms, as the first order perturbation terms do not vanish. First:
\begin{eqnarray}
&H^{(3_1)}_{m m'}  (\kk, w_0,w_1) =\sum_{l,l' =3\ldots 12}\frac{1}{E_l E_{l'}} \langle \psi_m | H_{\text{perturb}} (\kk, w_0)  |\psi_{l} \rangle \langle \psi_l | H_{\text{perturb}} (\kk, w_0)  |\psi_{l'} \rangle \langle \psi_{l'}  | H_{\text{perturb}} (\kk, w_0)  |\psi_{m'} \rangle  \nonumber \\ &=\boxed{  \frac{4 k_x w_0 \left(w_0^2-3\right) \left(k_x^2-3 k_y^2\right)}{9 \left(1-3 w_0^2\right)^2 \sqrt{w_0^2+1}} \sigma_0  -\frac{8 w_0^2 \left(k_x^2+k_y^2\right) \left(\sqrt{w_0^2+1}-2 w_1\right)}{9 \left(1-3 w_0^2\right)^2}(\sigma_y+ \sqrt{3} \sigma_x)} \end{eqnarray}

Second:
\begin{eqnarray}
&H^{(3_2)}_{m m'}  (\kk, w_0, w_1)= -\frac{1}{2} \sum_{l=3\ldots 12}\sum_{m''=1,2}\frac{  \langle \psi_m | H_{\text{perturb}} (\kk, w_0, w_1)  |\psi_{l} \rangle  \langle \psi_l | H_{\text{perturb}} (\kk, w_0,w_1)  |\psi_{m''}  \langle \psi_{m''} | H_{\text{perturb}} (\kk, w_0,w_1)  |\psi_{m'}  \rangle + h.c.  }{E_l^2} \nonumber \\ &= -\frac{2 \left(17 w_0^2+9\right) \left(k_x^2+k_y^2\right) \left(\sqrt{w_0^2+1}-2 w_1\right)}{9 \left(1-3 w_0^2\right)^2} (\sigma_y+ \sqrt{3} \sigma_x)
\end{eqnarray} (where $h.c.$ is the Hermitian conjugate)

The total third order Hamiltonian then reads: 
\beq
\boxed{ \frac{4 k_x w_0 \left(w_0^2-3\right) \left(k_x^2-3 k_y^2\right)}{9 \left(1-3 w_0^2\right)^2 \sqrt{w_0^2+1}} \sigma_0 -\frac{2 \left(7 w_0^2+3\right) \left(k_x^2+k_y^2\right) \left(\sqrt{w_0^2+1}-2 w_1\right)}{3 \left(1-3 w_0^2\right)^2}(\sigma_y+ \sqrt{3}\sigma_x) }
\eeq

\subsubsection{Fourth Order}

For the fourth order, there are now  four terms:

First, 
\begin{eqnarray}
&H^{(4_1)}_{m m'}  (\kk, w_0, w_1) =\nonumber \\ &-\sum_{l,l',l'' =3\ldots 12}\frac{1}{E_l E_{l'}E_{l''}} \langle \psi_m | H_{\text{perturb}} (\kk, w_0,w_1)  |\psi_{l} \rangle \langle \psi_l | H_{\text{perturb}} (\kk, w_0)  |\psi_{l'} \rangle \nonumber \\ & \times\langle \psi_{l'}  | H_{\text{perturb}} (\kk, w_0,w_1)  |\psi_{l''} \rangle \langle \psi_{l''} | H_{\text{perturb}} (\kk, w_0,w_1)  |\psi_{m'} \rangle  \nonumber \\ &= \boxed{ \frac{8  w_0 \left(7 w_0^2+3\right) k_x \left(k_x^2-3 k_y^2\right) \left(2 w_1-\sqrt{w_0^2+1}\right)}{27 \left(3 w_0^2-1\right)^3} \sigma_0 }\nonumber \\ &+ \boxed{ \frac{4 w_0^2 \left(k_x^2+k_y^2\right) \left(2 \left(w_0^4+16 w_0^2-9\right)  \left(k_x^2+k_y^2\right)+\left(w_0^2+1\right) \left(5 w_0^2-7\right)  \left(2 w_1-\sqrt{w_0^2+1}\right)^2\right)}{27 \left(w_0^2+1\right)^{3/2} \left(3 w_0^2-1\right)^3}(\sigma_y + \sqrt{3} \sigma_x)}\end{eqnarray}

Second,
\begin{eqnarray}
&H^{(4_2)}_{m m'}  (\kk, w_0) =\nonumber \\ &=\sum_{l,l' =3\ldots 12} \sum_{m''=1,2} \nonumber \\ &\frac{1}{E_l E_{l'}}(\frac{1}{E_l } +\frac{1}{ E_{l'}}) \langle \psi_m | H_{\text{perturb}} (\kk, w_0, w_1)  |\psi_{l} \rangle \langle \psi_l | H_{\text{perturb}} (\kk, w_0)  |\psi_{m''} \rangle \nonumber \\ & \times \langle \psi_{m''}  | H_{\text{perturb}} (\kk, w_0, w_1)  |\psi_{l'} \rangle \langle \psi_{l'} | H_{\text{perturb}} (\kk, w_0, w_1)  |\psi_{m'} \rangle = \nonumber \\ &= \boxed{   \frac{16 w_0^2 \left(17 w_0^2+9\right) \left(k_x^2+k_y^2\right)^2}{27 \sqrt{w_0^2+1} \left(3 w_0^2-1\right)^3} (\sigma_y+ \sqrt{3} \sigma_x) }\end{eqnarray}

Third, we have, adopting the notation $ \langle \psi_m | H_{\text{perturb}} (\kk, w_0)  |\psi_{l} \rangle = H'_{ml}$, etc:

\begin{eqnarray}
&H^{(4_3)}_{m m'}  (\kk, w_0,w_1)=-\frac{1}{2} \sum_{l,m'',m'''} \frac{1}{E_l^3} (H'_{mm''} H'_{m''m'''} H'_{m''' l} H_{l m'} + H'_{ml} H'_{lm''} H'_{m'' m'''} H'_{m''' m'})  \nonumber \\ &=\boxed{ -\frac{8 w_0^2 \left(35 w_0^2+23\right) \left(k_x^2+k_y^2\right) \left(\sqrt{w_0^2+1}-2 w_1\right)^2}{27 \sqrt{w_0^2+1} \left(3 w_0^2-1\right)^3} (\sigma_y+ \sqrt{3} \sigma_x)}
\end{eqnarray}

\begin{eqnarray}
&H^{(4_4)}_{m m'}  (\kk, w_0,w_1)=\frac{1}{2} \sum_{l,l',m''} \frac{1}{E_l E_{l'}} \left(\frac{1}{E_l}+ \frac{1}{E_{l'}}\right)  (H'_{ml} H'_{ll'} H'_{l' m''} H_{m'' m'} + H'_{mm''} H'_{m''l} H'_{ll'} H'_{l' m'})  \nonumber \\ &= \boxed{ \frac{32 k_x w_0 \left(w_0^2-15\right) \left(k_x^2-3 k_y^2\right) \left(\sqrt{w_0^2+1}-2 w_1\right)}{27 \left(3 w_0^2-1\right)^3}\sigma_0} \nonumber \\ &+\boxed{ \frac{4 \left(25 w_0^4+28 w_0^2+27\right) \left(k_x^2+k_y^2\right) \left(\sqrt{w_0^2+1}-2 w_1\right)^2 }{27 \left(1-3 w_0^2\right)^3 \sqrt{ w_0^2+1}}(\sigma_y+ \sqrt{3}\sigma_x) }
\end{eqnarray}

The full $4$'th order Hamiltonian reads

\begin{eqnarray}
&\boxed{-\frac{8 k_x w_0 \left(w_0^2+21\right) \left(k_x^2-3 k_y^2\right) \left(\sqrt{w_0^2+1}-2 w_1\right)}{9 \left(3 w_0^2-1\right)^3}\sigma_0+} \nonumber \\ & \boxed{\frac{4 \left(k_x^2+k_y^2\right) \left(2 w_0^2 \left(35 w_0^4+68 w_0^2+9\right) \left(k_x^2+k_y^2\right)-9 \left(w_0^2+1\right) \left(10 w_0^4+9 w_0^2+3\right) \left(2 w_1-\sqrt{w_0^2+1}\right)^2\right)}{27 \left(w_0^2+1\right)^{3/2} \left(3 w_0^2-1\right)^3} (\sigma_y+ \sqrt{3}\sigma_x)}
\end{eqnarray}

%\textbf{Zhida and Biao, I would appreciate some checking of these, although I understand it might not be possible for all}

If $w_1=\frac{\sqrt{1+w_0^2}}{2}$, then the expressions reduce to our previous Hamiltonian. We can label the two band Hamiltonian as:
\beq
\boxed{H_{\text{2band}}^{\text{Hex}}(\kk, w_0, w_1)= d_0(\kk ,w_0,w_1) \sigma_0+ d_1(\kk ,w_0,w_1)(\sigma_y+\sqrt{3} \sigma_x) } \label{hexagon2bandprojection1}
\eeq
where
\begin{eqnarray}
& \boxed{ d_0(\kk, w_0,w_1)=\frac{4 k_x w_0 \left(w_0^2-3\right) \left(k_x^2-3 k_y^2\right)}{9 \left(1-3 w_0^2\right)^2 \sqrt{w_0^2+1}}-\frac{8 k_x w_0 \left(w_0^2+21\right) \left(k_x^2-3 k_y^2\right) \left(\sqrt{w_0^2+1}-2 w_1\right)}{9 \left(3 w_0^2-1\right)^3}\sigma_0  }\label{hexagon2bandprojection1-d0}
\end{eqnarray}
and 
\begin{eqnarray}
&\boxed{  d_1(\kk, w_0,w_1)=( \frac{\sqrt{w_0^2+1}}{2}-w_1 )-\frac{4 w_0^2 \left(k_x^2+k_y^2\right)}{3 \sqrt{w_0^2+1} \left(3 w_0^2-1\right)}-\frac{2 \left(7 w_0^2+3\right) \left(k_x^2+k_y^2\right) \left(\sqrt{w_0^2+1}-2 w_1\right)}{3 \left(1-3 w_0^2\right)^2}+}\nonumber \\ &  \boxed{ \frac{4 \left(k_x^2+k_y^2\right) \left(2 w_0^2 \left(35 w_0^4+68 w_0^2+9\right) \left(k_x^2+k_y^2\right)-9 \left(w_0^2+1\right) \left(10 w_0^4+9 w_0^2+3\right) \left(2 w_1-\sqrt{w_0^2+1}\right)^2\right)}{27 \left(w_0^2+1\right)^{3/2} \left(3 w_0^2-1\right)^3} } \label{hexagon2bandprojection1-d1}
\end{eqnarray}
where the perturbation is made on the zero energy eigenstates of $H_{\text{Hex}}(\kk=0, w_0,  w_1=\frac{ \sqrt{1+w_0^2}}{2})$.

Notice that so far, remarkably the eigenstates are not $\kk$-dependent, they are just the eigenstates of $(\sigma_y+ \sqrt{3}\sigma_x)$. We did not obtain the fifth order for this Hamiltonian: due to the fact that the first order Hamiltonian does not cancel, this is not easy to do. 

\subsection{Calculations of the B1 shell first order perturbation}\label{app-B1shellapprox}
We now compute the shell $B1$ perturbation Hamiltonian: %$ - H_{A1,B1}  H_{kB1}^{-1}H_{A1,B1}^\dagger$:

\beq
\begin{split}
&- H_{A1,B1}  H_{kB1}^{-1}H_{A1,B1}^\dagger(\kk, w_0, w_1)=  \\ &\qquad\qquad  -\begin{pmatrix}
\frac{T_1 (k-2q_1)\cdot \sigma T_1}{ |k-2q_1|^2}&0 &0  & 0 & 0 &0 \\
0 & \frac{T_3 (k+2q_3)\cdot \sigma T_3}{| k+2q_3|^2} &0  & 0&  0&0 \\
  0 &0  &\frac{T_2 (k-2q_2)\cdot \sigma T_2}{|k-2q_2|^2} &0  &0& 0  \\
   0 & 0  &0&\frac{T_1 (k+2q_1)\cdot \sigma T_1}{|k+2q_1|^2} &  0 &0  \\
0 & 0  &0 &0 &\frac{T_3 (k-2q_3)\cdot \sigma T_3}{ |k-2q_3|^2} &0 \\
0 & 0  & 0 & 0 & 0 &\frac{T_2 (k+2q_2)\cdot \sigma T_2}{|k+2q_2|^2}  
\end{pmatrix}. \label{BShell1Perturbation1}
\end{split}
\eeq

We now compute the perturbation Hamiltonian:

\begin{eqnarray}
&H^{(B1)} (\kk, w_0,w_1) =\langle \psi_m | - H_{A1,B1}  H_{kB1}^{-1}H_{A1,B1}^\dagger(\kk, w_0, w_1) |\psi_{m'} \rangle \nonumber \\ &=\frac{1}{ \prod_{i=1,2,3} |k-2q_i|^2 |k+2q_i|^2 }( \widetilde{d}_0(\kk, w_0, w_1) \sigma_0+\widetilde{d}_x(\kk, w_0, w_1) \sigma_x + \widetilde{d}_y(\kk, w_0, w_1)\sigma_y+ \widetilde{d}_z(\kk, w_0, w_1)\sigma_z) \label{seq-B1shellprojected1}
\end{eqnarray}

where
\begin{eqnarray}\label{eq-B1-d0}
&\widetilde{d}_0(\kk, w_0, w_1)= \frac{4 k_x \left(k_x^2-3 k_y^2\right) \left(k_x^2+k_y^2+4\right) \left(\left(k_x^2+k_y^2\right)^2-4 \left(k_x^2+k_y^2\right)+16\right) w_0  \left(\sqrt{w_0^2+1}+w_1+1\right) \left(\sqrt{w_0^2+1}+w_1-1\right)}{\sqrt{w_0^2+1}} 
\end{eqnarray}

\begin{eqnarray}
& \widetilde{d}_z(\kk, w_0, w_1)= \frac{64 k_x k_y  \left(k_x^2-3 k_y^2\right) \left(3 k_x^2-k_y^2\right) w_0 \left(\left(\sqrt{w_0^2+1} w_1+w_0^2\right)^2+w_0^2\right)}{\left(w_0^2+1\right)^{3/2}}\end{eqnarray}

\begin{eqnarray}
&\widetilde{d}_x(\kk, w_0, w_1)=\nonumber \\ & -\frac{16 \left(\sqrt{3} \sqrt{w_0^2+1} \left(-\left(k_y \left(3 k_x^2-k_y^2\right)\right)^2+\left(k_x \left(k_x^2-3 k_y^2\right)\right)^2+64\right) \left(w_0^2-w_1^2\right)-2 k_x k_y \left(k_x^2-3 k_y^2\right) \left(3 k_x^2-k_y^2\right) \left(\sqrt{w_0^2+1} w_1^2+2 w_0^2 w_1+\sqrt{w_0^2+1} w_0^2\right)\right)}{w_0^2+1}\end{eqnarray}

\begin{eqnarray} \label{eq-B1-dy}
&\widetilde{d}_y(\kk, w_0, w_1)=\nonumber \\ & -\frac{16 \left(\sqrt{w_0^2+1} \left(-k_y^2 \left(3 k_x^2-k_y^2\right)^2+k_x^2 \left(k_x^2-3 k_y^2\right)^2+64\right) \left(w_0^2-w_1^2\right)+2 \sqrt{3} k_x k_y \left(3 k_x^2-k_y^2\right) \left(k_x^2-3 k_y^2\right) \left(\sqrt{w_0^2+1} w_1^2+2 w_0^2 w_1+\sqrt{w_0^2+1} w_0^2\right)\right)}{w_0^2+1}\end{eqnarray}

This gives the first order term of $H_{\text{Approx}1}(\kk)$ projected into the zero energy bands in the Hexagon model on the second magic manifold.

\subsection{Exact eigenvalues of the 1 shell model at $\Gamma_M$ point}\label{app-1shellexact}

At $w_0=0$, we find the $\Gamma_M$ point eigenenergies of the Hamiltonian $H_{\text{Approx} 1}= H_{kA1}+ H_{A1,A1} - H_{A1,B1}  H_{kB1}^{-1}H_{A1,B1}^\dagger$ in Eq.~\ref{Happrox1} to be the following:
\begin{eqnarray}
&\frac{ \left(-w_1^2+4 w_1-2\right)}{2}, \frac{\left(w_1^2-4 w_1+2\right)}{2} , \frac{\left(-w_1^2+2 w_1-2\right)}{2} , \frac{\left(-w_1^2+2 w_1-2\right)}{2} , \frac{ \left(w_1^2-2 w_1+2\right)}{2}, \frac{\left(w_1^2-2 w_1+2\right)}{2} , \nonumber \\
&\frac{\left(-w_1^2-2 w_1-2\right)}{2} , \frac{\left(-w_1^2-2 w_1-2\right)}{2} , \frac{\left(w_1^2+2 w_1+2\right)}{2} , \frac{\left(w_1^2+2 w_1+2\right)}{2} , \frac{ \left(-w_1^2-4 w_1-2\right)}{2}, \frac{ \left(w_1^2+4 w_1+2\right)}{2}
\end{eqnarray}
One sees the $\Gamma_M$ point has zero bandwidth at $w_1 =2- \sqrt{2}$, the same as that of the zero-bandwidth manifold $w_1=2 \sqrt{w_0^2+1}-\sqrt{3 w_0^2+2}= 2- \sqrt{2}$ in Eq. (\ref{eq-zerobandwidth}) for the two band model at $w_0=0$.

Furthermore, in the chiral limit $w_0=0$, the value $w_1=2 \sqrt{w_0^2+1}-\sqrt{3 w_0^2+2}= 2- \sqrt{2}$ for which the bandwidth is $0$ in our two band model is in fact \emph{exact} for the no-approximation Hamiltonian of the $n=1$ shell Hamiltonian (of $A1,B1$ subshells). We find its eigenvalues at $\Gamma_M$ to be 
\begin{eqnarray}
&\frac{\left(-\sqrt{5 w_1^2-6 w_1+9}-w_1-1\right)}{2} ,\frac{\left(-\sqrt{5 w_1^2-6 w_1+9}-w_1-1\right)}{2},  \frac{\left(-\sqrt{5 w_1^2-6 w_1+9}+w_1+1\right)}{2} , \frac{ \left(-\sqrt{5 w_1^2-6 w_1+9}+w_1+1\right)}{2}, \nonumber \\ 
&\frac{\left(\sqrt{5 w_1^2-6 w_1+9}-w_1-1\right)}{2} , \frac{\left(\sqrt{5 w_1^2-6 w_1+9}-w_1-1\right)}{2} , \frac{\left(\sqrt{5 w_1^2-6 w_1+9}+w_1+1\right)}{2} , \frac{\left(\sqrt{5 w_1^2-6 w_1+9}+w_1+1\right)}{2} , 
\nonumber \\ 
&\frac{ \left(-\sqrt{5 w_1^2+6 w_1+9}-w_1+1\right)}{2}, 
\frac{ \left(-\sqrt{5 w_1^2+6 w_1+9}-w_1+1\right)}{2}, \frac{\left(-\sqrt{5 w_1^2+6 w_1+9}+w_1-1\right)}{2} , \frac{ \left(-\sqrt{5 w_1^2+6 w_1+9}+w_1-1\right)}{2}, \nonumber \\ 
&\frac{ \left(\sqrt{5 w_1^2+6 w_1+9}-w_1+1\right)}{2}, \frac{\left(\sqrt{5 w_1^2+6 w_1+9}-w_1+1\right)}{2} , 
\frac{ \left(\sqrt{5 w_1^2+6 w_1+9}+w_1-1\right)}{2}, \frac{\left(\sqrt{5 w_1^2+6 w_1+9}+w_1-1\right)}{2} , \nonumber \\ 
&\frac{\left(-\sqrt{8 w_1^2-12 w_1+9}-2 w_1-1\right)}{2} , \frac{ \left(-\sqrt{8 w_1^2-12 w_1+9}+2 w_1+1\right)}{2},
\frac{\left(\sqrt{8 w_1^2-12 w_1+9}-2 w_1-1\right)}{2} , \frac{\left(\sqrt{8 w_1^2-12 w_1+9}+2 w_1+1\right)}{2} ,  \nonumber \\ 
&\frac{\left(-\sqrt{8 w_1^2+12 w_1+9}-2 w_1+1\right)}{2} , \frac{\left(-\sqrt{8 w_1^2+12 w_1+9}+2 w_1-1\right)}{2} , \frac{\left(\sqrt{8 w_1^2+12 w_1+9}-2 w_1+1\right)}{2} , \frac{ \left(\sqrt{8 w_1^2+12 w_1+9}+2 w_1-1\right)}{2}.
\end{eqnarray}
Therefore, we see that the active bands have zero bandwidth at $w_0=0, w_1= 2- \sqrt{2}$ in the $n=1$ shell model.

\end{document}